# ESO-*Athena* Synergy White Paper

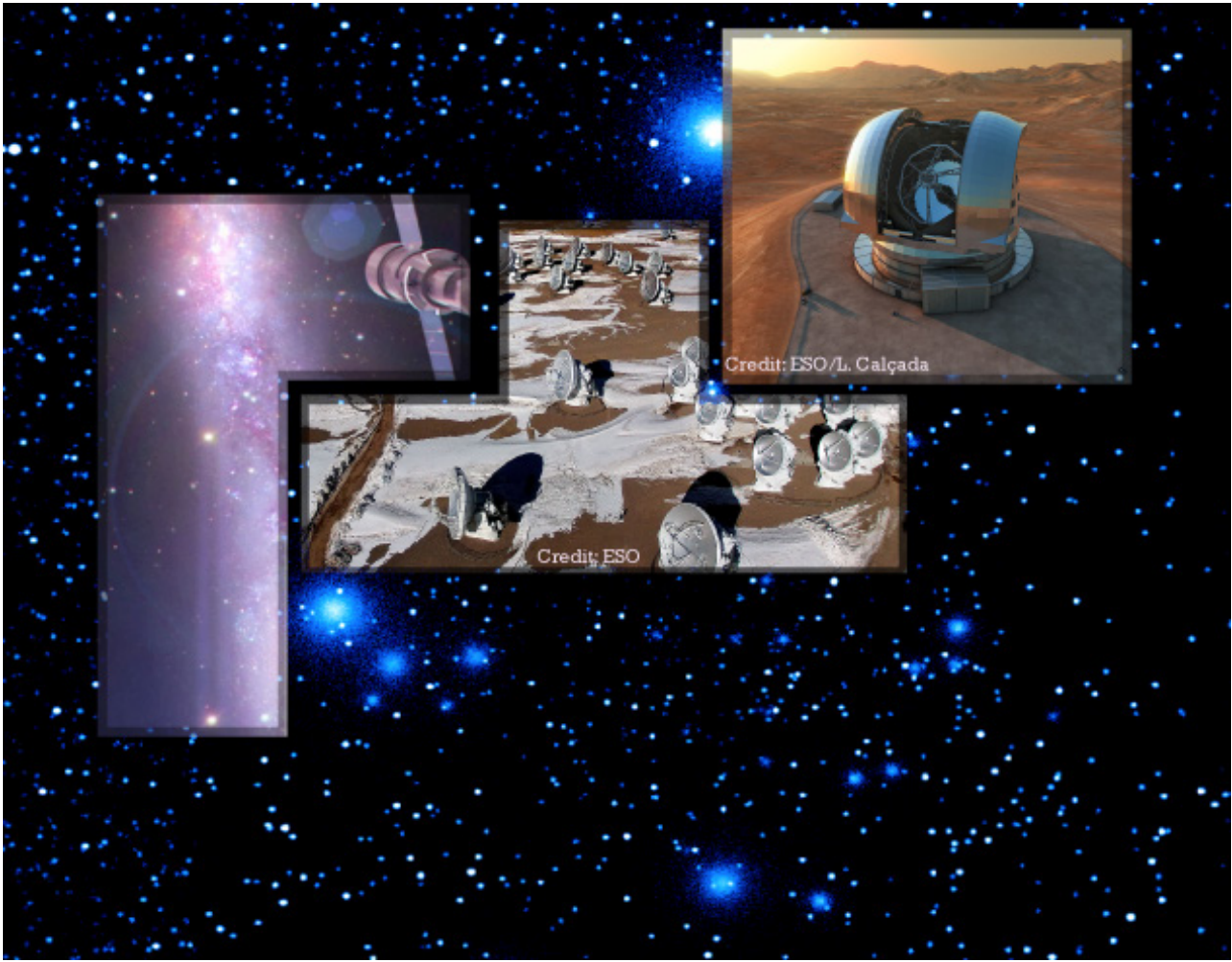

# Authorship

## Authors

P. Padovani, ESO
F. Combes, Observatoire de Paris, France
M. Díaz Trigo, ESO
S. Ettori, INAF-OABO, Italy
E. Hatziminaoglou, ESO
P. Jonker, SRON, The Netherlands
M. Salvato, MPE, Germany
S. Viti, UCL, UK

## Contributors

C. Adami, LAM, France
J. Aird, Durham University, UK
D. Alexander, Durham University, UK
P. Casella, INAF-OAR, Italy
C. Ceccarelli, IPAG, France
E. Churazov, MPA, Germany
M. Cirasuolo, ESO
E. Daddi, CEA Saclay, France
A. Edge, Durham University, UK
C. Feruglio, Scuola Normale Superiore, Italy
V. Mainieri, ESO
S. Markoff, University of Amsterdam, The Netherlands
A. Merloni, MPE, Germany
F. Nicastro, INAF-OAR, Italy
P. O'Brien, University of Leicester, UK
L. Oskinova, Postdam University, Germany
F. Panessa, INAF-IAPS, Italy
E. Pointecouteau, IRAP, France
A. Rau, MPE, Germany
J. Robrade, Hamburger Sternwarte, Germany
J. Schaye, Leiden Observatory, The Netherlands
F. Stoehr, ESO
L. Testi, ESO
F. Tombesi, NASA/GSFC & University of Maryland, USA

# Contents

# 1. Introduction

*Athena* (Advanced Telescope for High ENergy Astrophysics) is the X-ray observatory mission selected by the European Space Agency (ESA), within its Cosmic Vision 2015-2025 programme, to address the Hot and Energetic Universe scientific theme. It is the second L(large)-class mission within that programme and it is due for launch in 2028.

*Athena* will consist of a single large-aperture grazing-incidence X-ray telescope with 12m focal length and 5´´ half energy width on-axis angular resolution. The focal plane contains two instruments. One is the Wide Field Imager (WFI) providing sensitive wide field of view (FoV) imaging and spectroscopy, as well as bright source observation capability. The other one is the X-ray Integral Field Unit (X-IFU) delivering spatially resolved high-resolution X-ray spectroscopy over a limited FoV (see *Athena*: The Advanced Telescope for High-Energy Astrophysics, Mission Proposal, available at http://sci.esa.int/cosmic-vision/54013-athena-the-advanced-telescope-for-high-energy-astrophysics/). *Athena* will address a variety of science topics, ranging from the study of groups and clusters of galaxies, to the physics of accretion on compact objects, to finding the earliest accreting supermassive black holes (SMBHs) and investigating how they influence the evolution of galaxies and clusters through feedback processes. *Athena* will also have a fast target of opportunity observational capability (responding within four hours), enabling studies of $\gamma$-ray bursts (GRBs) and other transient phenomena. As an observatory, *Athena* will offer vital information on high-energy phenomena on all classes of astrophysical objects, from Solar System bodies to the most distant objects known (Nandra et al., 2013).

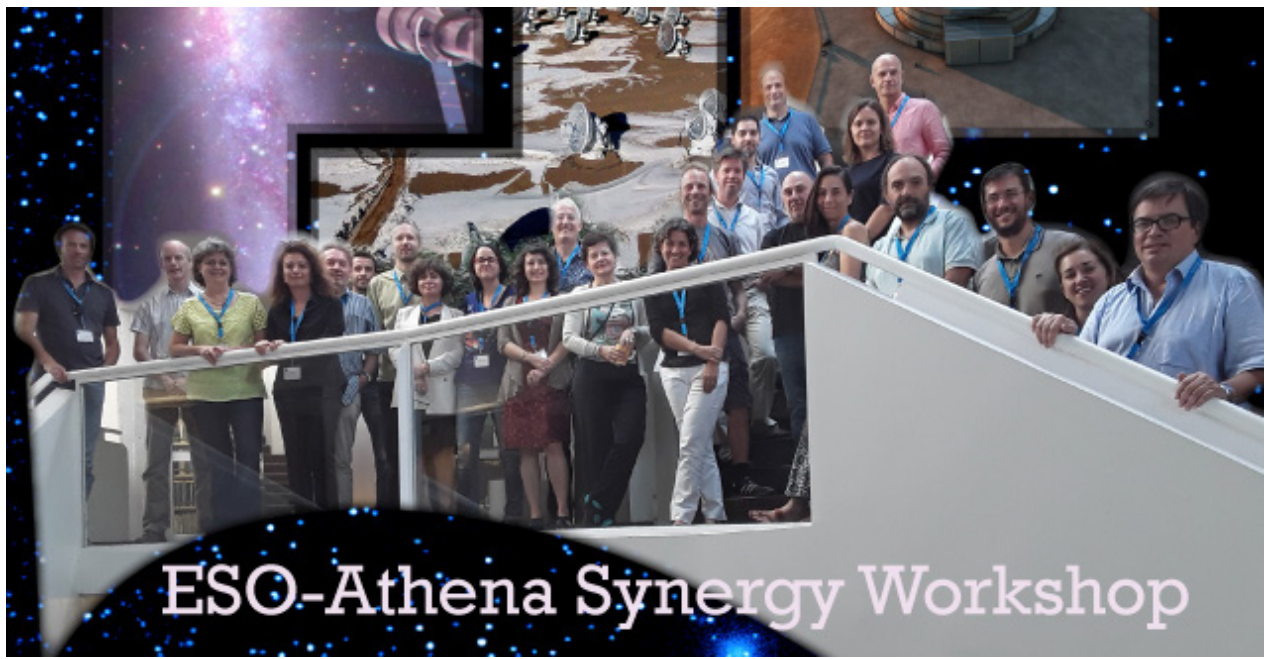

ESA has established the *Athena* Science Study Team (ASST) to provide guidance on all scientific aspects during the Assessment Phase for the *Athena* mission. One of the ASST's tasks is to identify and elaborate synergies with various astronomical facilities, which will be available in the late 2020s.

The ESO-*Athena* Synergy Team (EAST) has been tasked by the ASST and ESO to single out the potential scientific synergies between *Athena* and optical/near-infrared (NIR) and sub-mm ground based facilities, in particular those of ESO (i.e., the Very Large Telescope [VLT] and the Extremely Large Telescope [ELT], the Atacama Large Millimeter/submillimeter Array [ALMA] and the Atacama Pathfinder Experiment [APEX]), by producing a White Paper to identify and develop the:

- needs to access ESO ground-based facilities to achieve the formulated *Athena* science objectives;

- needs to access *Athena* to achieve the formulated science objectives of ESO facilities contemporary to *Athena*;

- science areas where the synergetic use of *Athena* and ESO facilities in the late 2020s will result in scientific added value.

Community input to the process happened primarily via a dedicated ESO - *Athena* Synergy Workshop (by EAST invitation only) that took place on Sept. 14 – 16, 2016 at ESO, Garching (https://indico.ifca.es/indico/event/247/). EAST worked mostly via email interactions. This White Paper presents the results of the EAST's work, sorted by synergy area.

Disclaimer: The views expressed in this paper are purely those of the individual members of EAST.

# Acknowledgements

We thank A.C. Fabian and R.K. Smith for reviewing the document, ESO for hosting the Workshop, ESO and ESA for funding EAST activities, and all participants in the EAST Workshop for enlightening discussions.

# 2. The Hot Universe: Early groups and clusters and their evolution

By 2030, major astrophysical questions related to the formation and evolution of galaxy clusters and groups, the largest collapsed structures in the Universe with a significant mass fraction of baryons in form of stars in galaxies and hot plasma, will still remain. Some of the most fundamental questions, such as "What is the interplay of galaxies, SMBHs, and intergalactic gas?", "What are the processes that drive the evolution of the chemical enrichment of the X-ray emitting gas?", "How and when did the first collapsed groups form?" will need dedicated efforts from the future astronomical facilities to be answered.

Among these facilities, the next large X-ray observatory *Athena* and ESO instruments on the VLT, ELT, as well as ALMA will be able to jointly address some specific related issues, such as: to obtain redshifts for the collapsed structures out to the time of their formation; to resolve the Sunyaev-Zel'dovich (SZ) effect both spatially and associated with high – z clusters; to study the physics (spectral energy distribution [SED], age, metallicity, star formation rate [SFR]) of the member galaxies; to evaluate the cold gas content of the cluster galaxies and their efficiency in forming stars. More details on this science case can be found in Ettori et al. (2013) and Pointecouteau et al. (2013).

## 2.1. Spectroscopic surveys of galaxies in clusters

To define membership in galaxy structure and unveil galaxies dynamics, an accuracy on redshift ~ 0.001 is needed. This is beyond what photometric redshifts can do in practice and therefore medium resolution spectroscopy is needed, assisted by high spatial resolution to resolve the member galaxies of the structures at higher redshifts.

Assuming a WMAP9 cosmology (Hinshaw et al., 2013) and a Tinker et al. (2008) cluster mass function, about 8 clusters more massive than $10^{14}$ $M_\odot$ are predicted per deg$^2$ at z < 2. Since *Athena* is expected to cover 50 deg$^2$ per year, in 10 years about 4000 massive clusters will be observed at z < 2. To provide a complete survey of massive clusters ( ~ $10^{14}$ $M_\odot$) up to redshift of 2, the following combination of ESO instruments could be considered: VISTA/4MOST (Section 13.3) for nearby structures, then VLT/MOONS (Section 13.1) up to z ~ 1.75, and ELT/multi-object spectrograph (MOS: Section 13.2) for the most distant ones (see also Figure 1).

- At z < 1.3, we expect about 3800 clusters. VISTA/4MOST can follow-up such targets, both in terms of collecting power and spectral coverage (see Figure 1). An area of 500 deg$^2$ can also be covered with 125 exposures in a survey-mode. Such a strategy (given the VISTA/4MOST multiplex capability) would imply 30 clusters of galaxies per VISTA/4MOST field, leaving therefore 50 fibers per cluster. This number is large enough to match the number of available and measurable galaxies per cluster at these redshifts. Note that we are not limited by fiber-fiber minimum possible distance: 50 galaxy targets in such clusters imply a distance between two fibers 1.7 times larger than the 15´´ quoted in the ESO documentation. No other packing limitations are considered. Assuming 4 hours per exposure to reach R=22.5, this implies a total exposure time of 500 hours over 10 years.

- In the redshift range z = [1.3, 1.75], we need to use VLT/MOONS for magnitude limitation and spectral coverage reasons (the VLT/MOONS spectral coverage is redder than that of VISTA/4MOST). We expect 180 clusters in this redshift range. A pointed strategy with VLT/MOONS is then well suited. With an exposure time of 4 hours per pointing, this would amount to 700 hours over 10 years, albeit the spectroscopic confirmation of passive galaxies from absorption features could require longer integrations.

- A dozen galaxy structures more massive than $10^{14}$ $M_\odot$ and over 500 deg$^2$ are expected in the redshift range z = [1.75, 2]. This redshift range could be studied by VLT/MOONS, but the galaxy targets of the considered clusters are too faint to be efficiently measured by this spectrograph. We therefore need the ELT collecting power. The expected performances of the ELT/MOS can measure these candidates within less than 50 hours.

A complete survey of massive galaxy clusters at z < 2 detected by *Athena* over a 10 year period and over a 500 deg$^2$ area would then require a possible but significant effort using ESO telescopes. This would imply for each semester, over a 10 year period (20 semesters, assuming a standard 15% overhead) 30 hours of VISTA/4MOST time, 80 hours of VLT/MOONS time, and half a night of ELT/MOS.

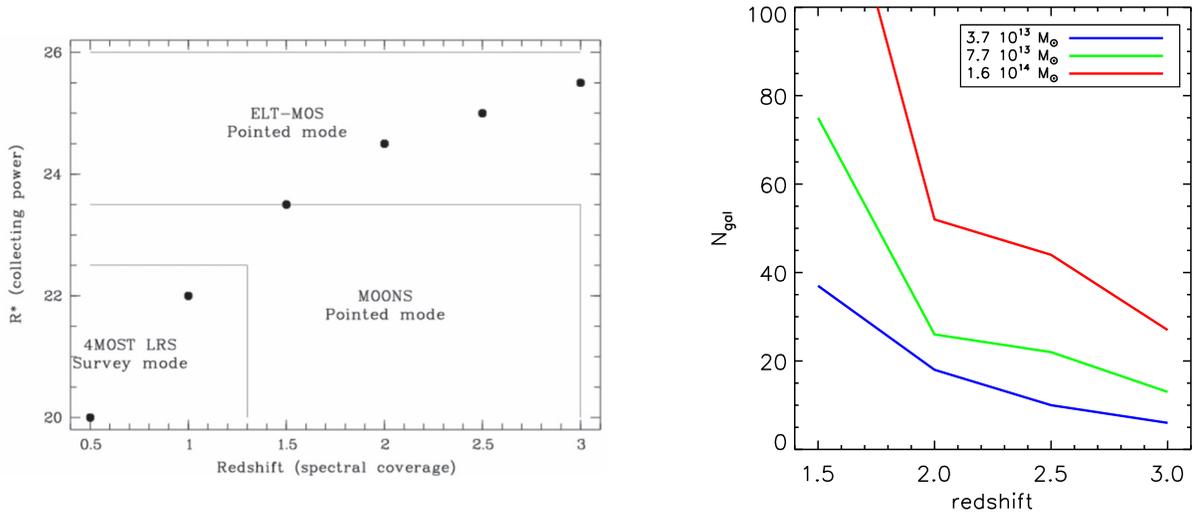

Figure 1: Left: The sensitivity of ESO instruments for multi-object spectroscopy is here compared with R* band magnitude limit associated with cluster members (from Ilbert et al., 2006, and assuming a Schechter galaxy luminosity function with α = -1.3). The absolute magnitude is here translated into apparent magnitudes up to z = 1.5; for higher redshifts, one magnitude increase between z = 1.5 and 2, and then a 0.5 magnitude increase between z = 2 and 2.5 and between z = 2.5 and 3 are considered. Right: Number of cluster galaxy members (measurable with the ELT/MOS instrument and about 4 hours of exposure) as a function of redshift. A mass-richness relation from Bauer et al. (2012) is used with $M_{200} = (1/1.029) \, e^A \, (N_{200}/20)^B \, 10^{14} \, M_\odot/h$, with A = 0.5, B = 1.05, where $M_{200}$ is the cluster gravitational mass within $R_{200}$, the radius within which the mean cluster density is 200 times the critical density at the cluster's redshift, and $N_{200}$ that represents the number of galaxies in the Colour Magnitude Red Sequence brighter than 0.4L*. The lines refer to different cluster masses, $M_{200}$, as indicated.

## 2.2. Proto-cluster formation

The first collapsed structures at z ≈ 2.5, and above, with masses a few times $10^{13}$ $M_\odot$ are expected to be hosted in a single dark matter halo with associated thermal X-ray emission, and lots of ongoing action, from strong star formation activity to quenching, from active galactic nuclei (AGN) feedback to the morphological transformation of galaxies and the formation of ellipticals. At those redshifts, AGN activity becomes much more prominent, following in parallel the rise of SFR and gas content in the Universe. In these conditions, interactions between the intracluster medium (ICM) and the cluster's environment through the inflow of pristine cold gas, the outflows originated from stellar and AGN's winds, and the deposition of warm plasma will shape the baryon distribution, the metal content and the energy budget of the structures.

Present observations at $z \approx 2$ already show evidence of a collapsed, cluster-sized halo and exhibit a high concentration of quiescent galaxies in the core, with a well-defined red sequence (Newman et al., 2014). Some of them still contain a substantial number of star-forming galaxies, and a few show clear evidence of enhanced star formation in cluster members with respect to field galaxies. The accretion of pristine gas from the surroundings can be responsible for the lower metallicities of the star-forming galaxies in clusters, as indicated by a significantly lower [N II]/Hα ratio (and a higher observed equivalent width [EW] of Hα; see Figure 2) associated with the cluster galaxy stacked sample with respect to the mass-matched field sample. In this context, to clearly establish a link between the structure formation and the process of galaxy assembly, high spatial resolution (diffraction limited) spectroscopy of the emission lines to trace EW enhancement and metallicity deficit will be required.

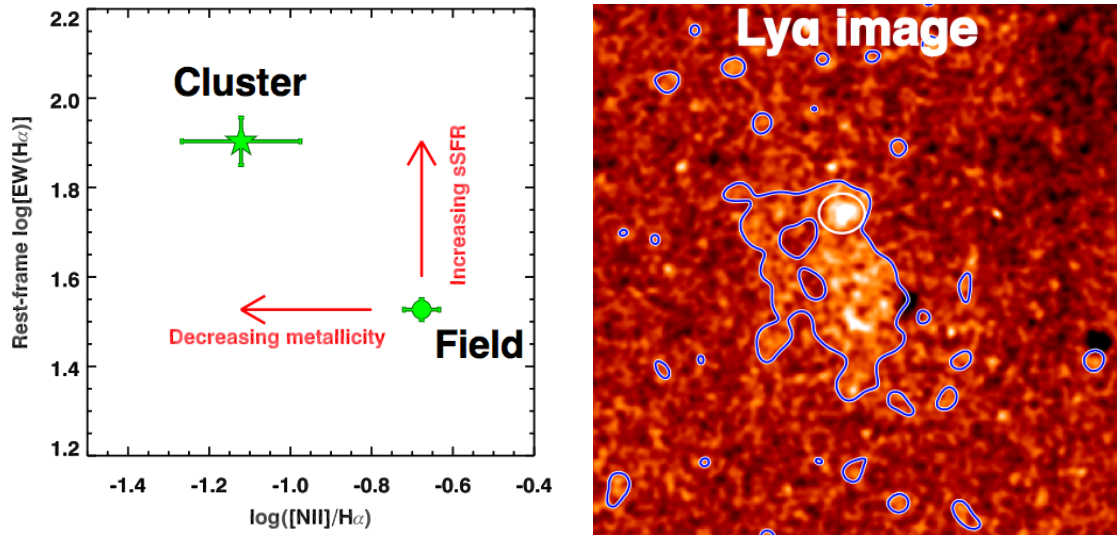

Figure 2: CLJ1149+0856 at z = 1.99 is one of the most distant galaxy cluster with an associated X-ray emission (Gobat et al., 2011). Constraints on its [N II] and Hα measurements (left; from Valentino et al., 2015) and on its Lyα nebula centred on the cluster's core (right; from Valentino et al., 2016) provide robust proxies of the ongoing star formation activity and deposition of cold gas, respectively.

At $z \geq 2.5$, objects like CLJ1001 (Wang et al., 2016) could represent an important transition phase between proto-clusters and mature clusters, showing evidence of massive star-forming galaxies in the core of a collapsed cluster-sized halo that still have to undergo the quenching process that will transform them in the quiescent population typical of such massive structure at later times.

At these redshifts, *Athena* is expected to detect about twenty galaxy groups with $M_{500} > 5 \times 10^{13} M_\odot$ in ten years (Pointecouteau et al., 2013). This would require about 80 hours (in total over 10 years, corresponding to an average of half a night per semester) of dedicated ELT/MOS follow-up.

*Athena* will be then able to constrain the global physical properties of the thermal emission from local clusters out to redshift 2, and beyond, down to the group regime where the ICM is more affected by any input of non-gravitational energy, as the feedback provided by stars and AGN. ESO synergies will be exploited to:

1. measure redshifts/membership (with VISTA/4MOST, VLT/MOONS, ELT/MOS);
2. obtain imaging, galaxy SEDs, and stellar masses (with ELT/MICADO and HARMONI: Section 13.1);
3. probe the galaxy morphology and mergers (with ELT/METIS: Section 13.1);
4. trace metal enrichment and gas in galaxies (with ELT/HARMONI, ELT-MOS and VLT/ MOONS);
5. recover the SFRs (with ELT/METIS, HARMONI);

6. trace any outflow of molecular and ionized gas (with ELT/HARMONI);

7. study the diffuse Lyα gas and its connection to the surrounding environment (with either an IFU-like instrument [ELT/HARMONI] or narrow band filters [ELT/MICADO]);

8. analyze the shock tracers (with ELT/METIS, HARMONI).

## 2.3. The SZ effect in galaxy clusters

The hot gas in groups and clusters of galaxies can be probed through two independent observables: Bremsstrahlung emission at X-ray wavelengths and the SZ effect due the inverse Compton scattering of cosmic microwave background (CMB) photons by the electrons of the hot intra-cluster gas. They have different dependencies on the ICM physical properties which makes them complementary: the X-ray brightness is a function mainly of the gas density, i.e., $I_x \propto n_e^2$, whereas the SZ effect is proportional to the integrated thermal pressure along the line of sight, i.e., $I_{SZ} \propto P_{th} \propto kT \times n_e$.

In the last two decades, the SZ measurements have gone from detection to detailed mapping. From the first significant mm measurements (e.g., Church et al., 1997; Désert et al., 1998), the first multi-band analysis (Holzapfel et al., 1997; Lamarre et al., 1998) and the first maps (e.g., Grainge et al., 1996; Pointecouteau et al., 2001) to the era of catalogues of SZ detected clusters by large surveys like ACT (Hasselfield et al., 2013), SPT (Bleem et al., 2015) and *Planck* (*Planck* Collaboration et al., 2016). These surveys have also pushed the detection of the SZ effect towards the distant Universe, beyond $z = 1$ (Bayliss et al., 2014). Moreover, their combination with X-rays in several analyses have led to new constraints on scaling and structural properties of clusters (*Planck* Collaboration et al., 2013a; Tchernin et al., 2016; Adam et al., 2016).

The need is now for higher spatial resolution. The new generation of SZ instruments mounted at the focus of large single dish telescopes, i.e., NIKA (Monfardini et al., 2010) at IRAM 30m and MUSTANG (Dicker et al., 2006) at GBT 100m, provide resolutions of 18′′ and 9′′ full-width at half maximum (FWHM) respectively. They probe spatial scales similar to X-rays and produce optimized joint SZ and X-ray imaging (Adam et al., 2015; Ruppin et al., 2016), which allow the constrain of the physical properties of the ICM without relying on X-ray spectroscopy (Pointecouteau et al., 2002; Kitayama et al., 2004). These SZ machines are already upgraded to their second generation.

### 2.3.1. Past and current SZ observations with ESO facilities

The SZ instrument on the APEX telescope has only imaged a few tens of clusters at 150 GHz at arcminute scales (Schwan et al., 2012), which had already been observed in the past at higher resolution (10-20′′ FWHM: Pointecouteau et al., 2001; Komatsu et al., 2001). The use of LABOCA (Nord et al., 2009; Lindner et al., 2015) at 345 GHz with about 20′′ resolution is the actual added value. It demonstrated the feasibility of imaging clusters through the SZ increment from the ground. Although such observations are challenging as they are easily hampered by contamination by large scale dust emission and sub-mm sources, they bring more leverage in the extraction of the intra-cluster gas properties through multi-band analysis.

The first results of SZ observations with ALMA have just been published demonstrating the feasibility to map the SZ signal with this large interferometer. Kitayama et al. (2016) have produced a mosaic image of the SZ effect towards the intermediate redshift cluster RX J1347-1145 in Band 3 making use of 10×7 m and 40×12 m antennas in their most compact configuration. The reconstructed map at 94 GHz extend over 1.5′ recovering scales up to ~ 40′′ with a resolution of 5′′ FWHM (the highest achieved to date in imaging the SZ effect). It is fully consistent with previous observations at slightly lower spatial resolution but across wider spatial scales by large single dish telescopes (Pointecouteau et al., 2001; Komatsu et al., 2001; Kitayama et al., 2004; Korngut et al., 2011; Adam et al., 2014, see Figure 3).

ALMA produced another interesting showcase result with a tentative detection in the El Gordo cluster at

z = 0.89 (Menanteau et al., 2012) of the SZ signature by the shock seen in radio and X-rays (Basu et al., 2016). Band 3 was used with 35 × 12 m antennas in their most compact configuration achieving a resolution of 3.5´´ on the "dirty" map. This result provides a hint of the potential capabilities of ALMA in resolving features such as shocks, edges, bubbles as predicted, e.g., by Yamada et al. (2012).

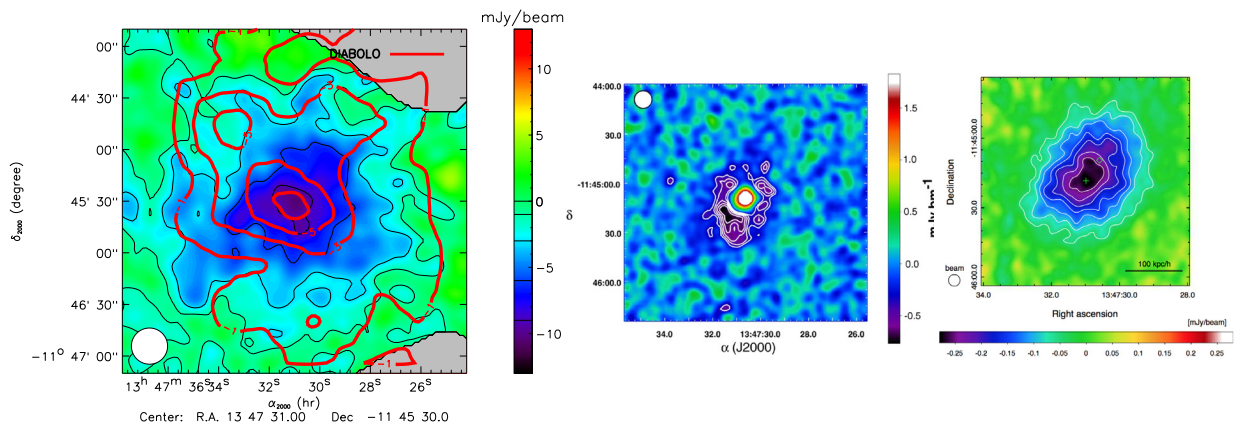

Figure 3: SZ maps of RX J1347-1145 (z = 0.45) scaled according to their spatial extent, from left to right: NIKA at IRAM 30m, MUSTANG at GBT 100m and ALMA. The resolution is 18´´, 9´´, and 5´´ FWHM respectively. Figures are reproduced from Adam et al. (2014, Figure 12), Korngut et al. (2011, Figure 12) and Kitayama et al. (2016, Figure 54). Note that the MUSTANG map is not point source subtracted whereas the two others are.

### 2.3.2. Synergies between *Athena* and SZ observations with ESO facilities

SZ observations will be extremely complementary to *Athena* observations in order to probe the hot Universe side of the science case. SZ imaging of the ICM pressure distribution will nicely complement the constraints expected from *Athena* on the bulk motions and turbulence of the ICM (Ettori et al., 2013) from the statistics of gas density fluctuations (Zhuravleva et al., 2014) and direct measurements of line shift and broadening from high-resolution X-ray spectroscopy by *Athena* X-IFU. Statistics of the pressure inhomogeneities also directly relate to the power spectrum of the gas velocity field (Khatri & Gaspari, 2016) and the kinetic SZ effect (a doppler shift of the CMB spectrum due to the peculiar motions of the gas) can be used to infer the velocity field of internal bulk motions (Adam et al., 2016). The mapping of pressure inhomogeneities via the SZ effect at the centre of clusters with a high spatial resolution, together with X-ray spatially resolved spectroscopy, will further probe features such as bubbles, ripples, etc. signing the feedback of the central AGN on the ICM, leading to a better understanding of its physics and its global energetics (Croston et al., 2013). SZ observations will also bring a complementary view to characterize the physics of shocks, fronts, edges and other features linked to the dynamical evolution of massive halos (Yamada et al., 2012).

Joint X-ray and SZ imaging allows the recovery of the physical characteristics of the ICM, compensating for low X-ray statistics preventing detailed spectroscopy (Kitayama et al., 2004; Ruppin et al., 2016). At the present time this concerns the study of the evolution of massive cluster properties in the redshift range of 0.5 < z < 1.0. In the era of *Athena*, we shall investigate the volume far beyond z = 1 out to the epoch of groups and clusters formation to quantify the evolution of scaling and structural properties of clusters such as their distribution in entropy (Pointecouteau et al., 2013). Furthermore, in the hunt for the first groups trapping hot gas in their already formed potential wells (Pointecouteau et al., 2013), follow-up detection in SZ will be an undeniable confirmation of the presence of a hot gaseous atmosphere and important for their first physical characterization (Mantz et al., 2014).

### 2.3.3. Perspectives on the SZ observations with ESO facilities

ESO will certainly play a major role in synergy with *Athena* for the observation of the hot Universe combining SZ and X-ray observations. The ability to map the SZ signal over scales of a few arcminutes with ALMA has now been demonstrated. The prominent perspective is for the very distant Universe where the Maximum Recoverable Scale (MRS) in a given band shall be larger than the extended SZ emission associated with high-z galaxy clusters. At present Band 3 provides the best choice with joint observations with the 7m and 12m arrays in their most compact configuration. The upcoming Band 1 will improve this ability still providing an exceptional angular resolution although operating at ∼ 30 GHz where the amplitude of the SZ effect is largely dimmed and might be subject to contamination by diffuse synchrotron emission. The perspective of Band 2 and its coupling with Band 3 will be a real opportunity, providing a wide photometric band to cover the SZ decrement. An angular resolution of the order of 5´´ and an MRS reaching the arcminute and beyond should allow detailed mapping of the gas pressure in features such as shocks, edges bubbles, rims, etc.

Interferometric measurements remain nonetheless sub-optimal for imaging extended emission. In this view, ESO is missing a large single dish (sub-)mm telescope (i.e., 40 - 50m) equipped with a wide FoV ( ∼10´) photometric camera operating at least at ∼2 mm, and, optimally, simultaneously at ∼1 mm and ∼850 $\mu$m. Such a configuration would provide spatial resolution ∼10´´ and ∼5´´ at 2mm and 850 $\mu$m respectively and strong leverage on the determination of the intra-cluster pressure through multi-band constraints. This would allow one to address all the aforementioned science cases, complementing the capabilities of ALMA.

The combination of such ESO facilities with *Athena* observations will lead to a deeper insight of the physics governing the ICM and of the link between its various components: SMBHs, galaxies, and the hot surrounding gas.

# 3. The Hot Universe: Physics of the ICM

The hot ICM is an important repository of baryons. In addition to providing a powerful tool to find clusters via X-rays, it is of great interest for plasma physics, beyond standard hydrodynamics or magnetohydrodynamics (MHD), and is necessary to understand AGN feedback. While the X-ray domain is the major window for ICM studies there are many areas where a combination of X-ray, (sub-)mm and optical/infrared (IR) data provides crucial insights into ICM physics. An incomplete list of such areas is outlined below.

## 3.1. Combining X-ray and SZ data

While we already know the global structure of many galaxy clusters, i.e., their radial profiles in density, temperature and pressure, the important information on the ICM physics is "hidden" in the small-scale deviations of the ICM thermodynamic properties from the mean values. A powerful diagnostic of such deviations is possible using *Athena* together with ALMA and APEX. ESO facilities can provide accurate line-of-sight-integrated electron pressure maps with high angular resolution, while *Athena* offers projected emissivity (square of the gas density) and temperature maps. Such studies are just beginning to appear (Basu et al., 2016; Kitayama et al., 2016), but they will be greatly boosted by *Athena*. A combination of X-ray and SZ data is especially useful to constrain the nature of ICM perturbations, e.g., distinguishing sound waves from internal waves. Such analysis can be performed either on resolved angular scales or even on unresolved scales, since the overall normalizations of the SZ and X-ray signals depend on the correlation between density and temperature fluctuations (Khedekar et al., 2013). For the former case ALMA resolution is useful, while for the latter case single-dish facilities like APEX are more suitable.

Studies of variations of the electron pressure across the shocks and the cold fronts (contact discontinuities) offer a unique opportunity to probe the contribution of non-thermal components to the total gas pressure and electron-ion equilibration times upstream the shocks. Equally attractive is the possibility to constrain the amount of thermal plasma in the bubbles of relativistic plasma inflated by AGN in the ICM, through joint X-ray and SZ analysis (Pfrommer et al., 2005; Prokhorov et al., 2010).

A powerful diagnostic is also possible from a combination of X-ray data on the gas line-of-sight velocities with data on the kinetic SZ effect, which probes peculiar velocities relative to the CMB frame.

## 3.2. Gas velocities

While measuring ICM velocities is one of the prime objectives of *Athena*, it is interesting to explore alternative possibilities of measuring velocities of X-ray emitting gas using ESO facilities from the ground. One such possibility is to use hyperfine splitting of the ground state of heavy element isotopes, whose nuclei have nonzero magnetic moment (e.g., Sunyaev & Churazov, 1984), although it might be difficult to detect these lines in emission (Chatzikos et al., 2014). The two most promising transitions are in hydrogen-like $^7$N (53 GHz) and $^{57}$Fe (97 GHz). The former transition is characteristic for $10^6$ K gas, while the latter is more important at $10^7$ K. The optical depth is $10^{-4}$ and the best hope is to detect such lines in absorption against very bright radio sources in the core of rich clusters.

Gas motions are especially important since their corresponding kinetic energy can make a non-negligible contribution to the total gas pressure. A possible way of measuring the pressure of the non-thermal component is by estimating the gravitating mass from the hydrostatic equilibrium equation (taking into account only thermal gas pressure $P = nkT$) and comparing it with the mass derived from stellar kinematics. The discrepancy between these two estimates serves as a proxy for non-thermal pressure. Since gas velocities can be measured by *Athena*, their contribution can be accounted for, leaving only magnetic fields and cosmic rays (CRs) as contributors to non-thermal pressure. With additional information on the magnetic fields and

## 3.3. Filaments of warm/cold gas and emission lines

The cores of many clusters host filaments of warm/cold gas with temperatures much lower than the hot ICM. Such filaments are (remarkably) powerful sources of molecular and atomic lines (e.g., H$\alpha$). The nature of the filaments and the source of energy that powers their emission have been a matter of debate for many decades. In the *Athena* band these filaments can reveal themselves as sources of $10^6$ - $10^7$ K gas emission surrounding the filaments and as low energy absorption features caused by weakly ionized gas. Measurements of the velocities of the cold (CO with ALMA, H$\alpha$ with VLT/X-Shooter to cover a range of redshifts) and hot gas (X-rays with X-IFU) in adjacent regions would clarify whether filaments and the ICM are moving together or not, placing constraints on the origin of filaments. One can also search in X-rays for the fluorescent lines of heavy elements, e.g., iron, arising when weakly ionized gas in filaments is exposed to the X-ray radiation of the hot gas (Churazov et al., 1998). The main goal of this exercise is the estimate of the total mass of cold gas (more precisely - the mass of heavy elements) which is insensitive to the properties of filaments as long as they are optically thin to X-rays. These results could be further compared with the molecular/atomic line measurements to understand the excitation mechanisms and overall energetics of the filaments.

The presence of bright emission lines like H$\alpha$ opens the possibility to get independent constraints on the ICM temperature and its optical depth for Thomson scattering by searching for a weak and diffuse emission in the form of very strongly broadened lines (Khedekar et al., 2014).

## 3.4. AGN feedback, inflows and outflows in the central region

While the importance of AGN feedback is well established in galaxy clusters and individual elliptical galaxies (e.g., Churazov et al., 2000; see also Fabian 2012 for a review), many important ingredients for a complete model are still missing. In particular, two aspects of AGN feedback could be studied with X-ray, optical and (sub-)mm data mass and energy budget of outflows and black hole (BH) feeding mechanisms. These issues still qualify for "ICM Physics", since the cooling instability and the transport coefficients might affect the behaviour of the gas. The first issue can be addressed by detailed mapping (spatially resolved and unresolved) of the velocity field of the inner few kpc region near the central AGN. Made across all energy bands, this will yield full information on the energy content and net flow rates of cold, warm and hot phases. The second issue requires information on the gas properties in the very central region ( ~100 pc and below) where the gravitational influence of the central BH becomes dominant, setting the rate of accretion. The potential of ESO and ALMA facilities, yielding the amount of cold gas and its velocity, coupled with X-ray data (density, temperature, and velocity field), has been already demonstrated for clusters and groups (e.g., McNamara et al., 2014; David et al., 2014; Russell et al., 2016). Especially interesting are detections of lines in absorption (against the bright central AGN), since they can help to determine the direction of the flows (Tremblay et al., 2016; see Figure 4).

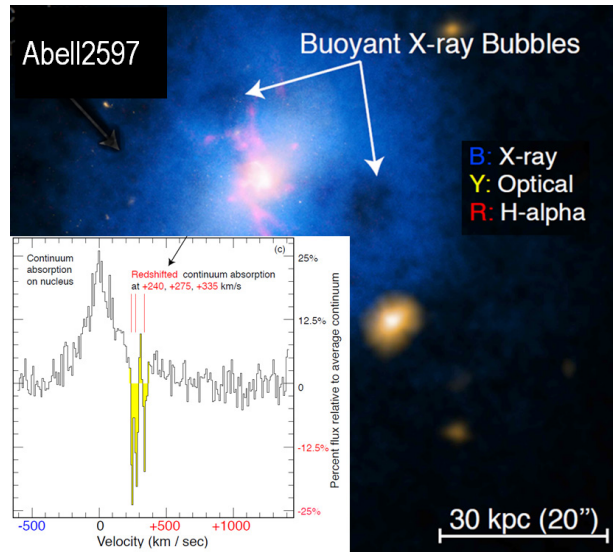

Figure 4: The hot X-ray gas is cooling into Hα and cold gas filaments in the galaxy cluster Abell2597, where the central AGN and its radio jet are excavating hot gas cavities, forming buoyant bubbles. The ALMA spectrum in the insert shows only redshifted absorption, in addition to CO(2-1) emission, indicating cold gas inflow, fueling the AGN (Tremblay et al., 2016).

## 3.5. Molecular gas around brightest cluster galaxies

Molecular gas plays a central role in the process of AGN feedback by providing the fuel required to generate energy close to the central BH (see also Section 9). In groups and clusters of galaxies, the effect of the AGN feedback on the gas in the ICM can be determined with reasonable accuracy from the cavities and shocks that the activity creates, and which are visible in X-ray imaging. The "calorimetry" that these observations provide allow us to calibrate the scatter in current AGN activity to the integrated activity over the last 1–100 Myr.

### 3.5.1. Key science questions

■ What fraction of the gas cools and fuels the central AGN? This can be addressed from a combination of X-ray observations to trace the amount of gas cooling from the initial temperature of the ICM, ALMA observations of cold molecular gas (largely through CO), VLT/MUSE observations of ionized gas, and VLT–ELT NIR and *JWST*[1] mid-IR (MIR) IFU observations of warm molecular gas. This comprehensive census of the total gas content of the system at $10 < T < 10^7$ K allows us to assess the balance between the gas cooling, the cold gas reservoir and the gas that goes into star formation, AGN accretion or is reheated. In any individual cluster this balance could be skewed by a cluster merger or powerful AGN outburst, but when calculated for a large sample should return a consistency between the gas cooling and gas consumption.

■ How much cooled gas is reheated or ejected? Through observations of the excitation and velocity of molecular and ionized gas, especially the ro-vibrational lines of molecular Hydrogen in the NIR and MIR, we can determine the location and likely heating rate of molecular gas. Future observations with *JWST* and ELT/HARMONI will be central to this work. The location of any gas heating is important as it may be affected by AGN jets and/or winds.

---

1 Although *JWST* (http://www.*JWST*.nasa.gov) does not fall within our remit, which is limited to ground-based facilities, we mention it a few times in this document when we discuss future observations relevant to some of the science topics discuss here.

- How do stars form from cooled gas? We have a number of different ways to determine the SFR of a brightest cluster galaxy (BCG), from UV, optical, MIR and far-IR bands. Recent work shows that these different tracers give consistent results and that these SFRs are consistent with a gas-star scaling relation different from the field. The role of the AGN in promoting star formation in the path of any jet or wind is a very open question and, given additional constraints on the energy and long-term orientation of jets provided by the X-ray observations in groups and clusters, mapping out the location of current and recently formed stars will allow us to draw very general conclusions about how prevalent "jet-induced" star formation is.

- How does cooled gas reach the central BH? The presence of a bright, unresolved core in the radio and sub-mm in many BCGs allows us to identify individual clouds along the line of sight to the centre of the galaxy in atomic Hydrogen (HI) and/or CO (David et al., 2014; Tremblay et al., 2016). The fraction of sightlines that show HI absorption is relatively high (>60%) in the bright sources observed to date (Hogan, 2014) so the density of gas clouds to the core is substantial. Creating a matched sample of sources with both HI and CO absorption will be transformational and provide information about the dynamics, excitation and chemistry of individual clouds close to the SMBH in the core of the BCG.

- How rapidly do BHs in BCGs grow? The advantage of selecting a sample of clusters from their extended X-ray emission is that it is a property that is completely independent of the instantaneous accretion rate of the BCG in each cluster. Making a selection on properties related to the BCG alone (e.g., radio power) will miss the least active systems that may dominate the population. From multiwavelength observations it is possible to infer the likely accretion rate of each BCG and determine the collective mass accretion rate of the ensemble of BHs. This allows us to constrain their growth rate and match that to the likely gas cooling rate of the surrounding cluster. One important aspect to this work is to determine the BH mass in a subset of this BCG sample through measurement of gas and/or stellar dynamics. The next generation of adaptive optics (AO)-assisted IFU observations will make this possible and *JWST* will extend the reach of these observations to higher redshift. Armed with these BH mass estimates then the accretion rates can be more accurately calibrated.

### 3.5.2. The role of ESO

There are several important sets of observations with the VLT, ALMA and *JWST* that can be made in the next decade that will define a key set of *Athena* observations to be performed to link gas cooling directly to the cooled gas traced at longer wavelengths.

Most of these observations can be made with existing or planned facilities:

- VLT/MUSE: optical IFU data to map ionized gas;

- VLT/SINFONI – KMOS: NIR IFU data to map warm molecular Hydrogen;

- ALMA: CO maps of cold molecular gas;

- ELT/*JWST*: NIR and MIR IFU data to map warm molecular Hydrogen.

Overall the current suite of ESO and ALMA instruments is capable of addressing most of the above questions and the ESO community will be able to take the lead in creating this observational and theoretical platform for future X-ray observations. When combined with *JWST* and ELT, the majority of the pre-*Athena* aspects of this theme will have been explored.

The addition of a large single dish to complement ALMA would add considerably to detecting the most extended CO emission missed by even the shortest baselines.

There are quite a few interesting time domain observations in the radio and optical of the AGN in BCGs that in the case of NGC 1275 and several other systems are known to be variable on month to decade timescales (Dutson et al., 2014; Hogan et al., 2015b). The Large Synoptic Survey Telescope (LSST: https://www.lsst.org) offers us the opportunity to search for optical variability of the AGN core in all BCGs where the central core contributes >2 - 3% of the total light.

### 3.5.3. Complementary observations

- X-ray flux of cluster emission. Since we want to observe mainly the brightest clusters, selected because of their strong X-ray flux, it is likely that we have already catalogued >90% of the *Athena* targets we might want to observe (at least at z < 0.5).

- X-ray morphology. The presence of peaked X-ray emission is a prerequisite for deeper follow-up given the observed connection to optical lines, radio emission and cold molecular gas. Having *Chandra* data for a large sample of clusters with a line luminous BCG would be an important foundation for all future work in this area. *Chandra* has a finite operational lifespan so ensuring that an X-ray selected sample of clusters is fully observed before the end of the mission needs some consideration and effort.

- Radio properties. We have a fairly complete radio imaging at 1 – 10 GHz for a large fraction of the brightest X-ray clusters (done with ATCA, JVLA, and VLBA: Hogan et al., 2015a,b) and this will improve in the next 5 – 10 years with the completion of the VLA Sky Survey (https://science.nrao.edu/science/surveys/vlass/), ASKAP EMU (http://www.atnf.csiro.au/people/Ray.Norris/emu/), and Westerbork Synthesis Radio Telescope (WSRT) WODAN surveys that will provide a more uniform coverage of the sky. These radio data will be used to identify groups and clusters from BCGs, with a wide range of radio powers and morphologies to study.

- Current AGN activity. The radio core and X-ray point source flux from the BCG can be used to identify the systems with the highest current accretion rates. By studying BCGs in a very large sample of X-ray selected clusters (>10,000 as will be possible from eROSITA), we can determine the intrinsic distribution of the (normalized) accretion rate of these systems that should predominantly be significantly less than $10^{-4}$ $L_{Edd}$. The integral of this accretion rate distribution can be used to predict the average growth rate of the SMBHs in BCGs.

# 4. The Hot Universe: Missing baryons in cosmic filaments

Baryons trace the large scale structures of the Universe that form under the gravitational action of the dominant dark matter. Theory predicts that most of these baryons reside in vast unvirialized filamentary structures that connect galaxy groups and clusters (the "Cosmic Web"). This is confirmed by measurements of the baryon density in the Ly$\alpha$ forest at z > 2. At the current epoch (z < 1), however, about one half of the baryons are missing (e.g., Shull et al., 2012) and the fraction of detected baryons falls short of the cosmic expectation at all mass scales in gravitationally-bound structures (from galaxies to clusters of galaxies: see, e.g., McGaugh et al., 2010), with most of them expected to reside in the photoionized and shock-heated intergalactic medium (IGM), in the circumgalactic medium (CGM), and in a hot and diluted gaseous phase distributed over large scales, the so-called warm-hot intergalactic medium (WHIM; see Figure 5).

Because the majority of these baryons are supposed to emit X-rays, sensitive high-resolution X-ray observations such as the ones that will be provided by *Athena* will teach us about the formation, structure and evolution of these systems, which are the largest in the Universe. ESO's facilities are mandatory to investigate how WHIM filaments correlate with metal and Ly$\alpha$ absorbers and how they are traced by the SZ signal, how the metal distribution evolves, how matter accretes and clumps along filaments towards massive virialized structures. More details on this science case can be found in Kaastra et al. (2013).

## 4.1. Physics of the IGM

In the IGM, gravity heats the gas and, combined with radiative cooling which is also enhanced from metal pollution, drives "cold" inflows. The feedback from star formation and AGN injects energy and/or momentum through the associated outflows. A balance between inflows and outflows can be reached through the self-regulation of star formation and BH growth.

Absorption and emission associated with the CGM will be highly biased towards regions with high density, high metallicity, and at some particular temperatures. Absorption measurements also suffer from confusion due to projection effects. Moreover, observations without sufficient angular resolution will smooth out clumpy emission, misinterpreting the large-scale filamentary emission. To avoid these biased measurements, both high spectral and spatial resolution are required. Numerical simulations, along with virtual observations, can support the interpretation of the observed distribution of the gas properties (see e.g., Figure 5).

The synergy between *Athena* and ESO telescopes will be efficiently probed by detecting and obtaining redshifts for galaxies, including the faint and obscured ones down to 0.01 L$^*$ at z < 1 and, at higher densities, z < 0.5.

In particular, to characterize the population of the galaxies responsible for the metal pollution of the filaments detected with X-ray absorption, dedicated redshift surveys at ±500 km s$^{-1}$ with respect to the filament's redshift and across R ∼ 5 Mpc ( ≈30′ × 30′ at z = 0.5) will be requested and possible with instruments like VLT/MOONS and VISTA/4MOST. It will be extremely relevant to the characterization of the IGM to look for correlations between properties of the CGM and galaxies in regard to their physical parameters (such as mass, SFR, chemical composition, morphology), the orientation and the the kinematics. Moreover, considering that the galaxies' interstellar medium (ISM) will outshine the surrounding CGM and IGM in X-ray emission lines, this emission will trace the large-scale structure and will therefore also be filamentary when viewed at lower angular resolution. Thus, the detection, and redshift determination, of the faint galaxies will allow observers to identify, and mask, their 3-D locations to observe gas flows in galaxy haloes or to test the intergalactic interpretation of extended, low surface brightness X-ray line emission.

Then, the detection, and redshift determination, of the faint galaxies will permit to identify, and mask, their 3-D locations to observe gas flows in galaxy haloes or to test the intergalactic interpretation of extended, low surface brightness X-ray line emission.

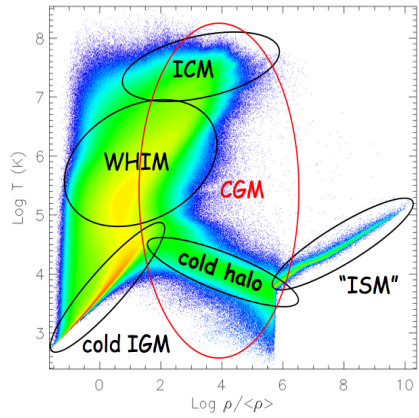
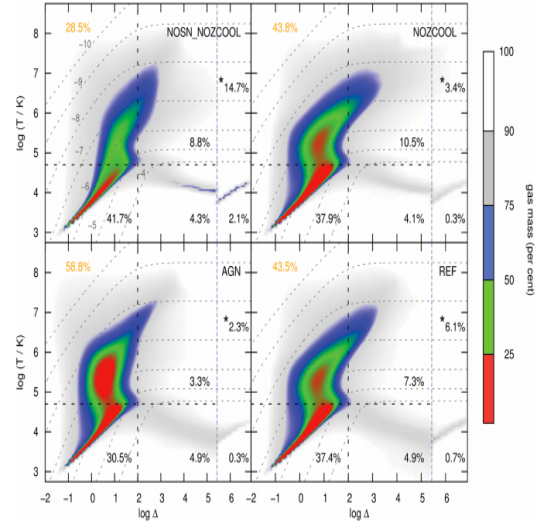

Figure 5: Left: Phase space diagram of the cosmic baryons at redshift 0. Right: Distribution of gas mass over various phases predicted by different models, indicated on the top right-hand corner of each panel (see Table 1 in Tepper-García et al., 2012). The vertical (horizontal) dotted line indicates the density (temperature) threshold at $\Delta = 100$ (T = $5 \times 10^4$ K) which separates unbound (cool) from collapsed (warm-hot) gas, where $\Delta \equiv \rho_b / \langle \rho_b \rangle$, where $\langle \rho_b \rangle$ is the mean baryonic density. The region to the right-hand side of the blue dashed line shows the high-density, star-forming gas with physical densities $n_H \geq 0.1$ cm$^{-3}$. The coloured areas show the cumulative gas mass (in %) indicated by the colour bar on the right-hand side. The percentages in each panel show the baryon mass fraction in the corresponding phase defined by the temperature/density thresholds. The baryon mass fraction in WHIM is labelled in orange. The starred percentage indicates in each case the baryonic mass fraction in stars. The dotted contours, which are identical in all panels, give the neutral hydrogen fraction ($n_{HI}/n_H$) of gas in ionization equilibrium as a function of $\Delta$ and T. The left-hand panels show the effect of two most extreme scenarios, that is, including both feedback by type II supernovae (SNeII) and AGN with respect to neglecting feedback altogether. The right-hand panels show the effect of neglecting radiative cooling by heavy elements. Clearly, feedback by SNeII (and AGN) heats a significant amount of gas above temperatures T = $5 \times 10^4$ K, with the WHIM fraction increasing from 28% (top left-hand panel) to 59% (bottom left-hand panel). Interestingly, the IGM fraction only changes by ~ 10%, indicating that feedback shifts a large fraction of the ISM from haloes into intergalactic space (from Tepper-García et al., 2012, reproduced with permission).

## 4.2. Observational Studies of the WHIM

A multiwavelength approach is needed to address in its entirety and complexity the characterization of the physical properties of the WHIM. Observations of the CGM at other wavelengths (H$\alpha$, rest-frame UV) will complement the X-ray view. For faint sources, statistical detections through stacks centred on/in between galaxies, cross correlations with galaxy number density and the SZ and lensing signals will be necessary. High-resolution spectroscopy in the far-UV (FUV), X-ray, and mm bands constrains the amount of cosmic baryons, measuring $\Omega_b$ from the cool (5.0 < log T < 5.5), warm (5.5 < log T < 6.7) and hot (log T > 6.7) phases of the IGM, through column density measurements of their main tracers: O VI (in the FUV and X-rays) and broad (with a width parameter $b$ > 40 km s$^{-1}$) Ly$\alpha$, for the cool phase; O VII-VIII, Ne IX, N VII, Mg XI and the rare isotopes of $^{27}$Al XI and $^{13}$C VI (X-rays and mm), for the warm phase; O VIII, Ne X, $^{27}$Al XI (X-ray, mm) for the hot phase. The hotter and denser phase can be also seen in metal emission lines (mainly O VII and O VIII) in the X-rays (e.g., Branchini et al., 2009). Moreover the ratio of line intensities in the X-rays and FUV will provide useful diagnostics on the absolute metallicity (ratio of metal [X-rays] to HI [FUV lines]), temperature (ratio of metal lines from different ions of the same element), turbulence (ratio of at least two lines from the same ion and a line from a different ion of the same element) of the WHIM. The line ratio between different isotopes of the same element (X-rays and mm) will give valuable information on nucleosynthesis processes at z < 2.

In mm bands, the detection of SZ signal in WHIM filaments is very challenging. However, it could be achieved for its densest parts such as the surrounding medium within super clusters or bridges of matter between coalescing pairs of clusters (*Planck* Collaboration et al., 2013b). Added to the constraints from the X-rays with *Athena*, this would improve the determination of the nature and properties of the phase of the hot gas (Kaastra et al., 2013). Moreover, hyperfine splitting of the emission lines of rare isotopes and molecules associated to the WHIM cold interface can be detected (Sunyaev & Docenko, 2007). For example, ALMA follow-ups of *Athena*'s candidates for WHIM filaments can constrain $^{27}$Al XI and $^{13}$C VI lines with Bands 2–3, and CO (1-0) and (6-5) molecules with Bands 2–10.

Deep UV and optical/IR (OIR) imaging and IFU spectroscopy (with, e.g., ELT/HARMONI at least down to 0.01 L* at the filament's redshift) are the necessary synergy to identify galaxies surrounding filaments and measure their relative distances to fully understand the galaxy environment of WHIM filaments, their location in the large-scale structure, and classify any galaxy-WHIM association.

ELT/HARMONI and ALMA (Bands 2–10) can detect and survey atomic and molecular outflows in galaxies surrounding *Athena*'s targets, allowing one to measure their energetic and metal budget, and to assess the feedback between virialized (galaxies, galaxy groups and clusters) and non-virialized (IGM) structures and the processes that regulate the metal pollution of the IGM.

# 5. The Energetic Universe: SMBH history

## 5.1. Galaxy evolution: relationship between SMBH growth, SFR, galaxy bulge and disk, and cold gas

There are strong indications that the growth of galaxies and that of the SMBHs residing at their centres are connected. Globally, galaxy and SMBH growth are seen to track each other with redshift (i.e., from a comparison of the SFR and SMBH accretion rate cosmic histories: e.g., Aird et al., 2013; Madau & Dickinson, 2014; Brandt & Alexander, 2015). Relic evidence for this growth is also seen from the tight connection between SMBH mass and various galaxy properties (spheroid luminosity, mass, and velocity dispersion) in local systems (e.g., Graham & Scott, 2013; Kormendy & Ho, 2013). Given the large difference in size scale between the SMBH and the galaxy it seems likely that some process(es) regulate this apparent joint growth (e.g., gas inflow mechanisms; AGN or star formation outflow; stellar winds; star-formation winds). However, it is currently unclear what these processes are and whether SMBH-galaxy growth is concordant in all systems or just a subset of systems. With high and uniform sensitivity across a large FoV, *Athena* will provide a huge increase in the source statistics of X-ray AGN over current facilities, allowing for unprecedented source population studies and detailed measurements of the connection between the growth of SMBHs and galaxies (see also Section 5.3). However, *Athena* will only provide one half of this connection (the SMBH growth: the AGN) and other facilities are crucially required to understand the connection between the growth of SMBHs and galaxies. We review here the synergy between *Athena* and ESO facilities, focusing on distant (z > 0.5) X-ray AGN to understand the cosmological evolution of SMBH and galaxy growth. More details on this science case can be found in Georgakakis et al. (2013).

### 5.1.1. ESO – *Athena* synergies

Assuming that the problem of the identification of the multiwavelength counterparts to the *Athena* sources is solved (Section 5.2), there is a huge synergy between ESO facilities and *Athena* for characterising the properties of the AGN host galaxy: the structure and stellar components of the host galaxy, the star formation, and the chemical abundance and kinematics of the gas in the host galaxy (including outflows driven by the AGN and star formation). For this the requirements are:

1. sensitive high spatial resolution imaging (from optical through to mm wavelengths);
2. sensitive spatially resolved spectroscopy (i.e., IFU).

High spatial resolution imaging is required for measurements of the host-galaxy starlight, the structure of the galaxy, and the distribution of star formation. With these data it will be possible to calculate the mass of the host galaxy, decouple bulge components from disk components, identify galaxy interaction/merger components, and to map out the regions of star formation and measure SFRs. At rest-frame far-IR to mm wavelengths, ALMA provides the potential for sub 0.1´´ resolution, which allows for measurements of the dust-obscured star-formation continuum on sub-kpc scales and constrain, for example, whether the star formation is compact or extended. At shorter wavelengths, the host-galaxy starlight can be mapped, including regions of unobscured star formation (i.e., young stars) and the global structure of the galaxy (bulges; disks; interaction/merger remnants). HST, and shortly *JWST*, are currently the work horses for such analyses, although it is not clear that either will be operational by the launch of *Athena*. However, with AO, the ELT will provide both higher sensitivity and higher spatial resolution imaging than that achieved by either HST or *JWST*. For example, the first generation ELT instruments MICADO (0.8 - 2.4 $\mu$m, imaging and spectroscopy) and METIS (imaging at L, M, N, and Q bands) (Section 13.1) will allow for sub-kpc measurements of the rest-frame UV and optical starlight for distant X-ray AGN and measurements of the host-galaxy structure. The longer wavelength ELT/METIS bands will be mostly sensitive to the brighter mid-IR sources but will provide a view of the short-wavelength dust-obscured star formation, in addition to constraints on the AGN itself.

IFU spectroscopy is required for measurements of the chemical abundance and kinematics of the gas in the host galaxy, including outflows driven by the AGN (see Section 7). Again, ALMA is likely to be the work horse for longer-wavelength observations and will provide, for example, maps of the cold gas components in CO; however, such measurements require a lot of observing time and it will be expensive to obtain spatially resolved CO constraints for a large sample of X-ray AGN. At shorter wavelengths, the ELT first-generation instrument HARMONI will provide IFU observations at 0.45 - 2.45 $\mu$m and allow for the mapping of the host-galaxy kinematics (e.g., through rest-frame optical emission lines: H$\alpha$ and [OIII]) to distinguish between rotation signatures, irregular kinematics (e.g., potential signatures of mergers), and high-velocity components (i.e., potential signatures of outflowing gas, either from the AGN or star formation). However, since ELT/HARMONI is a single IFU, obtaining constraints for a large sample of X-ray AGN will be observationally expensive. The second generation ELT/MOS is a multi-IFU and will allow for the kinematic components for more than 10 X-ray AGN within a single FoV (~ 7´ in diameter) at 0.8 - 1.8 $\mu$m, providing the potential to build up detailed constraints for the overall X-ray AGN population. For the brightest systems it will also be possible to directly constrain the starlight kinematics, through measurements of the rest-frame optical absorption lines.

## 5.2. Determination of OIR counterparts and redshifts of distant X-ray sources

Redshift and SED are important characteristics of any extragalactic object which cannot be determined unless the precise sky position is available for the spectroscopic follow-up and/or for the assembling of the multiwavelength data. While sources identified in the optical and NIR regime usually have the required localization accuracy (sub-arcsecond level for distant sources), this is rarely the case for the sources identified in the X-ray band. Here, the positional measurement error is not constant and depends on the astrometry, on the X-ray counts, and the spatially varying PSF and usually can reach up to ~ 5 - 7´´ for off-axis observations (for *Chandra* and XMM-*Newton*). In turn, a circle of radius 5 - 7´´ can contain up to a few tens optical/NIR sources and the problem to solve becomes "which of these sources is the most probable counterpart to the X-ray emission?". This pairing process is done by means of ancillary data catalogues that can differ in depth and definition of "source detection", which add an additional difficulty to an already complicated problem.

In the specific case of *Athena*, the positional accuracy requirement for WFI is set to 1´´ at 3$\sigma$. This will enable a counterpart identification for at least 90% of the sources at the point source sensitivity limit of $2.4 \times 10^{-17}$ erg s$^{-1}$ cm$^{-2}$ in the deepest pointings. The WFI will achieve this requirement due to its large FoV (40´ × 40´) that will include a large number of X-ray bright sources in one pointing, thus improving the astrometry. In addition, the PSF is expected to be more regular than is currently available over the entire FoV allowing an improved centering also for the faint sources.

Among the scientific goals of *Athena* is the detection of a few hundred z > 6 AGN (Section 5.3) and a few tens of Compton-Thick (CT) AGN in the redshift range 1 < z < 4, expected to be hosted in galaxies as faint as AB ~ 28 in the optical and NIR bands. At these depths, the CANDELS/*Chandra* Deep Field South (CDFS) catalogue (Guo et al., 2013) suggests a source density ~ 1 arcsec$^{-2}$. Thus, even with the localization accuracy of the WFI, it will be challenging to identify which of the 3 sources falling in the circle with radius equal to the positional uncertainty is the correct counterpart to the X-ray emission.

In the last 20 years, Maximum Likelihood (Sutherland & Saunders, 1992) has been the most used method for quantifying the likelihood for a source to be associated with the X-ray emission, rather than being a chance coincidence. Applied consecutively to catalogues from different photometric bands (optical, NIR/MIR), and by comparing (and often complementing via visual inspection) the results from the various associations, the most reliable counterpart is defined (e.g., Brusa et al., 2010; Luo et al., 2010; Marchesi et al., 2016; Nandra et al., 2015; LaMassa et al., 2016). A detailed review of the method is provided in Naylor et al. (2013) who clearly pointed out how the wrong estimate of the background population affects the probability of finding the correct counterpart if this is faint (see also Brusa et al., 2007). This problem can be eliminated by the use of Bayesian statistics (Budavári & Szalay, 2008; Pineau et al., 2016) which supports also the assumption of priors (e.g., Salvato et al., 2017; Dwelly et al., 2017) for determining whether or not a source is the counterpart of an X-ray emitter. Granted, the correct prior needs to be built and adopted. Currently, we are missing a large area of relatively deep X-ray and optical data needed to build the prior but in the next decade, the new eROSITA all-sky X-ray survey (Merloni et al., 2012) combined with the Dark Energy Survey (https://www.darkenergysurvey.org/) and EUCLID (http://sci.esa.int/euclid/) should start filling this gap.

### 5.2.1. The role of ESO

The few tens of square degrees that *Athena* WFI will map searching for high-z and obscured AGN will rely on the deep ancillary data provided by Subaru-HSC (http://www.naoj.org/Projects/HSC/), LSST, and WFIRST (https://www.nasa.gov/wfirst) observations. Those ancillary data will be used to design suitable priors and identify the most probable counterpart.

After this identification phase, ESO telescopes and instruments will have a crucial role in the characterization and confirmation of the AGN nature of the sources via spectroscopy. First of all, the access to the rich ESO spectroscopy archive will be of paramount importance in rejecting low redshift interlopers and false associations. The screening process will then continue by means of MOS instruments like VISTA/4MOST, which will identify sources brighter than $r_{AB} \sim 23$. For the follow-up of the cleaned samples of 60 CT AGN over 6 deg$^2$ (Georgakakis et al., 2013) and 400 z > 6 AGN over 40 deg$^2$ (Aird et al., 2013), access to ALMA will be crucial, where the redshift of the high-z AGN will be measured via [CII] as well as to ELT/HARMONI, which will be used for targeting the CT AGN for an estimated total of 50 hours. The latter instrument, together with the IFU VLT/MUSE, will not only allow the study of the AGN, but will be the only means to identify the counterparts to X-ray sources that are so obscured that they remain undetected in the deepest broad band optical and NIR images. In fact, these sources manifest themselves exclusively through the presence of emission lines that only an IFU instrument can detect.

## 5.3. SMBHs at high redshift

Up to the highest known redshifts, most, if not all, galaxies harbour a SMBH. Despite their abundance, however, we know neither what drives the growth of SMBHs nor what triggers AGN activity at any redshift. The situation is even less clear at high redshifts, with several questions still unanswered, namely:

1. What types of galaxies host AGN at z > 6;

2. How rapidly are SMBHs growing in different galaxies;

3. What are the physical mechanisms that bring gas into the centres of galaxies at such high redshifts, driving thus SMBH growth;

4. Whether the accreting SMBHs have an influence on their host galaxy.

One of *Athena*'s main science goals is indeed to answer the above questions. X-ray surveys are very efficient at finding AGN over a wide range of luminosities, as they are less affected by obscuration than e.g., optical/UV, and can find fainter AGN that are usually missed by optical surveys. But while current X-ray observatories are sensitive enough to find z > 6 quasars, they lack the capability to cover large areas efficiently. The known population of z > 6 objects consists of extremely luminous quasars, powered by SMBHs with $M_{BH} \sim 10^9 \, M_\odot$, comparable to the most massive local SMBHs but at a time when the Universe was less than one Gyr old. What is really needed in order to answer the above-mentioned questions is to identify large samples of more typical, low-to-intermediate luminosity AGN at z > 6, as they are crucial for probing the epoch when the first SMBHs formed. Detection of low-luminosity AGN at such high redshifts will provide information on the minimum mass of SMBH seeds (see Section 9.4) and, at the same time, constrain the mechanisms by which SMBHs grow.

*Athena* will enable X-ray surveys to be carried out two orders of magnitude faster than with *Chandra* or XMM-*Newton*, thanks to the combination of its large collecting area, and the sharp PSF over a substantial fraction of its large FoV, allowing for the exploration of a new discovery space. A multi-tiered *Athena* WFI survey of a total observing time of 28Ms (Aird et al., 2013), is expected to identify over 600,000 AGN, among which more than 400 at z = 6 - 8 and around 30 at z > 8 (see Figure 6 for typical *Athena* sensitivities compared to those of other facilities, including ALMA and the ELT, for a z = 7 source).

These X-ray surveys will pinpoint active SMBHs within samples of z > 6 galaxies identified by the large, state- of-the-art optical and NIR imaging surveys that will be available in the late 2020s. Deep ALMA follow-up observations of high-redshift X-ray AGN will probe the cold dust component and obscured SFRs via sub-mm spatially resolved continuum measurements, as well as molecular gas mass and dynamics using the [CII] 158 µm line. Further spectroscopic follow-up with ELT will confirm their redshifts and will provide the BH masses for type 1 AGN, while AO-assisted NIR/MIR imaging (ELT/MICADO and METIS) will yield stellar masses and SFRs, respectively. We will thus be able to determine the physical conditions within the host galaxies of early SMBHs, vital for understanding the formation of SMBHs, their subsequent fuelling, and to finally assess their impact on the early Universe.

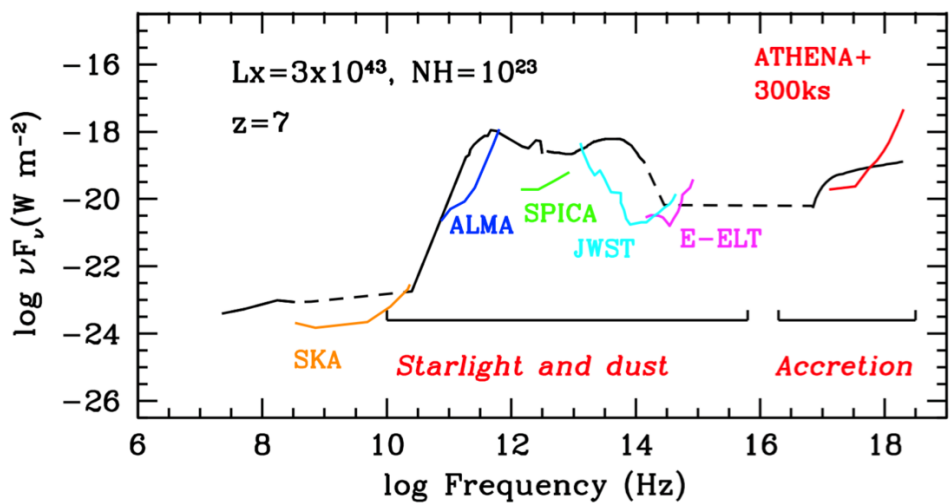

Figure 6: Broad-band SED of a moderate- luminosity obscured AGN at z = 7 that will be observable in *Athena* surveys. The 3σ sensitivities of ALMA (in blue) and the ELT (in magenta), as well as SKA, SPICA and the *JWST* are shown (Aird et al., 2013).

# 6. The Energetic Universe: SMBH accretion disks

## 6.1. The relationship between X-ray and optical/UV emission of AGN

To study the cosmic history of accretion on to BHs, three measurements are crucial: BH mass, bolometric luminosity, and radiative efficiency. For standard, optically thin, geometrically thick disks, BH spin is directly related to the radiative efficiency, η, which can be estimated from the bolometric luminosity, $L_{Bol}=\eta(dM/dt)c^2$, assuming the mass accretion rate, $dM/dt$, is known. The latter can be measured from models of the continuum SED (i.e., after removing the galaxy component) at long enough wavelengths (rest-frame optical), provided the BH mass is accurately determined. Thus, observational tests of fundamental properties of relativistic accretion theory demand, as a pre-requisite, accurate measure of BH mass and of the full, multiwavelength SED. More details on this science case can be found in Dovciak et al. (2013).

## 6.2. Black Hole Masses

Reliable measures of BH mass require access to the dynamical information (i.e., resolved absorption line features) well within the gravitation sphere of influence of a SMBH. The most common SMBHs in the local universe have masses $\sim 10^8 M_\odot$ and a sphere of influence (Bondi radius) of the order of a few parsecs. Observations with ELT at the diffraction limit in the NIR band will allow resolution of the Bondi radius up to z $\sim 0.1$ (i.e., 10pc, $\sim$ 6 mas), i.e., over a volume of the Universe more than $10^3$ times bigger than that accessible to HST (which reaches z $\sim 0.01$).

For more distant SMBHs, only time-resolved spectroscopy (reverberation mapping) will allow SMBH mass estimates via studies of the broad line region of AGN. Here the limitation is programmatic, rather than physical, as 8m class telescopes are powerful enough to detect the signal up to the highest redshifts, but intensive use of telescope time is needed to monitor the variations over the relevant timescales ($\sim$ a few 100(1+z) days for a SMBHs of $10^8 M_\odot$). Flexible use of MOS on 8m class telescopes (e.g., VLT/MOONS) will allow large, statistically significant samples of AGN to be sampled and accurate BH masses to be derived.

## 6.3. AGN SEDs: Accretion disks and coronae

Once BH masses are known, an accurate determination of the SEDs of the active core is needed to unveil the processes taking place in the innermost region of the accretion flow, allowing constraints to be put on the radiative efficiency and, possibly, BH spin.

The gravitational energy of matter dissipated in the accretion flow around a BH is primarily converted to photons of UV and soft X-ray wavelengths. The lower limit on the characteristic temperature of the emerging radiation can be estimated assuming the most radiatively efficient configuration: an optically thick accretion flow. Taking into account that the size of the emitting region is r $\sim 10 R_G$ (where $R_G = GM_{BH}/c^2$ is the gravitational radius) and assuming a black body emission spectrum one obtains $kT_{bb}=(L_{bol}/\sigma_{SB}\pi r^2)^{1/4} \approx 14(L_{bol}/10^{44})^{1/4}(M_{BH}/10^8)^{-1/2}$ eV. For typical SMBH masses in AGN and luminosities (and thus accretion rates), the expected spectrum of the accretion disk peaks in the optical-UV bands.

Standard accretion disk theory needs to be supplemented by a description of the disk vertical structure and, in particular, of its atmosphere, in order to accurately predict spectra. This, in turn, depends on the exact nature of viscosity and on the micro-physics of turbulence dissipation within the disk. A second complicating effect is that the real physical condition in the inner few hundreds Schwarzschild radii of an AGN might be more complex than postulated in the standard accretion disk model. For example, density inhomogeneities resulting in (unstable) cold, thick clouds which reprocess the intrinsic continuum have been considered at various stages as responsible for a number of observed mismatches between the simplest theory and the observations. All of the above-mentioned problems are particularly severe in the UV part of the spectrum, where observations are most challenging.

The second "universal" component of the accretion flow emission emerges in the X-ray band. The upper end of the relevant temperature range reached by accretion flows on to BHs is achieved in the limit of optically thin emission from a hot plasma, possibly analogous to the solar corona, hence the name of accretion disk "coronae" (Galeev, 1970). The virial temperature of particles near a BH, $kT_{vir} = GM_{BH}m/r \propto mc^2/(r/R_G)$, does not depend on the BH mass, but only on the mass of the particle, m, being $T_{vir,e} \sim 25(r/10R_G)^{-1}$ keV for electrons and $T_{vir,p} \sim 46(r/10R_G)^{-1}$ MeV for protons. As the electrons are the main radiators that determine the emerging spectral energy distribution, while the protons (and ions) are the main energy reservoir, the outcoming radiation temperature for optically thin flows depends sensitively on the detailed micro-physical mechanisms through which ions and electrons exchange their energy in the hot plasma. Indeed, the values of the electron temperature typically derived from the spectral fits to the hard spectral component in accreting BHs, $kT_e \sim 50 - 150$ keV, are comfortably within the range defined by the two virial temperatures. The observed hard X-ray spectral component from hot optically thin plasma is believed to be produced by unsaturated Comptonization of low frequency seed photons from the accretion disk itself (when present), with characteristic temperature $T_{bb}$. Such a spectrum has a nearly power-law shape in the energy range from $\sim 3 kT_{bb}$ to $\sim kT_e$, as observed. For the parameters typical for BHs in AGN this corresponds to the energy range from about a few tens of eV to about 50 - 100 keV. The photon index $\Gamma$ of the Comptonized spectrum depends in a rather complicated way on the parameters of the Comptonizing media, primarily on the electron temperature and the Thomson optical depth. In fact, the emerging power-law slope depends more directly on the Comptonization parameter $y$, which is set by the energy balance in the optically thin medium: the ratio of the energy deposition rate into hot electrons and the energy flux brought into the Comptonization region by soft seed photons are critical for the resulting spectral shape.

Broadly speaking, a significant part, if not the entire diversity, of the spectral behaviour observed in accreting BHs, can be explained by changes in the proportion in which the gravitational energy of the accreting matter is dissipated in the optically thick and optically thin parts of the accretion flow. One of the key features of the corona/disk system is the observed non-linear relation between $L_X$ and $L_{UV}$, which shows no redshift dependence and low intrinsic dispersion of $\sim 0.2$ dex (Jin et al., 2012; Lusso et al., 2016: see Figure 7). The fact that UV bright quasars emit fewer X-rays (the $\alpha_{ox}$ - $L_{UV}$ correlation) suggests that the corona/disk energy coupling depends on both accretion rate and BH mass, but a full understanding of the observed trends on sound physical principles is still lacking. In very general terms, this might point towards a connection between accretion disk physics and the mechanism(s) of corona generation in AGN (Merloni et al., 2003; Wang et al., 2004).

In short, the physical origin of the tight coupling between cold (accretion disk) and hot (corona) phases remains elusive. Despite significant progress in MHD simulations of the accretion disk achieved in recent years there is no accepted global model of accretion on to a compact object able to fully explain all the different SEDs observed, nor the transitions among them. In fact, the main open questions regarding the origin of the X-ray emitting coronae and the reflection component in AGN (and in X-ray binaries [XRBs]) are intimately connected with those left open by the classical theory of relativistic accretion disks. The main ones concern:

1. the physical nature of the viscous stresses and their scaling with local quantities within the disk (pressure, density);

2. the exact vertical structure of the disk and the height where most of the dissipation takes place;

3. the nature of the inner boundary condition.

Here, the usefulness of the combination of *Athena* and ESO facilities will be at least twofold: on the one hand, *Athena* WFI surveys will provide large statistical unbiased samples of X-ray AGN across a wide range of evolutionary stages (Section 5.3). Massive multi-object spectroscopic follow-up (with 8m or bigger telescopes) will be necessary to unveil the fundamental properties of the optical/UV emission in such systems, with NIR spectrographs (e.g., VLT/MOONS) necessary to push the study of AGN SEDS to z >1. On the other hand, high-resolution X-ray spectroscopy with *Athena* X-IFU will probe in detail the physical process in the inner accretion disk regions of a large population of AGN. The combination of time-resolved, sensitive optical/UV (VLT/X-Shooter and ELT/HARMONI) and X-IFU observations of these AGN will be particularly useful to build a complete, detailed picture of the accretion-disk corona system by constraining temperature, ionization state and relativistic dynamics of the accretion flow.

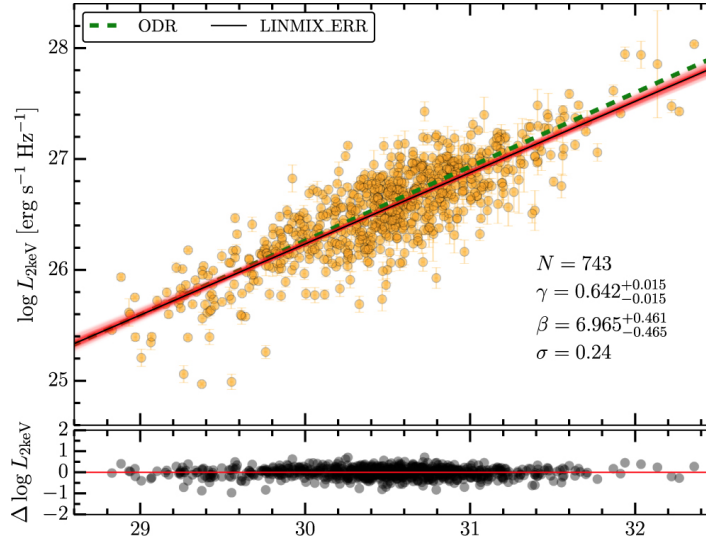

Figure 7: Rest-frame monochromatic X-ray and UV (2500 Å) luminosities for a UV-selected sample of QSOs with good quality X-ray detections and minimal dust extinction (orange circles). Dashed green and solid black lines show the results of a regression analysis. The lower panel shows the residuals of log $L_{2keV}$ and log $L_{2500}$ with respect to the LINMIX$_{ERR}$ best-fit line. A non-linear correlation is observed with high significance and low intrinsic scatter; it remains unexplained so far. From Lusso et al. (2016).

# 7. The Energetic Universe: AGN feedback – Molecular outflows

Massive molecular outflows are a common phenomenon, at least in the local Universe (e.g., Feruglio et al., 2010, 2013, 2015; Aalto et al., 2012; Sturm et al., 2011; Combes et al., 2013; García-Burillo et al., 2014; Cicone et al., 2014; Dasyra et al., 2015, etc.). They are often associated with ionized atomic outflows and neutral (HI) outflows (Veilleux et al., 2013; Morganti et al., 2016). They show line of sight velocities of up to $10^3$ km s$^{-1}$, carry a few percent of the AGN bolometric luminosity, and their mass loading factor is significantly larger than that found in starburst galaxies (see Fiore et al., 2017, for an almost complete compilation of galaxy wide outflows). Given these properties, it is likely that massive outflows can regulate and modulate galaxy and BH growth over the lifetime of the Universe from the first galaxies after the epoch of reionization, through the peak epoch of galaxy/AGN assembly z ~ 2, and to the local Universe (see also Sections 3.4, 3.5, 9).

## 7.1. ESO – *Athena* synergies

*Athena*, in synergy with ESO telescopes, can provide fundamental contributions in at least four main topics:

- Nuclear winds (ultrafast outflows [UFOs]) connection to molecular outflows. Optical line ratios will allow one to measure electron temperatures, ~$10^4$ K from [OIII] λ4363, 4959, 5007 at z ~ 2, electron densities (~ 500 cm$^{-3}$) and extinction ($A_V$ ~ 1 - 4 mag). Ionized gas masses can be estimated from the Hα luminosity (e.g., using VLT/SINFONI: Nesvadba et al., 2008).

- Shocks: how outflows propagate in/modify the ISM/CGM. The excitation of the gas can be determined by BPT diagrams with VLT/MUSE (Hamer et al., 2015) and the NIR lines may further characterize excitation and dynamics ($H_2$, [FeII], Paschenα, with VLT/SINFONI: e.g., Dasyra et al., 2015 see Figure 8).

- Extended hot/warm haloes and CGM: the fate of outflowing gas, do molecular outflows reach the CGM? While they can be detected through X-ray lines (O VII, O VIII), they are also kinematically determined in detail though their correlated Hα filaments (e.g., Cecil et al., 2002).

- Effects of massive outflows on heating of first groups and clusters (z > 2; see also Section 2.2). The shocks are traced both in X-rays or the NIR $H_2$ line (e.g., Guillard et al., 2009) and through IFU Hα observations (Rodríguez-Baras et al., 2014).

Today the UFO - galactic outflow connection could be observed only in a few sources (Mrk 231, IRAS 11119 and APM 08279: Feruglio et al., 2015; Tombesi et al., 2015, and Feruglio et al., in prep.), by exploiting the synergy of *Chandra* and XMM-*Newton* and IR/mm telescopes (Herschel/NOEMA). For these sources the momentum boost measured for the galactic outflow versus UFO is consistent with energy conserving galactic winds (see King & Pounds, 2015, for a review). More data will be available to confirm this relation in the next few years, including ALMA data on PDS456, which is one of the main sources for this science topic. Additional constraints will come from the many nuclear semi relativistic winds that will be detected by *Athena*.

## 7.2. ALMA – *Athena* synergies

Even if we will be able to constrain this relation between UFOs and extended cold outflows, this will still be an indirect argument because we measure the kinetic energy of the UFO on scales of $10^{15}$ cm, of outflows on scales of hundreds of parsecs, but we are not able to capture the physics at work, i.e., observe the interface between hot and cold, how the cold wind forms, or the shock front that is predicted by theoretical models. The so-called two-stage acceleration model (King & Pounds, 2015) tries to connect the nuclear wind with the galactic wind, and predicts an internal shock on scales of $10^{17}$ cm, and a forward shock on the kpc scale where the wind impacts the quiescent ISM. *Athena* and ALMA synergy can provide information on this scenario. With *Athena* X-IFU it will be possible, for nearby sources, to map the forward shock region. With ALMA it is possible to capture the process of reformation of molecules in the post shock region, where one should observe different molecular abundances with respect to the molecular disk (in particular a larger HCN/CO ratio).

As for the fate of the outflowing gas, whether it reaches the CGM and the galactic halo, and the energy, entropy and metal transport into the CGM, *Athena* X-IFU will be able to measure spatially resolved abundances of metals in the CGM to understand how the CGM was enriched and connect it to extended outflows observed at other wavelengths (i.e., sub-mm with ALMA), at least in nearby galaxies.

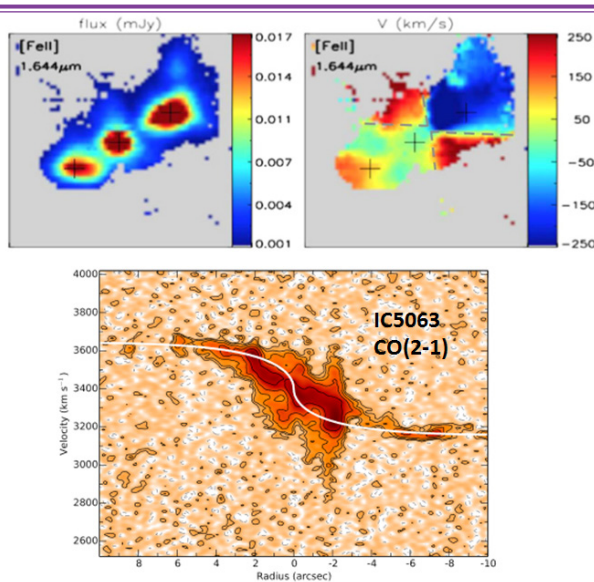

Figure 8: The X-ray AGN IC5063 is driving atomic and molecular outflows in the host galaxy, seen here in the [FeII] line at 1.64µm in the NIR with SINFONI on the VLT (Dasyra et al., 2015, top left: total flux, top right: velocity field), and also in the position-velocity CO(2-1) diagram with ALMA (Morganti et al., 2015, bottom).

### 7.2.1. A prototype: NGC 6240

One example of synergy between X-ray and mm observations in this topic is the case of the NGC 6240 outflow. There, a CO outflow with velocity -400 km s$^{-1}$ spatially coincides with H$\alpha$ filaments. *Chandra* spectra provide evidence for shocked gas at the position of the H$\alpha$ emission and suggest that a shock is propagating eastwards compressing the molecular gas while crossing it (Feruglio et al., 2013). Stronger line emission is expected in non-equilibrium (shock) models, because of the broader ion distribution with respect to thermal equilibrium models at the same temperature and metal abundances. In particular, shock models, like XSPEC PSHOCK are known to produce spectra with prominent line emission from ionized species as Mg, Si and S. Feruglio et al. (2013) show what is achievable today with a CCD spectrum with spectral resolution of 150 eV. *Athena* X-IFU, with 2.5 eV resolution, will be able to find shock tracers (strong emission lines from ionized species) much better, and in many nearby galaxies. In general, *Athena* X-IFU will provide spatially resolved high-resolution spectroscopy of shocked regions, their distribution, their spatial coherence, and ALMA can map the molecular phase in this regions, to capture the physics and effects of AGN winds on the multiphase gas. *Athena* might help determine whether the interaction between nuclear hot and extended cold winds happens in most of the sky as seen from the AGN, or only in some parts, and in general how spatially coherent these shocks are. This would help us to understand how BHs quench their own mass supply and how the M-σ relation is (most likely) established. The theoretical thinking, supported by numerical simulations (Zubovas & Nayakshin, 2014; Gaibler et al., 2012), is that some of the dense material is not pushed away by these energetic outflows and can still accrete on to the SMBH, but getting observational data to confirm or deny this would be particularly useful.

## 7.3. High redshift synergy

One obvious goal is to push these studies up to the peak epoch of AGN and galaxy assembly, z = 1 - 2. This can be done in spectroscopy with *Athena* (without spatial resolution). As of today, the multiphase galactic winds and their effect on the ISM could be observed in only a handful of z = 2 - 4 quasars (Cano-Diaz et al., 2012; Carniani et al., 2016), by exploiting the synergy between VLT/SINFONI and ALMA. These quasars are not currently studied nor detected in X-rays, but *Athena* will be able to detect many AGN and their nuclear winds at these redshifts. Together with ALMA studies of the molecular and or atomic ([CII], [CI]) phase, this will provide information on multiphase outflows at this epoch. In addition, through ELT/HARMONI and MOS it will be possible to study also the ionized phase of outflows in hundreds of AGN down to a bolometric luminosity of $10^{44}$ erg s$^{-1}$ at z = 1 and a few $10^{44}$ erg s$^{-1}$ at z = 2.

With regard to the role of AGN feedback in the heating of the cluster ICM, X-rays uniquely probe the relative importance of gravitational and non gravitational heating of the ICM in proto-clusters and -groups. In particular, in local clusters we know that the energetic input of radio AGN (radio mode feedback) is important in regulating the entropy budget and profile of the ICM. An additional and early energy input is, however, required to account for the ICM entropy profile. At high redshift, when the first proto-groups and clusters are formed, quasar mode feedback may contribute to this heating for two main reasons:

1. at z ~ 2 the AGN fraction is higher than at z ~ 0 (it can reach 10-30%);
2. the typical AGN is more luminous (a few $10^{44}$ erg s$^{-1}$ at z ~ 2).

*Athena* will be able to find proto-clusters and proto-groups thanks to its survey speed. In these *Athena* will measure thermodynamics, and detect AGN as point sources. Follow-up studies of AGN with ALMA (plus VLT/SINFONI and ELT/HARMONI, MOS) will be able to investigate the presence and properties of multi phase outflows in these AGN.

# 8. The Energetic Universe: Ultra-fast outflows

In the X-ray band UFOs are observed in the Fe K band through blue-shifted Fe XXV and Fe XXVI absorption lines (Chartas et al., 2002; Pounds et al., 2003; Tombesi et al., 2010, 2014; Gofford et al., 2015). These UFOs are highly ionized, with ionization parameter log $\xi \sim 4 - 6$ erg s$^{-1}$ cm, they have high column densities in the range $N_H \sim 10^{22} - 10^{24}$ cm$^{-2}$, and most importantly they show mildly relativistic velocities, from $v_{out} \cong 10{,}000$ km s$^{-1}$ up to $v_{out} \sim 0.3 - 0.5c$, where c is the speed of light (Chartas et al., 2009; Tombesi et al., 2015; Nardini et al., 2015). Such outflows seem to be common, being detected in about half of local radio-quiet Seyfert galaxies and quasars, and in bright radio galaxies as well (Tombesi et al., 2010, 2014; Gofford et al., 2015). More details on this science case can be found in Cappi et al. (2013).

## 8.1. Signature of the UFO/host interaction

Even if recent years have seen significant improvements in the study of X-ray winds and their implication on AGN feedback, there are still several compelling open questions. The microphysics of the winds, such as their density, turbulence, and ionization/velocity structure are still largely unconstrained. The driving mechanism of such winds, either radiation or MHD, is still unclear. The launching radii and geometry of UFOs are loosely constrained. Most importantly, these uncertainties propagate into the estimate of the mass outflow rate and kinetic power, which currently allow us to have only order of magnitude estimates. Last but not least, the signatures of the interaction of UFOs with the host galaxy environment, such as shocks, hot bubbles and galaxy-scale winds, are only becoming evident now thanks to the synergy with large facilities in other wavelengths.

The unprecedented sensitivity combined with the spatial/spectral resolution of *Athena* will allow us to conclusively address most of the issues regarding UFOs in both local and high-z AGN. In particular, simulations of PDS 456 from Cappi et al. (2013) have shown that the blue-shifted Fe XXV-XXVI lines from UFOs will be detected with very high significance, thereby allowing estimates of the velocity, column, ionization and turbulence of the gas with very high precision. Subsequently, this will allow one to conclusively constrain the location, geometry, and energetics of such outflows with only a few percent uncertainties. These estimates will be fundamental to reliably quantify the power of AGN winds to inflate the hot shocked bubble expected from simulations of AGN feedback.

## 8.2. ESO – *Athena* synergies

The detailed comparison of winds detected both in the optical/UV and X-rays in a sizeable sample of broad/narrow absorption line quasars, now restricted only to a handful of objects due to the limited signal-to-noise ratio (SNR), will then allow to systematically test the predictions of different wind acceleration mechanisms, and likely conclude if they are radiation or MHD driven (Proga & Kallman, 2004; Fukumura et al., 2010). The synergy between spectroscopic instruments on *Athena* and VLT and ELT instruments (X-Shooter for nearby sources and HARMONI at higher redshifts) will allow us to connect the hot flows close to the AGN to the colder, slower flows further out, and test the underlying physical mechanisms (energy or momentum driven).

As shown in Figure 9 for the representative case of the ULIRG NGC 6240, *Athena* X-IFU will provide high SNR spectra from 5´´ × 5´´ spaxels, thereby allowing us to simultaneously study both the nuclear X-ray source, even if highly absorbed, and the extended regions of photo-ionized or shocked gas. This will open the possibility to map the interaction between the AGN wind and the ISM, and compare it with regions of starburst emission.

However, the best results will be achieved only combining datasets at different wavebands, because they trace the gas at different locations, and with different ionizations, velocities, and densities. In particular, the comparison of *Athena*'s estimates of the UFO power with extended, cold molecular gas with ALMA will allow to conclusively demonstrate if the putative "quasar-mode" feedback due to AGN winds is indeed responsible for the origin of large-scale molecular outflows, and the subsequent quenching of star formation (e.g., Tombesi et al., 2015; Feruglio et al., 2015). Moreover, simultaneous X-ray and mm observations will allow us to investigate the complex mixing and interplay between the hot and cold phases. Spatially resolved observations with VLT/SINFONI and MUSE, will allow us to explore the ionized atomic phase and to investigate regions dominated by shock or photoionization emission.

The exquisite sub-arcsec spatial resolution of ALMA and the ELT/MICADO, HARMONI, and MOS instruments will permit the exploration of the presence of extended emission, bubbles and galaxy-scale winds to high redshift sources, especially at the peak of AGN and starburst phases at $z \simeq 1 - 4$. The synergy with the spectroscopic capabilities of *Athena* to study point sources at such high redshifts will enable the investigation of the presence of accreting BHs even in highly obscured cases, and the presence and power of X-ray winds, therefore allowing us to understand the build-up of SMBHs and galaxies at the heyday of the Universe (e.g., Georgakakis et al., 2013).

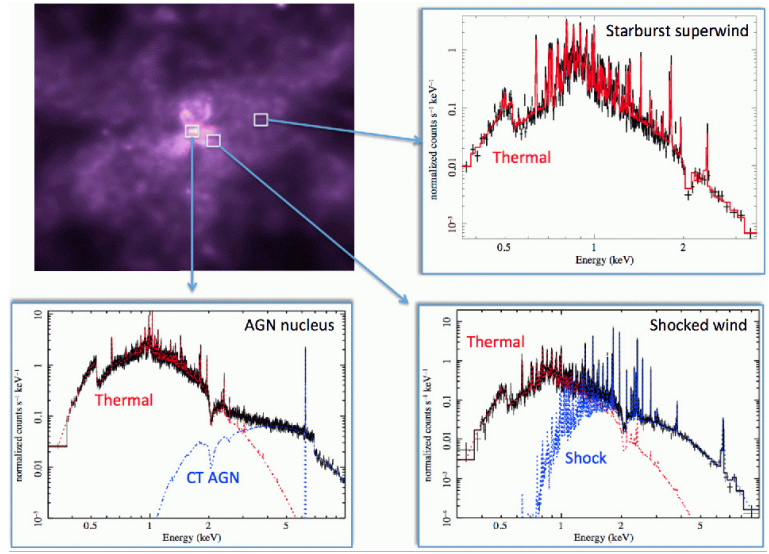

Figure 9: Top left: *Chandra* X-ray image of the nearby ultraluminous IR galaxy (ULIRG) NGC 6240 (Nardini et al., 2013). Top right: *Athena* X-IFU simulated thermal spectrum from a star-formation driven diffused superwind emission. Bottom left: X-IFU simulated spectrum of the nucleus of NGC 6240 including a soft thermal component plus a buried reflected component from the CT AGN double nuclei. Bottom right: X-IFU simulated spectrum of part of the plume of ionized emission south of the nucleus and attributed to 25% of starburst superwind (thermal) emission plus 75% of shocked emission. Figure adapted from Cappi et al. (2013). Model parameters from Feruglio et al. (2013) and Nardini et al. (2013).

# 9. The Energetic Universe: Accretion Physics

Accretion is a central theme of modern astrophysics. Most astronomical objects, from galaxies to stars and planets, are formed by accretion processeses and accretion on to neutron stars (NSs) and BHs powers the most luminous phenomena in the Universe. Associated with accretion on to compact objects is the existence of outflows, in the form of uncollimated winds or highly collimated relativistic jets. They are observed in XRBs and SMBHs at the centre of galaxies, and, in particular, jets are thought to be a fundamental ingredient of GRBs. BH jets and winds are uniquely capable of carrying away a significant fraction of accretion energy (potentially more, if the reservoir of BH spin energy can be tapped). While the extent of winds is not clear yet, there is evidence that jets carry energy from near event horizon scales out to scales hundreds of billions of times larger from the cavities that they carve into the ISM or IGM, which also give us a measure of the total energy dumped over long periods. The mechanical energy, as well as particles and electromagnetic fields, deposited can greatly impact the surrounding material, from "simple" heating to ionizing, both of which change the gas state and can affect star formation. The blanket term for this is "BH feedback". Specifically, when jets are involved it is termed "radio mode" or "maintenance mode" feedback and it is often seen in relatively nearby clusters of galaxies. BH feedback is thought to play a role in regulating the growth of the most massive galaxies. In particular, winds are thought to be fundamental for clearing out material and halting accretion during the quasar phase, which is short lived and can be super-Eddington. Even stellar mass BHs can inflate bubbles in the ISM with their jets, and all jets are copious accelerators of high-energy particles, i.e., CRs. We would like to understand the impact that BH winds and jets have on these two scales. However, we are still completely lacking an end-to-end, predictive theory for what comes out of a BH as a function of what goes in.

As discussed in the next sections, this field has enormously benefited and will continue to benefit from multiwavelength studies. These, however, have so far been scarce and very challenging because of the dearth of observing time made available for target of opportunity (ToOs) or more generally flexible scheduling of the telescopes or time-allocation-committees with experience in transient phenomena (see e.g., Middleton et al., 2016). This is aggravated by the fact that (near-) simultaneity is often requested for observations at different wavelengths, especially for XRBs. Therefore, the fantastic observatory capabilities expected at (sub-)mm, optical, IR and X-ray wavelengths in the next decade should be accompanied by new observing strategies and observatory policies to make such (near-) simultaneous multiwavelength campaigns and the consequent scientific advancement possible.

## 9.1. Jets

Notwithstanding the existence of relativistic jets in both XRBs and SMBHs, it is not clear yet how, where and why they are formed, whether the energy channeled into the jet originates from the accretion energy or is tapped from the spin of the BH, which is the acceleration and collimation process, and what is their connection to uncollimated winds and the accretion flow. The jet launching process itself is arguably the most hotly debated question in accretion physics today. The process of jet formation and acceleration in the extreme gravitational fields of BHs is likely the same for BHs of all masses. Indeed, jets are launched by BHs of all scales when in low-luminosity or super-Eddington states. For different reasons both of these accretion regimes result in a geometrically thick accretion disk, which seems to be a key ingredient for the launching of magnetized jets of plasma.

In recent years, simultaneous observations at radio/X-ray wavelengths have revealed a tight, nonlinear correlation between luminosity in these two bands in the low-luminosity accretion state of XRBs at which jets are present ("hard state"). The correlation, seen to hold over orders of magnitude during outbursts, has a very specific slope that has constrained the processes creating the X-rays to be radiatively inefficient, such as synchrotron and inverse Compton emission. Somewhat more surprising is the fact that SMBHs in configurations similar to the hard state (low-luminosity AGN [LLAGN]/LINERs, Fanaroff-Riley I radio galaxies,

BL Lacs) also follow the same correlation over cosmological timescales. In fact there seems to be a plane in the 3D space of radio luminosity, X-ray luminosity, and mass on which all BHs in these steady-jet states fall, referred to as the Fundamental Plane (Merloni et al., 2003; Falcke et al., 2004; Plotkin et al., 2012; but see also Panessa & Giroletti, 2013). The discovery of this plane would not have been possible without the radio/X-ray synergy. However, in the last decade it has become increasingly clear that the (sub-)mm and IR bands, in conjunction with the X-ray band, are key to fully characterize the jet emission.

Recent studies of the continuum synchrotron emission from supermassive and stellar-mass BHs have revealed a strong coupling between the (sub-)mm–IR bands and the X-ray band. For XRBs the (sub-)mm is a key band where a spectral break, which is associated with the first location of particle acceleration in the jets, moves through during outbursts (Russell et al., 2014). The frequency of this break anti-correlates with the accretion power as measured by the X-ray band, probably due to MHD-driven structures within the jets, providing critical constraints for self-consistent jet models (e.g., Polko et al., 2014). For SMBHs, we have seen now that the regions closest to the event horizon are brightest in the (sub-)mm, making nearby SMBHs like Sgr A* and M87 key targets with the Event Horizon Telescope, whose core is comprised of the phased up ALMA array. Even for systems where mm-VLBI is not the aim, continuum measurements of (sub-)mm and X-ray together (ideally simultaneously, with polarization!) provide the best constraints on models of jet launching and accretion physics than any two other bands.

While the regions closest to the event horizon, where jets are launched, are probed in the (sub-)mm band in SMBHs, in XRBs such regions are probed at OIR wavelengths, albeit on different timescales. In AGN, because of the very long dynamical timescales, we are limited to population-statistics studies to understand the evolution of what we observe. In contrast, in XRBs the accretion rate varies on timescales from years to ms. Such a variable accretion flow, on timescales which are very close to the light travel time of the BH event horizon (or NS surface) size scale (~ 0.1 ms), is presumably powering the observed relativistic jets (Malzac, 2013, 2014). This makes jets from stellar-mass BHs in XRBs (ranging from bright discrete plasma ejections to steady, continuously replenished jets depending on the accretion rate) very constraining for models of jet formation taking into account time-dependency.

### 9.1.1. ESO – *Athena* synergies

Observations of BH XRBs at high time resolution simultaneously in X-rays with *Athena* and at longer wavelengths with ESO facilities can probe jet physics over the whole expected range of timescales, offering possibly the best tool to study how the variability in the accretion flow is transferred into the relativistic jet. Changes in the jet speed and brightness are usually explained in terms of internal shocks, caused by variability in the accretion flow, but this has never been observed directly. *Athena* will expand on current studies on the accretion flow variability in X-rays, studying how/when the variability is driven by the soft disk, and transferred into and through the hard hot inflow.

What is needed, in order to obtain a breakthrough in this field, is to match this with similar time resolution and statistics in the IR, in order to complete the observational picture and allow a detailed modelling. IR fast timing is now starting to become available, and it has recently proved its potential, with the unambiguous detection of sub-second variability from a relativistic jet (Casella et al., 2010; Kalamkar et al., 2016, see Figure 10), following similar studies in the optical (e.g., Gandhi et al., 2008). These first data, obtained on a bright BH with VLT/ISAAC simultaneously with the X-ray observatories RXTE and XMM-*Newton*, provided us with an upper limit to the jet launching timescale, and allowed one to estimate the jet speed, as well as the size of the IR-emitting region, harnessing a time resolution as high as tens of ms. A simultaneous monitoring campaign with *Athena* and ESO (8m or even 4m class) IR facilities, aimed at monitoring the variable jet in a bright BH at high time resolution, would provide key information on how changes in the accretion flow influence the jet properties, thus helping to constrain the launching and powering mechanism. On the other hand, we stress that an existing ESO instrument such as VLT/HAWK-I, with its recently implemented fast-timing mode, if mounted on the ELT, would guarantee a high SNR, sufficient to provide the necessary information on a large number of sources, thus allowing population studies and generalising the results. Similarly, outstanding scientific results could be obtained by implementing a fast-timing mode on existing or planned IR detectors (e.g., on ELT/MICADO).

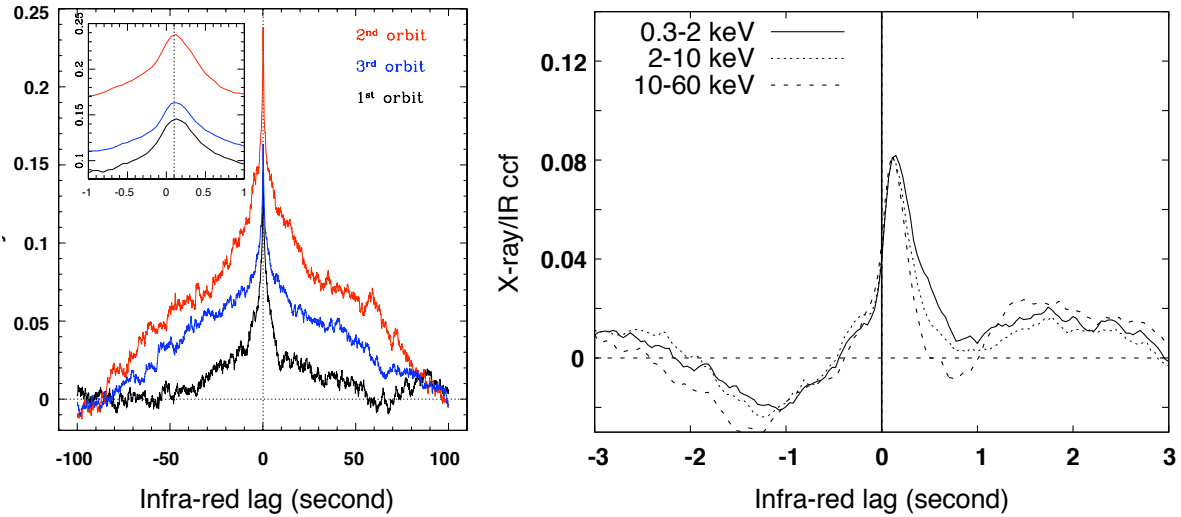

Figure 10: Left: Cross-correlation function between the X-ray and NIR light curves for the BH GX 339-4. The small positive delay (~0.1 s) indicates a jet origin for the IR variability and yielded an estimate of a semi-relativistic jet speed. This was the first detection of fast jet variability (adapted from Casella et al., 2010). Right: Cross-correlation function of NIR relative to hard (power-law dominated) and soft (disk-dominated) X-ray bands for the BH GX 339-4 (adapted from Kalamkar et al., 2016). The NIR variability is strongly correlated with the disk emission.

## 9.2. Accretion disk winds

Winds in accretion disks have become a major topic in studies of both XRBs and SMBHs in recent years. Similarly to jets, there are still key problems to be solved including the amount of energy and mass carried away, their launching mechanism (magnetic, thermal or radiative pressure), which could be different for different types of sources (XRBs or SMBHs) or even at different accretion states, and the impact of the winds on the systems themselves (e.g., on the disk structure) and on their environment.

In XRBs only recently has it been realized that winds could be a key ingredient for our understanding of these systems. Following the discovery that equatorial photoionized plasmas could be ubiquitous to all XRBs (Boirin et al., 2005; Díaz Trigo et al., 2006), statistical studies showed that such plasmas were detected as a wind in 83% of BHs and in 33% of NS XRBs and as a static atmosphere in the rest of the cases (Díaz Trigo & Boirin, 2016, Table 1), indicating that the amount of mass expelled could be very different for NSs and BHs, even at different accretion rates. Moreover, since the mass outflow rate was shown to be of the order of or above the mass accretion rate (e.g., Ueda et al., 2004, 2009; Ponti et al., 2012, Figure 6), this could lead to instabilities in the accretion flow and even trigger accretion state changes (Shields et al., 1986).

### 9.2.1. ESO – *Athena* synergies

In XRBs, winds have been mostly studied in the X-rays. In this band, *Athena* spectra will advance the field by precisely characterising the wind outflow velocities, column densities and ionisation state. The synergy between *Athena* and OIR instruments at ESO facilities could be fundamental to understand if there is a connection between the high velocity winds detected in two NSs via P-Cygni profiles of the Brackett γ lines (Bandyopadhyay et al., 1999) and the X-ray winds, and if so, whether winds are accelerated at large radii. The OIR band could also prove important to make a connection between winds and accretion state changes or length of the accretion outburst (Muñoz-Darias et al., 2016). Other examples of synergies between X-ray and OIR frequencies for wind studies include simultaneous SED fitting at such frequencies to recognise the presence of optically thick winds that block the X-ray light but whose effect can be indirectly seen by matching the irradiated part of the disc visible at optical frequencies and the inner disc visible at X-ray frequencies (Shidatsu et al., 2016).

In AGN, the X-ray evidence of winds and fast outflows is increasing in all types of AGN (Ballantyne, 2005; Reeves et al., 2009, 2010; Torresi et al., 2010, 2012; Tombesi et al., 2010; Longinotti et al., 2015). *Athena* spectra will be able to characterize these winds in great detail determining their properties over a wide range of ionizations, column densities and velocities. ALMA is already providing new discoveries of outflowing molecular gas in local AGN (Cicone et al., 2014; Combes et al., 2014; Morganti et al., 2015). The synergy with *Athena* will bridge nuclear and galactic winds shedding light on the formation and evolution of winds, their interaction with jets and their contribution to galaxy feedback (see Section 7 for details). The potential association of nuclear winds with water masers (Greenhill et al., 2003) could also be probed by searching for nuclear winds with *Athena* and for water masers with ALMA (Hagiwara et al., 2013) in the numerous sources that have already shown water masers at 22 GHz (e.g., Hagiwara et al., 2001).

## 9.3. Jet-disk-wind connection

Disentangling the different components that contribute to the SEDs of SMBHs and XRBs is key for our understanding of the interplay between what is going into a BH via disk accretion and what goes out in the form of jets and winds. Multiwavelength observations are once more fundamental for these studies.

Historically, AGN have been divided into radio-loud (RL) and radio-quiet (RQ) depending on the relative strength of the radio versus optical emission. RL AGN are able to launch powerful relativistic jets that emit from the radio to the high energy band, while in RQ AGN the origin of the weak, confined radio emission is still a matter of debate and the X-ray spectrum is totally dominated by the accretion disk and the hot corona emission. The X-ray spectra in RL AGN are characterized by weak reprocessing features with relatively flat photon indices and weak cold reflection signatures possibly due to a different disk ionization state, to lower column densities or covering fraction, or to a different illumination from the base of the jet (e.g., Grandi et al., 2002; Ogle et al., 2005; Ballantyne, 2007; Sambruna et al., 2009; Walton et al., 2013). Overall, it is difficult to decouple the accretion disk - corona radiation and the beamed jet radiation in RL AGN X-ray spectra. However, the separation and interpretation of the radiative processes in AGN is a major astrophysical problem. The high spectral throughput and resolution of *Athena*, combined with the unprecedented capabilities of ALMA, will revolutionize our understanding of the accretion and ejection phenomena on several grounds. For the first time, it will be possible to properly characterize the X-ray spectra of RL AGN. In particular the detection of the Fe K line will serve as a probe of the accretion power to be compared with the ALMA measurements at > 50 GHz, which will map the innermost regions at the base of the jet. Relativistically broadened Fe lines will be easily detected by *Athena* allowing a measurement of the BH spin (e.g., Brenneman & Reynolds, 2006; Goosmann et al., 2006). This will permit to test the spin paradigm which connects the high values of the BH spin with the production of collimated jets (Sikora et al., 2007), where large values of the spin should be predominantly found in RL AGN.

On the RQ AGN side, if present, a jet base should produce flat or inverted radio spectra dominating the mm bands ( ∼ 100 GHz). Several RQ quasars and Seyferts with high frequency excess in the radio continuum have been discovered (Antonucci & Barvainis, 1988; Behar et al., 2015) and this excess has been interpreted as self-absorbed synchrotron from a compact AGN core. However, Laor & Behar (2008) have suggested that this radio emission has its origin in accretion disk coronal emission. If the latter is confirmed, the detection of such emission by ALMA has the potential to characterize the magnetic field properties in the corona and determine its non-thermal content (Inoue & Doi, 2014).

### 9.3.1. ESO – *Athena* synergies

High-frequency radio spectra from ALMA observations obtained at high-angular resolution simultaneous with *Athena* observations will be able to discriminate between competing coronal models. Indeed, radio emission from synchrotron emission of power-law electrons in coronal models should produce a spectral break at around 300 - 1000 GHz, as the innermost corona becomes optically thin (Raginski & Laor, 2016), while a similar radio and X-ray spectral slope is expected if the X-rays are produced by comptonization from power-law electrons (Raginski & Laor, in preparation).

Both in RL and RQ AGN, the *Athena* and ALMA synergy will allow a characterization of energy ranges of the SED critical for inflow and outflow models, allowing the two contributions to the overall spectral emission to be disentangled. In particular, for LLAGN it will be possible to obtain high quality X-ray spectra for low luminosity objects (down to $10^{-15}$ - $10^{-14}$ erg cm$^{-2}$ s$^{-1}$ fluxes between 0.3 - 10 keV) and ALMA spectra in an energy range (50 - 300 GHz) so far uncovered. A compilation of LLAGN SEDs over a large range of Eddington ratios will trace accretion state transitions with unprecedented statistics and accuracy. This picture will be further completed by the advent of SKA in its complete phase by 2028 (Orienti et al., 2015).

The characterization of the SED is also of concern for XRBs, where we can track different accretion states in a single XRB, as opposed to AGN where such states are tracked in "different" objects (e.g., RL vs. RQ AGN). Besides the already mentioned radio/X-ray correlation in XRBs linking the jet to the innermost parts of the accretion disk and the corona, it has been recently suggested that the existence of jets and winds in XRBs could be anticorrelated (Neilsen & Lee, 2009). Simultaneous radio/X-ray campaigns targeted to observe transitional states where jets and winds appear and disappear have so far yielded controversial results: e.g., Díaz Trigo et al. (2014) concluded that changes in the SED during such transitional states were sufficient to make a wind "disappear" due to its full photoionization, while Hori et al. (2014) concluded the opposite. The discrepancy arises from different assumptions on the part of the SED that is not covered by the observations. This could be aggravated at the transitional states at which these studies are done since the jet could be significantly changing at OIR and (sub-)mm wavelengths due to the change in the spectral break frequency (see Section 9.1). Moreover, the presence of winds in states where jets are present could reveal itself at OIR wavelengths (Wu et al., 2001; Rahoui et al., 2014) instead of the X-ray wavelengths at which winds are typically tracked. In summary, the combination of *Athena* and ELT/HARMONI for wind studies with high-resolution spectroscopy at both X-ray and NIR frequencies and ELT/HARMONI and ALMA for spectral continuum measurements in the NIR and (sub-)mm bands for jet studies, should prove fundamental to disentangling contributions of the jet, wind and disk at a given accretion state and thus unveil the jet-disc-wind connection in XRBs.

## 9.4. ULXs

Ultra-luminous X-ray sources (ULXs) are defined as off-nuclear X-ray point sources with $L_x$ = $2\times10^{39-42}$ erg s$^{-1}$ (Colbert & Mushotzky 1999). Those ULXs with $L_x > 10^{41}$ erg s$^{-1}$ suggest the presence of intermediate mass ($10^{2-5}$ $M_\odot$) black holes (IMBHs) if they radiate isotropically at sub-Eddington levels as observed in stellar BHs and AGN. The detection of gravitational waves (GWs) from a binary BH merger, producing a final BH of 62 $M_\odot$ (Abbott et al., 2016) proves the existence of BHs more massive than the canonical value of 10 $M_\odot$ that is generally found in XRBs hosting a BH (Casares & Jonker 2014), although it is not clear that such binary BH mergers could produce ULXs.

IMBHs could be the building blocks of SMBHs (e.g., Ebisuzaki et al., 2001) and if they exist they could help explain the puzzling observation that, even at z > 6, SMBH with masses >$10^9$ $M_\odot$ exist (e.g., Bañados et al., 2016). Alternative to an IMBH explanation for ULXs is that their host BHs accrete above the Eddington limit, or at or below the Eddington limit but with their emission significantly beamed.

## 9.4.1. ESO – *Athena* synergies

A strong He II λ4686 emission line from the surrounding nebula is detected in a ULX, which seems to rule out that beaming in that ULX is responsible for the high luminosity (e.g., Kaaret et al., 2004). Continued VLT (e.g., HAWK-I, FORS2, and MUSE) and future ELT observations of ULX nebulae to study BH feedback and the amount of beaming are important for the question of whether BHs can accrete above the Eddington limit, and for the question if IMBHs exist. Planned ELT instrumentation such as HARMONI, MICADO and MOS will be important for these studies. As explained below, mass measurements can be obtained from time-resolved NIR spectroscopic observations of red-supergiant mass donors to ULX such as those identified by Heida et al. (2014). Given that the best X-ray positions afforded by even the *Chandra* satellite can yield multiple candidate counterparts NIR IFU observations can in several cases be necessary to detect orbital motion of the counterpart. The problem of multiple candidate counterparts in an X-ray localization error region will also exist for transient ULX sources newly discovered by *Athena*. *Athena* X-IFU studies of the ULX point source can help determine the nature of the nebulae by studies of outflows from the ULX (e.g., Pinto et al., 2016).

Important for the study of ULXs is the direct measurement of the BH masses. If we can identify the mass donor to the ULX, one has a single lined spectroscopic binary. If the orbital motion of this mass donor star can be determined one gets a lower limit to the mass of the BH. Identifying the mass donor in ULXs turned out to be difficult as the very bright accretion disk often outshines the companion star in all but a few cases (Motch et al., 2014). A campaign to obtain NIR images of all ULXs within 10 Mpc with the goal of identifying red-supergiants mass-donors to ULXs was recently started (Heida et al., 2014, see Figure 11). Those red-supergiants are bright in the NIR where they may dominate the emission. The 10 Mpc limit is brought about by the need to be able to take spectra of the potential counterparts with 10m class telescopes. Spectra are needed to secure the identification as a red-supergiant and to measure the radial velocity variations due to Doppler shifts in the location of the many spectral absorption lines (cf. Heida et al., 2015; Heida et al., 2016).

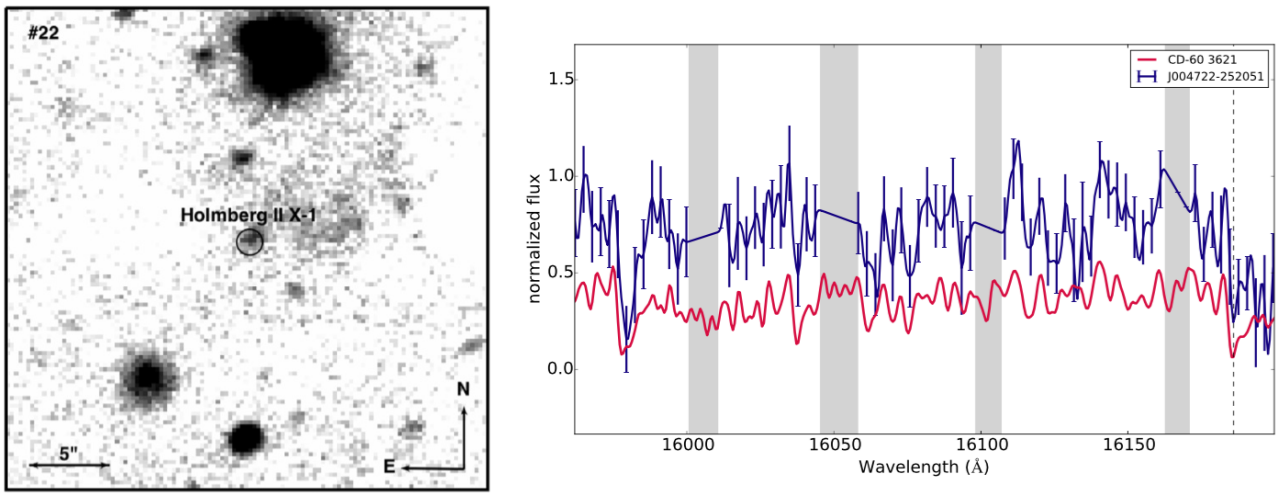

Figure 11:  Left: Figure from Heida et al. (2014) showing the NIR candidate counterpart to the ULX Holm II X–1. Right: Part of a NIR VLT/X-Shooter spectrum of a ULX in NGC 253 (top spectrum) showing the comparison to a high SNR X-Shooter spectrum of a red-supergiant from our own Galaxy (bottom spectrum). The grey-ed out areas are places where telluric emission from the Earth's atmosphere dominates the light.

# 10. The Energetic Universe: Transient Science

At the time of the launch of *Athena* and first-light of the ELT, transient science will be an important mainstream aspect of astronomy as transient science is (among) the main goals of facilities such as LIGO/Virgo, eROSITA, ZTF (http://www.ptf.caltech.edu/ztf), the LSST, and SKA. ESO is preparing for and contributing to this transition by the public survey PESSTO (http://www.pessto.org/) and the new NTT instrument SOXS. We highlight three aspects of transient science here: understanding GRBs, using GRBs as back-lights for studies of their host galaxies, and tidal disruption events (TDEs) for the study of the formation of SMBHs and the potential existence of IMBHs. For more background on this science case see Jonker et al. (2013a).

## 10.1. GRB studies

GRBs are the most luminous sources at all redshifts. They come in two generic types, short (SGRBs) and long (LGRBs) with a division around 1 - 2 seconds, as measured from the duration of the detected γ-ray light. In both GRB types the spectra are thought to be primarily non-thermal, and a variety of features are seen in the light curves: flares, plateaus, rapid decays etc. These are suggestive of common jet acceleration and emission processes. However, the progenitor is different as evidenced by the variety of locations and host galaxy properties seen.

LGRBs are associated with regions of recent star formation and several have been associated with SNe, supporting a massive star collapse origin (a so-called collapsar). SGRBs in contrast are not associated with SNe and are found in galaxies with older regions of star formation, or indeed no recent star formation, supportive of an older progenitor. This is most likely a binary merger involving at least one non-black hole (e.g., two NSs). Current GRB samples also show a median redshift difference between LGRBs and SGRBs, with the former having a significantly larger median redshift ($z \sim 2.2$ compared to $z \sim 0.8$). This is partly due to SGRBs having a lower luminosity but may also be related to the progenitor (binaries take some time to merge).

*Athena* has a high priority science case to use GRBs as a probe of the distant Universe, but the same data can also probe the progenitor and physics of the jet. To make full use of *Athena* data and to pre-select the best targets for *Athena*, it is essential to obtain complementary multiwavelength data on the GRB host and light curve. ESO has a long history of GRB studies and current and future facilities are well suited to the task.

### 10.1.1. ESO – *Athena* synergies

Here we note some synergy examples:

1. To identify high-redshift GRBs OIR data are required, as the γ-ray or X-ray data from the current generation transient finders do not provide high-fidelity redshift information (this may change for future transient facilities but even then high-resolution spectroscopy and imaging will only come from ESO complementary data). Once a GRB has been identified, redshift information can be obtained using either multi-band imaging (e.g., a GROND-like instrument [http://www.mpe.mpg.de/~jcg/GROND/]) or spectroscopy (e.g., VLT/X-Shooter). Early data are required to enable *Athena* to make best-use of the available X-ray light and hence a Rapid Response Mode (RRM) is required on the ground facility.

2. To understand GRB progenitors it is essential to combine γ-/X-ray data with OIR/sub-mm/radio imaging and spectroscopy of the local environment. For example: is the GRB in a host galaxy currently forming stars? What is the mass of the host? What is the local metallicity (study Pop I/II/III)? And what is the age of the host (e.g., a binary may take a long time to merge)? Identification of an associated SN through OIR photometry and spectroscopy using for instance ELT/MICADO and HARMONI observations can provide collapsar evidence. ELT/MICADO IR data a few days after the trigger can probe the so-called macronova due to the r-process in residual material from a binary merger (this can also be used to locate electromagnetic counterparts to GW sources). Whereas one could argue that ESO facilities do not require *Athena* observations for these goals, such observations would provide a much more coherent picture of the GRBs.

3. For jets we want to understand the particle acceleration and emission process(es). The prompt luminosity is dominated by γ-/X-ray emission but to probe the spectral shape, particle energetics, forward/reverse shock contributions etc., we require both early and late-time multiwavelength data. While synchrotron emission is thought to dominate the prompt and afterglow emission from GRBs, there is some evidence for thermal (photosphere) emission and there are multiple possible locations along the jets which can be involved. Time-dependent data are essential, so RRM is required along with flexible scheduling at late-times. We note that the CTA-South (https://www.cta-observatory.org) facility to be located at the ESO Paranal site will also provide complementary data.

4. Additional insights into jet physics can be provided by polarization data such as those currently obtained using VLT/FORS2 (Wiersema et al., 2014), although it is unclear whether there will be an ELT optical/NIR imager with similar capabilities (for example, the ELT planetary camera and spectrograph [PCS] instrument-concept is unlikely to be suitable for polarization measurements of the faint GRB afterglows). Limited data to date show evidence for a strong reverse shock contribution to optical polarization at early times and for (in one case) unexpectedly strong circular polarization. The availability of ESO instrumentation capable of obtaining polarization data in the *Athena* era would be very useful.

Finally, we note that GRB studies require the discovery of GRBs! We will be able to study existing GRB hosts as new powerful facilities, such as ELT, come on line. However, several of the science areas above require prompt data. Hence, a transient finder in space is a requirement. If such a facility had on-board redshift determination capability that would greatly assist prioritization and the accumulation of complete GRB samples (currently only about one third of GRBs have redshifts determined).

## 10.2. High-redshift GRBs illuminating the first star forming regions

For the brief moments of their existence, LGRBs and their afterglows are the most luminous beacons in the Cosmos. Their association with the deaths of massive stars allows us to pinpoint the locations of the first star forming regions in the very early Universe (z > 7) and to study the chemical fingerprints of the first generations of stars. By using the bright, but short lived, afterglow emission as temporary backlights, high-resolution spectroscopy probes the dust content, gas metallicities, abundance patterns, and HI fraction of their birthplaces and through sight lines of their host galaxies. Once the afterglow has faded, observations of high-redshift host galaxy light in emission can provide important constraints in particular on properties of galaxies at the faint end of the luminosity function, which are important for understanding the sources responsible for the re-ionization of the Universe.

### 10.2.1. ESO – *Athena* synergies

High-redshift GRB afterglows are high-priority science targets for *Athena* and ESO and harbour the possibility for strong synergies between the two facilities. Specifically, one of the science objectives of *Athena* is to observe at least twenty-five z > 7 GRB afterglows using X-IFU high-resolution spectroscopy. *Athena*'s ToO response time of < 4 hr will allow one to observe moderately bright X-ray afterglows (fluence of $10^{-6}$ erg cm$^{-2}$) with sufficient SNR. Observations in the X-ray band provide the unique capability of simultaneously probing all elements from C to Ni in all their ionization and binding states and thus offer in particular access to the ionized gas (for z > 7 C transitions are redshifted into the EUV; transitions of Si and heavier elements will still be in the *Athena* band though). In contrast, rest-frame UV-optical observations probe predominantly the neutral gas component but have access to the very important hydrogen transitions. The latter are needed for estimating metallicities. This complementarity will help distinguishing between the relative chemical abundances distinctive of primeval (Population III) explosions and evolved (Population II) stellar populations using ratios, such as [Si/O] vs. [C/O] and [Fe/C] vs. [Si/C] (e.g., Ma et al., 2016). For absorption systems with low metal column densities ELT/HARMONI or HIRES (Section 13.1) spectra could play an important role for *Athena* by providing guidance for identifying the corresponding absorption features in the X-IFU data. The high-resolution capabilities of the X-IFU and of ELT/HIRES will also allow studies of the kinematics (inflows, outflows) and turbulence of the ionized and neutral components and thus constrain the structure and geometry of the ISM along the sight line to the death location of the GRB progenitors.

Some optical/NIR afterglows can be very bright even at high redshift, and thus accessible for high-resolution spectroscopy with, e.g., VLT/X-Shooter. However, the majority of the z > 7 sources will require ground-based observations with the ELT in RRM. As observing time at the ELT and with *Athena* will be on extremely high demand, and in order to maximize the synergy, a coordination of target priorities between the two facilities would be important. A mechanism would be desirable by which high-redshift afterglows that trigger observations at one of the facilities would receive an appropriate priority also at the other facility. A first step in this direction would be to allow the submission of joint *Athena*-ELT proposals for ToO observations[2].

As also alluded to in Section 10.1.1 the most important information required before triggering *Athena* is the redshift of the afterglow. Current GRB detection facilities, such as *Swift*, provide very accurate afterglow positions but require ground-based follow-up activities for obtaining either photometric (e.g., GROND) or spectroscopic (e.g., VLT, GTC, Gemini, Keck) redshifts. The need for ground-based observations will depend on the capabilities of the next generation GRB satellite concepts (e.g., THESEUS: http://www.isdc.unige.ch/theseus) but it is expected that support from 2-8m class telescopes with rapid response mode capable instruments such as VLT/X-Shooter or NTT/SOXS will remain critical also during the ELT and *Athena* era.

GRBs are in principle very efficient in selecting high-z objects, however only 6(2) sources with redshift larger than 6(7) have been identified up to know. This is thought to be the result of the limited sensitivity and energy band choice of the current generation of satellites. The extrapolation of the peak flux distribution to lower fluxes suggest that up to 10% of all GRBs are at z > 6 (Salvaterra et al., 2011) and a mission with higher sensitivity, in particular at lower energies where the peak flux of high-redshift sources would be shifted into, will be necessary. With a significant sample of high-redshift GRBs at hand we can also look at the emission properties of their host galaxies. Despite significant efforts with very deep HST observations with H-band limits of up to $m_{AB} > 30.3$ in the case of the z = 8.23 GRB 090423 (Tanvir et al., 2012), only 2 of the z > 6 host galaxies, namely GRB 050904 at z = 6.295 and GRB 140515A at z = 6.327 have been detected so far (McGuire et al., 2016). This indicates, in agreement with numerical (e.g., Trenti et al., 2012) and semi-analytical modelling (e.g., Salvaterra et al., 2013), that the high-redshift host galaxies sample the faint end of the galaxy luminosity function and have SFRs of the order of 0.03 - 0.3 $M_\odot$ yr$^{-1}$. Even with the power of ELT/HARMONI only the brightest subset of the z > 7 host galaxies will be detectable and for only a few morphological studies with ELT/MICADO will be possible.

## 10.3. Tidal disruption events

SMBHs with masses larger than $10^9$ $M_\odot$ have been found already when the Universe was less than 1 Gyr old (e.g., Bañados et al., 2016). SMBHs can just form quickly enough out of continued Eddington-limited accretion of gas on to 100 $M_\odot$ seed BHs. However, feedback can disrupt the gas supply and slow the accretion rate (Silk & Rees 1998). In order to solve this problem, one can:

---

2 The problems for quasi-simultaneous ground- and space-based observations (e.g., communication between facilities, visibility and weather conditions) are recognized but coordinated programs on best-effort basis should be supported.

1. start with more massive seed BHs such as IMBHs;
2. grow partially due to merging together BHs;
3. allow mass to be accreted at a rate higher than the Eddington limit.

TDEs can help solving the conundrum of the formation of SMBHs as they might provide insight into accretion at super-Eddington accretion rates, they may point us to SMBHs that are on the verge of merging, and their rate is a probe for the mass of the seed BHs that evolved into SMBHs.

For several decades, astronomers have speculated that a hapless star could wander too close to a SMBH and be torn apart by tidal forces. It has only been with the recent advent of numerous wide field transient surveys that such events have been detected in the form of giant-amplitude, luminous flares of electromagnetic radiation from the centres of otherwise quiescent galaxies. The discoveries, spanning the whole electromagnetic spectrum from X-rays, over UV and optical events, to a small number of events launching relativistic radio jets, have caused widespread excitement, as we can use these TDEs to study SMBHs and their surroundings in quiescent galaxies. Whereas AGN (such as quasars) host SMBHs that are supplied by steady streams of fuel for periods much longer than a thousand years, TDEs offer a unique opportunity to study a single SMBH under feeding conditions that change over timescales of days or months. TDEs offer our only hope of studying the evolution of their accretion disks for a wide range of mass accretion rates and feeding timescales. In addition, because the rate of TDEs is massively enhanced in binary BH systems, TDEs are expected to point us to galaxies that are likely to host compact binary SMBHs. At a later stage in the binary SMBH evolution, LISA (https://www.elisascience.org) will study the GWs emitted when such binaries merge. In the next few years, the largest growth area will be in the greatly expanded surveys of the transient sky (such as the Gaia satellite, ZTF, LSST, the eROSITA satellite, SKA, etc.) that will reveal how SMBHs shine by ripping apart orbiting stars and swallowing the stellar debris.

A fast X-ray transient which is conceivably located at the distance of M86 has been recently discovered (Jonker et al., 2013b; see Figure 12). If true, the properties of the transient are consistent with tidal disruption of a white dwarf by an IMBH (e.g., MacLeod et al., 2016). Such an event would be direct dynamical evidence for the existence of IMBHs, which would help solve the problem of SMBH formation.

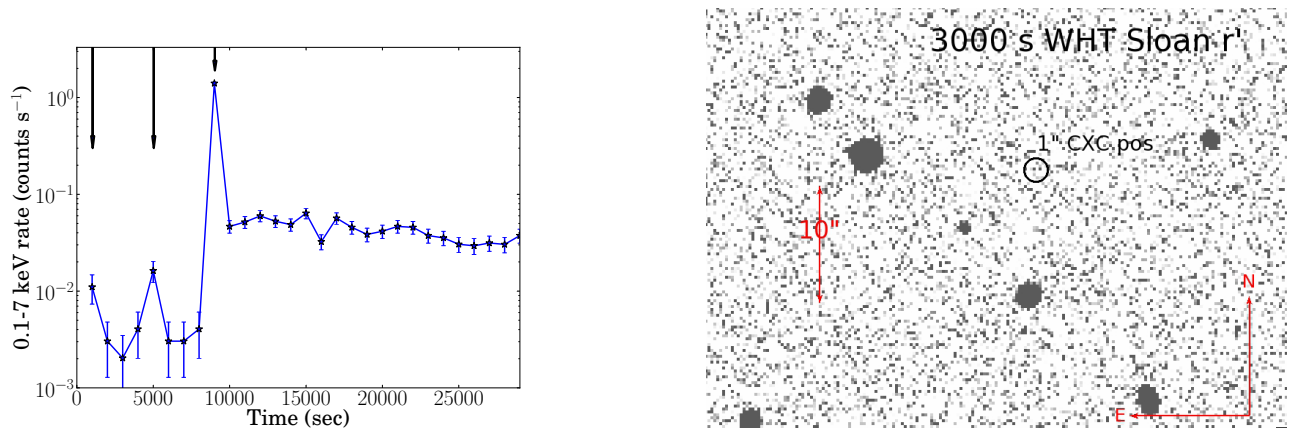

Figure 12: Left: X-ray light curve of the fast-X-ray transient found by Jonker et al. (2013b). Two precursor events are present, with a time delay that is similar in size to the orbital time scale expected for a white dwarf orbiting an IMBH. Figure from Jonker et al. (2013b). Right: Deep WHT r[1]-band observation of the field of the transient found in Jonker et al. (2013b).

Luo et al. (2014) and Glennie et al. (2015) found additional events with similar characteristics and the latter authors calculated event rates, showing they could be as high as $10^5$ over the whole sky per year with 0.3-7 keV fluxes above $2 \times 10^{-10}$ erg cm$^{-2}$ s$^{-1}$. Optical follow-up of the event found by Luo et al. with latency of about 18 days found no optical event down to deep limits (R=26.1; Treister et al., 2014) although a potential host (dwarf) galaxy was found nearby on the sky in deep HST observations (with AB magnitudes R ~ 30, z ~ 28, and $K_s$ ~ 26; Luo et al., 2014). This shows the need for fast and deep OIR follow-up of such newly discovered X-ray transients.

### 10.3.1. ESO – *Athena* synergies

*Athena* WFI with its more than 8 times larger FoV and its 100 times higher effective area at 1 keV than *Chandra* (ACIS-I detector) should thus find ~ 1000 such events serendipitously per year (conservatively assuming there are an equal number of sources at each flux level). *Athena* is investigating whether it can send out alerts after finding these. Follow-up with deep imaging and spectroscopy such as with VLT/FORS2 or ELT/MICADO or with IFUs such as VLT/MUSE and ELT/HARMONI when the source localisation (for instance in case of an *Athena*-discovered event) is insufficient for acquisition on a slit, is necessary to classify and extract astrophysical parameters such as BH mass and spin of these enigmatic events. This would be facilitated by an RRM on VLTs as well as perhaps on the ELT.

Besides these fast events, more "regular" TDEs require high quality multiwavelength data over multi-epochs to determine the emission mechanism for the optical and X-ray light observed in these systems. This is needed to reap the full potential of TDEs for addressing questions of SMBH growth, jet formation, stellar populations and their dynamics in galactic nuclei, and the physics of BH accretion under extreme conditions including the potential to detect relativistic effects near the SMBH. Whereas simultaneous multiwavelength observations may not be necessary, contemporaneous observations will be. Hence there is a need to plan observations together for *Athena* and ESO facilities.

# 11. Observatory Science: Star Formation

## 11.1. The early stages

Irradiation of the ISM from both X-rays and CRs leads to the creation of secondary energetic electrons. These electrons hit the H atoms and $H_2$ molecules, leading to the ionization and heating of the gas. High-energy irradiation is the only way to ionize dense molecular gas, where UV photons do not penetrate. Once the H atoms and $H_2$ molecules are ionized, they quickly react with $H_2$, the most abundant species in the gas, and form the molecular ion $H_3^+$. This ion then passes the positive charge to other species, such as $OH^+$, $H_2O^+$, $HCO^+$, and $N_2H^+$. By observing the abundance of these species one can then derive the flux of X-ray or CR irradiating the gas. Absorption spectra of species such as $H_3^+$ in the NIR can provide us with such rates. However such observations are limited to the presence of strong IR continuum background sources (Neufeld et al., 2010) and hence ionization maps cannot be derived with this method. On the other hand, observations of the low rotational lines of HCO+ and $N_2H^+$ and their deuterium (D)-containing isotopologues have transitions in the mm, and hence they are excited in cold gas. Their transitions can be observed with ground-based telescopes like ALMA; in particular Bands 2 and 3 will give us access to the J = 1-0 transition of these species.

Guelin et al. (1977) proposed to use the $R_D$ = $HCO^+$/$DCO^+$ abundance ratio to estimate the ionization of cold (≤ 30K) molecular gas, as molecular gas ionization is inversely proportional to $R_D$ and depends on the exponential of the gas temperature. In addition, Ceccarelli et al. (2014) proposed to use $R_N$ = $HCO^+$/$N_2H^+$ in warm environments where $R_D$ cannot be used.

### 11.1.1. ALMA – *Athena* synergies

For objects where large gas column densities prevent an X-ray detection, molecular observations, leading to the measurement of $R_D$ and $R_N$, will be complementary to *Athena*. Example of synergies between *Athena* and ALMA are:

- Detection of X-ray emission from deeply embedded Class 0 sources. During the first phases of the birth of a Sun-like star, represented by Class I-III sources, copious X-ray emission is observed. This emission is likely due to the flares caused by magnetic reconnection on the surface of convective stars and, to a lesser extent, to star-disk reconnection (Bouvier et al., 2013). On the other hand, X-ray emission has not yet been detected in younger objects, the so-called Class 0 sources. This is not because Class 0 sources do not emit X-rays but, rather, because their circumstellar envelopes have large column densities of gas and dust (≥ $10^{24}$ cm$^{-2}$), which absorb X-rays (Giardino et al., 2007). Determining the X-ray fluxes from Class 0 stars is important because the induced enhanced ionization might couple the gas with the magnetic fields contrasting the infall motion; the subsequent heating may result in an enhanced pressure contrasting the infall as well. Single dish observations (Ceccarelli et al., 2014) already led to the discovery of an intense high-energy source in a very young embedded proto-cluster in the Orion Molecular Complex (see Figure 13). ALMA, with its high sensitivity, will be able to carry out observations of $N_2H^+$ and $HCO^+$ in a large sample of embedded X-ray sources and, thus, will provide information complementary to what *Athena* will do.

- High spatial resolution maps of X-ray emission in systems of multiple protostars. Ground-based observations of molecular ions are essential even if *Athena* detects X-ray emission from a protostar and/or a proto-cluster. Stars are often formed in multiple systems: binaries and triple systems, and clusters. X-ray emission from one or more sources in the multiple system and cluster would greatly affect the evolution of the single components of the system/cluster as well as the system/cluster itself. It is, therefore, of extreme importance to disentangle what source is responsible for the X-ray emission detected by *Athena*: ALMA, with its great spatial resolution, will be able to map the molecular ion emission on sub-arcsec scales (smaller than the Solar System diameter), and, consequently, disentangle the contribution from the different components of the system/cluster emitting the signal detected by *Athena*.

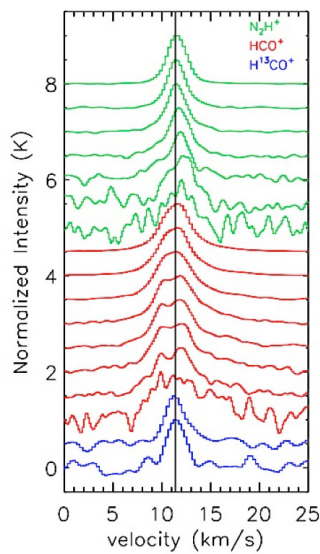

Figure 13: Herschel observations of the ions HCO$^+$ and N$_2$H$^+$ towards the proto-cluster OMC-2 FIR4 (Ceccarelli et al., 2014); the spectra are taken from Kama et al. (2013).

## 11.2. Protoplanetary disks

Protoplanetary disks, the birthplace of planets, surround pre-main sequence stars for long times (~$10^6$ years: Haisch et al., 2001), during which they evolve both physically and chemically. How dust and gas turn into planetary systems and how the diversity of planets depend on the disk chemical properties are among the open questions in protoplanetary disks studies. The answer to these questions requires a multiwavelength approach. Several processes affect the evolution of the disk and the subsequent planet formation, such as ionization and heating of the gas, as well as turbulence. Most processes are highly affected by the central source. In particular, X-ray emission is likely to be correlated with turbulence and the disk "dead" zones, i.e., regions of low ionization which impede angular momentum and mass transport leading to the build up of material in a ring-like structure (Armitage, 2011).

Young stellar objects are strong emitters of X-rays, which affect the chemistry of the protoplanetary gas, as well as ionize and heat it (disk photoevaporation due to X-rays shall be discussed in the next section; here we concentrate on disk heating and ionization). Disk heating will then affect both planet formation and migration, while an increase in ionization will lead to a stronger coupling between gas and magnetic fields, in turn affecting the disk evolution.

*Athena* will observe the rotational modulation of the iron Fe 6.7 keV line as well as other ions in order to constrain the X-ray fluxes from Young Stellar Objects (YSOs) (see Sciortino et al., 2013). X-rays are not the only emission affecting the chemistry and dynamic of disks: the reactions of CR ions will also create ionization and heating.

### 11.2.1. ALMA – *Athena* synergies

There are several sources of ionization in a protoplanetary disk, beside X-rays:

1. stellar and interstellar UV, only important in the more exposed parts of the disks;

2. decay of short-lived radionuclides (SLRs), most effective if the low mass stars are born in an environment where massive star formation has also occurred;

3. thermal;

4. CRs (Glassgold et al., 2012).

The latter penetrate the disks and hence are important throughout the gradient of densities and temperatures. X-rays and CRs, however, do lead to very similar chemistries and hence discerning between these two sources is not trivial. Moreover, the CR rate is rather uncertain in our own galaxy (Dalgarno, 2006) and X-ray ionization rates are certainly variable. The chemistry resulting from cosmic ionization is best studied in the sub-mm. In fact, an indirect way to measure the CR ionization flux is indeed via the study of the most abundant ions (see Section 11.1). Hence, three potential synergies can be envisaged between *Athena* and ALMA:

■ Simultaneous observations with *Athena* and ALMA. Although challenging, simultaneous observations with *Athena* and ALMA would be desirable. ALMA's high spatial resolution (0.1 - 0.2´´) capabilities will allow us to look close to the young star, where the timescales are very rapid. Observations of key ions ($N_2H^+$, $HCO^+$, $DCO^+$, $H_2D^+$) with ALMA together with X-ray flux measurements should allow us to determine, separately, the CR and X ray ionization rate as a function of time.

■ Observations in the X-rays and sub-mm of a sample of protoplanetary disks at different evolutionary stages will allow us to constrain the average CR ionization rate as well as the average X-ray ionization rate. Models (Cleeves et al., 2014) find that sensitive observations with ALMA of close-by (100pc) disks can distinguish between "high and low CR" disks; however at the moment these models are hampered by the fact that the X-ray flux's impact on ionization and chemistry is highly uncertain because the shape of the spectrum beyond E ~ 2 keV in both quiescent and flaring states is not known.

■ High spatial resolution observations with ALMA that allow us to constrain the CR ionization rate as a function of distance from the protostar will help us find the "dead zones", locations in the disk where there is an increase in accretion stress from a low ionization zone: the chemical structure of the dead zones will be significantly different from that of other regions within the disk. The location and extent of the dead zones may depend on the X-ray intensity because the X-ray ionization induces magnetorotational instability turbulence. Turbulence in turn affects the linewidth of CO lines, observed with ALMA (see Figure 12 in Flock et al., 2015). It is therefore clear that X-ray and sub-mm measurements are both necessary for the determination and characterization of such zones.

## 11.3. Young Stellar Objects

Low mass YSOs (M ≲ 1 M$_\odot$) are active sources of high energy radiation (Feigelson et al., 2007). The high energy radiation feedback that young stars produce on the parental molecular clouds, protoplanetary disks and young planets is thus much stronger than for normal main sequence stars. The key questions that need to be addressed are the extent to which this feedback determines the outcome of the star and planet formation process and what is the ability of primordial planetary environments to develop into life-supporting biospheres.

An initial attempt to determine the effect of X-ray feedback on molecular clouds and star formation was carried out by Lorenzani & Palla (2001), who focused on the suppression of star formation by X-ray-created ionized bubbles in molecular clouds. In recent years, much of the focus has shifted to the effect of X-rays on the chemical processing of gas and dust in the ISM (see also Section 11.1) and most of the progress has focused on laboratory measurements (e.g., Ciaravella et al., 2016). However, the availability of sensitive IR high-resolution spectrographs and ALMA will allow us, in synergy with X-ray measurements from *Athena*, to derive more stringent observational constraints on the different chemical composition as a function of X-ray irradiation in molecular clouds.

The effect of X-ray radiation on the immediate surrounding of each YSO is expected to be a determining factor for the chemical evolution and dissipation of the protoplanetary disk during the formation stage. X-ray photoevaporation is believed to be one of the main mechanisms that will determine the termination of planet formation and disk dispersal (Alexander et al., 2014, see also Section 11.2.1). Detailed comparisons of X-ray photoevaporation models and observations have been mostly of a statistical nature. We have only a very few cases in which a detailed comparison of different observables is possible, the best one probably being the TW Hya disk, where the combination of the X-ray emission from the star is well characterized and both the wind and disk characteristics can be probed with the necessary sensitivity and spatial resolution in the IR and millimetre (Ercolano et al., 2017).

### 11.3.1. ESO – *Athena* synergies

In the *Athena* era we do expect that continuum and CO observations with the upgraded ALMA (at a resolution ~0.05´´), as well as ELT/METIS and HIRES IR high spectral resolution observations of icy dust mantle features will allow us to probe in detail disk evolution for a significant sample (hundreds) of nearby star forming regions.

The chemical processing of solids, gas, and ices in the protoplanetary disk by high energy radiation is thought to be an essential step to develop the chemical complexity leading to pre-biotic molecules and eventually life-sustaining biospheres. Yet so far the observational constraints of these processes are scant and the theoretical interpretation is debated. In combination with high angular resolution and high sensitivity instruments in the IR and millimetre, *Athena* has the potential of linking the characteristics of X-ray emission (intensity, spectral hardness, variability) with the chemical processing in the inner regions of the disk. Another potential key development that will be made possible by ELT/HIRES high-resolution spectroscopy (potentially up to R = 100,000) of planetary atmospheres will be the study of the effect of X-ray illumination on planetary atmospheres and their evolution, since the main molecular absorbers (TiO, CO, $H_2O$, $O_2$, etc.) in such atmospheres will be affected by X-ray irradiation (see Maiolino et al., 2013).

# 12. Observatory Science: Stars

## 12.1. High energy emission from low mass stars

Flare activity is a highly dynamic phenomenon occurring in stellar atmospheres associated with the release of magnetic energy in reconnection events (Martens & Kuin, 1989; Tsuneta, 1997). Active stars are the best candidates to observe flares, since their occurrence rate correlates with the activity level of the star. A specific advantage of late-type stars like M dwarfs for flare studies results from their relatively cool photospheres, that provide a distinct contrast to the hotter flare plasma. Flare related emission is seen at virtually all wavelengths from radio to X-rays and thus multiwavelength observations are well suited to study stellar flares, their evolution, and to put them into context with models that are based largely on spatially-resolved flare studies from the Sun. High-energy emission from low-mass stars is ubiquitous through the main sequence. It originates in the million degree hot coronal plasma that is created by magnetic activity that is strongly coupled to the dynamo power of the star; activity declines with advancing stellar age due to the spin-down via magnetic braking. Its dependence on stellar mass and internal structure allows the probing of dynamos and emission mechanism in multiple stellar populations. Understanding the solar-stellar connection and addressing the potential habitability of exoworlds will be major topics of future astrophysical research. Activity is variable on all timescales and its connection to the surface features enables fruitful multiwavelength studies. The study of high-energy emission from low-mass stars and their flare activity is also addressed in Sciortino et al. (2013).

### 12.1.1. ESO – *Athena* synergies

There are two main topics in the field of low mass stellar studies where synergies between *Athena* and ESO instruments ought to be highlighted:

■ Stellar flares are transient phenomena that evolve quickly (timescales of seconds) and have quite short durations (minutes to hours). To achieve synergies in flare physics, multiwavelength simultaneous observations are required to capture significant individual events (Merloni et al., 2012). Large effective areas and high spectral resolution are important to track the temporal evolution of the flare with sufficient SNR and high sampling rate. The *Athena* X-IFU fulfils the requirements for obtaining emission line flux measurements with high cadence due to its high sensitivity and spectral resolution, thereby providing the coronal part of the data to measure physical properties like temperatures, densities, abundances and velocities of the plasma. Depending on the exact science case, the most suitable current ESO complementary facilities are those that provide high-resolution optical spectra (e.g., VLT/UVES) or medium resolution broad-band spectra (e.g., VLT/X-Shooter). Again, line fluxes and shapes provide the most important diagnostics to determine flare and atmospheric parameters. An even higher temporal sampling can be obtained with the future ELT/HIRES instrument: in this case, short exposure times combined with fast readout on timescale of seconds are mandatory technical requirements.

In flare and coronal studies, e.g., of very-low mass stars and brown dwarfs, radio emission is typically observed at cm wavelength (e.g., VLA, VLBI), where strong gyrosynchrotron emission from non-thermal electrons can be observed. From an ESO instruments perspective the future low frequency bands of ALMA are of largest interest, probably the SKA bandpass is more suited.

- To study the high-energy emission from low-mass stars, multiwavelength campaigns can provide multiple synergies. Strictly simultaneous observations are not always required, but again depending on the science case (near-)simultaneous observations might be preferable due to the often variable nature of the emission. As above, complementary high-resolution and broad-band spectroscopy (VLT/UVES, X-Shooter) is well suited to characterize targets in great detail. Spectral lines provide information on stellar parameters (elemental abundances, gravity, temperatures), they can be used as youth (Li absorption) or activity indicator (H$\alpha$, CaII H+K or IR-triplet), or to measure effective magnetic fields (FeH line broadening). Combined with suitable modelling, surface features like spots can be reconstructed from spectra via Doppler imaging or magnetic field configurations from spectropolarimetry via Zeeman Doppler imaging (ZDI). In the field of spectropolarimetry ELT/HIRES with its high sensitivity will be the instrument of choice, especially for ZDI maps of low-mass stars, since measurements of the full four Stokes parameters are best suited to reconstruct the often complex magnetic fields in these objects.

## 12.2. High energy emission from high mass stars

Massive stars ($M_{initial}$ > 8 $M_\odot$, see Figure 14) peak in the UV and hence produce a copious amount of ionizing photons. This strong radiation drives fast supersonic stellar winds. Feedback from massive stars is among the key astrophysical factors regulating cosmic matter. Massive stars emit X-ray radiation at nearly all stages of their lives. Since the majority are born, live, and die in binary systems, the collision of stellar winds in a binary leads to strong plasma heating and X-ray emission. Moreover, stellar magnetism plays an important role in shaping stellar interiors and winds, and may be responsible for the X-ray emission.

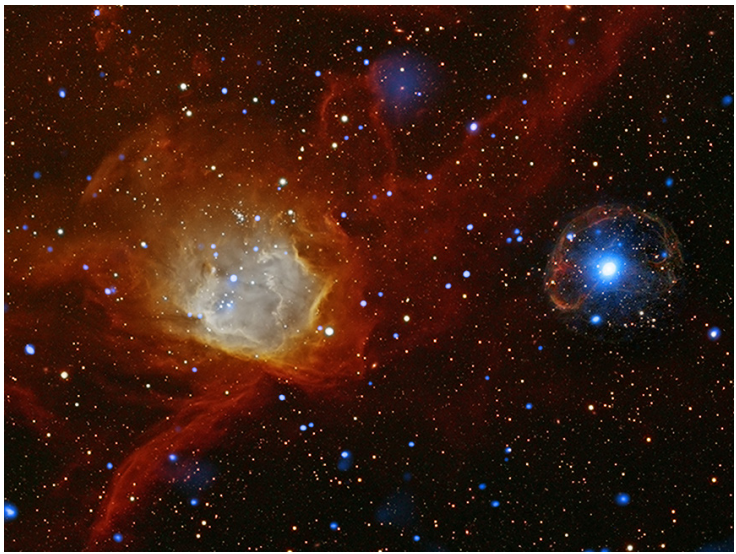

Figure 14: Composite image of a star forming region in the Small Magellanic Cloud. X-rays from *Chandra* and *XMM-Newton* are colored blue and optical data from the Cerro Tololo Inter-American Observatory in Chile are colored red and green. To the left is the young massive star cluster NGC 602 ionizing the HII region N90. To the right is the X-ray pulsar SXP 1062 embedded in the supernova remnant seen as a faint optical shell filled with diffuse X-rays. The numerous field point sources are predominantly background AGN and galaxy clusters. The image size is approximately 14 arcmin. Credits: X-ray: NASA/CXC/Univ.Potsdam/L.Oskinova et al. & ESA/XMM-Newton; Optical: AURA/NOAO/CTIO/Univ.Potsdam/L.Oskinova et al.

## 12.2.1. ESO – *Athena* synergies

Several synergies between *Athena* and ESO can facilitate our understanding of massive stars and their winds. In particular:

- Young massive stars are mainly located in clusters in our own as well as in other galaxies. Often the radiation from massive stars is nearly completely absorbed by the dust surrounding the clusters. Hence the fundamental parameters of massive stars can often just be obtained via observations in the IR and X-rays. Both *Athena* and the ELT can probe stellar feedback in the vicinity of nuclear BHs. For example, more than 30 evolved Wolf-Rayet (WR) massive stars are present within 12´´ of the SMBH in the centre of our Galaxy.

- Massive stars in the Galactic Centre are often in binary systems. X-ray spectra of the gas heated in the collision of their stellar winds are dominated by strong emission lines, including iron lines, and hence can probe the physics of colliding stellar winds (Rauw & Nazé, 2016). In particular, the changing orientation of the wind collision zone during the binary orbital motion is reflected in the changes of emission line shapes. The combination of high effective area and high energy resolution of *Athena* will enable us to record these spectral changes and provide sensitive probes of the inner parts of the colliding wind interaction. Besides producing strong X-ray emission, colliding wind binaries of certain types of WR stars are important cosmic dust makers. The pinwheel dusty outflows from the outer parts of colliding wind regions have been successfully observed in the IR (Tuthill et al., 2006). The synergy between *Athena* and ELT IR imaging cameras will be important to study in real time the interplay between dust formation and high-energy processes in massive star binaries.

- Synergy between IR and X-ray observations has already significantly increased the census of Galactic WR-type stars (Mauerhan et al., 2010; Nebot Gómez-Morán et al., 2015). However, to date, the only high-resolution X-ray spectrum available is for WR6 (Huenemoerder et al., 2015). Its analysis revealed strong broad emission lines that originate in the far out regions of stellar winds. The mechanism responsible for the production of X-rays in WR winds is not yet known. *Athena* will routinely collect high-resolution X-ray spectra of single WR stars, including those that represent the latest evolutionary stage of a massive star prior to its collapse. These objects have the fastest stellar winds among all non-degenerate stars and are X-ray sources (Oskinova et al., 2009). These fast winds and intense stellar radiation should lead to shocks in the nearby molecular clouds and influence star formation. High spatial resolution (< 0.5´´) ALMA observations in the vicinity of young massive stars of molecular species affected by X-ray irradiation (e.g., $HCO^+$) and shocks (e.g., SiO and $CH_3OH$) will be necessary to quantify such influence.

- Stellar winds of hot stars are radiatively driven and intrinsically unstable. Radiative 1-D hydrodynamic numerical models of non-stationary stellar winds predict that large stochastic X-ray variability on time scales of hours should result from strong shocks in stellar winds (Feldmeier et al., 1997). However, no large stochastic X-ray variability is observed in massive stars winds. Instead, all massive stars monitored in X-rays show a low level quasi-periodic variability attributed to the presence of corotating interacting regions in winds of rotating stars (Oskinova et al., 2001; Massa et al., 2014). These provide observational support to the model where X-rays from massive stars originate in large scale corotating interacting regions in their winds (Mullan, 1984).

The synergy between IFUs operating on *Athena* and ESO telescopes will enable us to monitor simultaneously emission lines in the optical and in the X-ray regime for a large sample of stars, e.g., in star clusters, and thus finally solve the mystery of X-ray emission from massive stars. The best emission lines for studying wind variability in hot stars are H$\alpha$ and He II $\lambda$4686 Å. To investigate large scale wind structures a spectral resolution R = 10, 000 is sufficient, while following small wind structures requires R = 50, 000 or higher. Future generations of optical VLT IFUs or ELT instruments will permit such studies.

- Mass-loss rates of massive stars across a wide range of metallicities are needed in order to understand stellar evolution, stellar populations, and their feedback in the early Universe. Moreover, the whole variety of core-collapse SN types and long GRBs, as well as GW events, may be explained only when the "true" mass-loss rates of massive stars are determined. However, at present, our knowledge of massive metal poor stars is severely incomplete. The synergy between X-ray and optical high-resolution spectroscopy provided by *Athena* and the ELT will help to constrain empiric mass-loss rates for stars in galaxies over a broad range of metallicities. The X-ray emission line spectra will be used to derive stellar wind properties, while optical spectra will provide reliable estimates of metallicity. Up to now no X-rays have been observed from single massive stars with metallicity lower than half-solar. *Athena* will obtain spectra of massive stars not only in the metal poor Small Magellanic Cloud but also in other nearby metal poor dwarf galaxies. At the same time, the optical spectra of massive stars in these galaxies will be measured by, e.g., ELT/HIRES. A spectral resolution of 50, 000 - 100, 000 is required to model optical spectra that are rich in metal lines; about 300 lines of C II, C III, C IV, N II, O I, O II, Ne I, Ne II, Mg II, Si III, Si IV, Fe II and Fe III are present in the 3500 Å – 6000 Å spectral range (e.g., Nieva & Przybilla, 2012).

- There is growing evidence that magnetism plays an important role in establishing the structure of the outer stellar layers and wind. Only 10% of massive stars have strong magnetic fields (Hubrig et al., 2013; Fossati et al., 2015). Many (but not all) magnetic massive stars display peculiar X-ray emission. Besides large scale organized field, smaller scale fields with a complex configuration are likely present in massive stars. These fields may play an important role in generating X-rays, but are difficult to measure using current instruments and techniques. Spectropolarimetry with the ELT will allow us to measure weak magnetic field, and use new methods, such as e.g., the Hanle effect (Ignace et al., 2011). It will also provide unprecedented measurements of structures in stellar winds and establish how strong these winds deviate from spherical symmetry. This is especially important for the fast rotating massive stars that likely evolve homogeneously and are the progenitors of massive BHs.

- The discovery of GWs from merging massive BHs poses important questions about the evolution of metal poor massive binary stars. Their evolution is largely determined by rotation and mass-loss via stellar winds. Deep *Athena* observations of metal poor galaxies are needed for the validation of binary evolution models. At later evolutionary stages, potential GW progenitors may host a compact companion, a NS or a BH – these systems will be also discovered by *Athena*. Follow-up spectroscopic studies in the optical, e.g., with ELT/HIRES, will measure radial velocities and chemical composition of massive binaries, elucidating masses and evolutionary status of these objects. The properties of compact companions, their masses and spins, will be obtained from X-ray spectroscopy. The *Athena* and ELT synergy will reveal the nature of ULXs (Section 9.4) and their donor stars, and establish evolutionary links between high mass XRBs, ULXs, and the GW progenitors.

- ELT instruments, such as HIRES, will allow the study of rotational velocities and binarity fraction of massive stars in embedded clusters and in the obscured regions of our own Galaxy (with V < 20 - 22). To study rotation, R > 50, 000 is required in order to resolve the line profiles. Similar requirements are needed for the radial velocity measurements. Because massive stars usually pulsate, the spectral resolution of ELT/HIRES is needed to obtain robust measurements of radial velocities. The radial and rotational velocities of massive stars are best studied in the optical, because in the NIR the spectral lines are formed in the wind and are highly variable. Some of the most important optical lines in this context are the Ca II $\lambda$8500 Å line as well as Paschen H-lines. The synergy between the ELT, able to access massive star populations in obscured regions, and *Athena*, able to penetrate high extinctions, will allow for the first time to obtain a complete census of the massive star population in the outermost regions of the Galaxy, and dramatically improve the sample of these stars towards the Galactic Centre.

To achieve these important synergies, it is crucial that the ELT instruments will have spectral coverage of H$\alpha$ and He II $\lambda$4686 Å lines which are the key stellar wind diagnostic lines in the optical. A spectropolarimeter on the ELT is required to significantly advance the usefulness of ELT for massive star studies. It is also desirable to have the possibility of simultaneous observations between *Athena* and ESO instruments.

# 13. The optical-NIR-sub-mm perspective

We discuss here the view from ESO's side by providing details on VLT/MOONS, the ELT instruments, in particular the MOS, VISTA/4MOST, the ESO and ALMA archives, future ALMA and ESO developments, and finally the (likely) ESO – *Athena* astronomical scene in the 2020s.

## 13.1. VLT/MOONS and ELT instruments

The next generation of highly-multiplexed optical and NIR spectrograph for the VLT called MOONS (Cirasuolo et al., 2014), which should be operational by 2019, will provide a unique capability to follow-up sources observed with *Athena*.

VLT/MOONS will be capable of observing 1024 sources over the largest FoV (500 sq. arcmin) offered by the 8.2m VLT. It will offer the possibility to select between two observing modes: a medium resolution mode, covering simultaneously the 0.64 - 1.8 $\mu$m range with a resolving power R ~ 4,000 - 6,000, mostly foreseen for extragalactic studies, and a high-resolution mode in which the region around the Calcium triplet and part of the H-band are observed at R ~9,000 and at R ~20,000 respectively, mostly designed for detailed chemical abundances in stars.

The medium resolution mode will be particularly suited to follow-up galaxy clusters selected with *Athena*. Thanks to the wide wavelength coverage it will be possible not only to obtain reliable redshift measurements by detecting multiple emission lines, but also to determine the physical properties of cluster members using multiple rest-frame optical line ratios (e.g., the BPT diagram). The ability to extend into the near-IR is also well suited to follow-up X-ray selected AGN, in particular the type 2 AGN which are notoriously weaker (relatively to the host galaxy) in the optical, but relatively bright in the NIR.

While VLT/MOONS will be capable of providing a large-scale follow-up of the *Athena* selected sources, for a more in-depth understanding of the astrophysical properties of these sources we will have to rely on the ELT. The ELT is the new flagship ESO project and with its 39m diameter primary mirror it will be the largest optical/NIR telescope in the world, with first light planned by the end of 2024. The ELT will have a suite of state-of-the art instrumentation which include: MICADO (Davies et al., 2016) a diffraction limited imager with a FoV ~ 1´ with its AO system MAORY (Diolaiti et al., 2016); HARMONI (Thatte et al., 2016) an integral field spectrograph with various spaxel scales (from diffraction to seeing limited); METIS (Brandl et al., 2016) an MIR imager, single slit and high-resolution integral field spectrograph; as well as a second generation multi-object spectrograph MOSAIC (Hammer et al., 2016, see also Section 13.2) and a high-resolution (R ~ 100,000) spectrograph HIRES (Marconi et al., 2016). The collecting power and angular resolution of the ELT combined with such a powerful and versatile suite of instruments will provide detailed follow-up of sources selected by *Athena*. In particular, deep imaging with ELT/MICADO will allow identification of optical/NIR counterparts down to AB ~ 30 mag while ELT/HARMONI and MOSAIC will permit detailed 3D, spatially-resolved spectroscopy to determine their physical and chemical properties.

## 13.2. Synergies between an ELT/MOS and *Athena*

We explore here the possible synergies between a MOS mounted on the ELT (see Figure 15) and *Athena*. We single out this ELT instrument for the reason that current and previous X-ray satellites had clear synergies with MOSs mounted on 8-10m ground-based telescopes, c.f. the spectroscopic follow-up of X-ray sources in the CDFS conducted with VLT/FORS and VIMOS (Szokoly et al., 2004). It is therefore plausible that several synergies will be in place when *Athena* and an ELT/MOS will be in operation.

ESO and the MOSAIC consortium signed in March 2016 a two year Phase-A contract for the study of a MOS for the ELT (Hammer et al., 2016). The current MOSAIC concept aims at using the full 40 arcmin$^2$ FoV of the ELT combining a high-multiplexing mode (≈200 fibers) with a high-definition mode (≈10 IFUs). The science cases of MOSAIC (Evans et al., 2016) have a wide overlap and clear synergies with those of *Athena*.

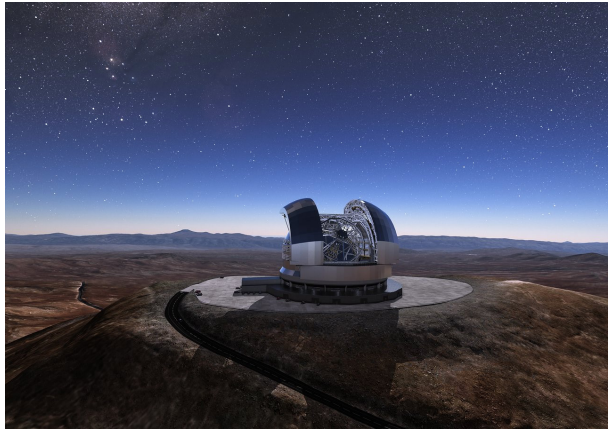

Figure 15: The Cerro Amazones mountain in the Chilean desert, near ESO's Paranal Observatory, will be the site for the ELT, which, with its 39-metre diameter mirror, will be the world's biggest optical/NIR eye on the sky. Here, an artist's rendering shows how the telescope will look on the mountain when it is complete in 2024. Credit: ESO/L. Calçada

We highlight and discuss here a few of them.

- Probing the Epoch of Reionization. The identification and the study of the first galaxies is one of the main goals of the ELT/MOS. At the same time the nature of the sources that produced the UV radiation that reionized the Universe is still elusive. In particular the AGN contribution to the reionization remains unclear. Current studies (e.g., Giallongo et al., 2014) rely on AGN luminosity functions at z > 4 based on photometric redshifts. *Athena* will provide a statistical sample of AGN at z > 6 that we will be able to spectroscopically identify with an ELT/MOS. This will allow us to construct AGN luminosity functions at z > 6 and finally quantifying the contribution that the AGN radiation may had have in reionizing the Universe.

- The environment of z > 6 AGN. In the current model of structure formation the brightest AGN at z > 6 are supposed to be located in the most massive dark matter halos with extended large-scale structures. In a MOSAIC FoV we can have one *Athena* z > 6 AGN candidate and several Lyα emitters at z > 6. We will be therefore able to characterize the environment of such high-z AGN combining these studies with high-redshift galaxy surveys.

- Spectroscopic follow-up of proto-clusters at z > 2. *Athena* will identify proto-cluster candidates at z > 2. An ELT/MOS will be perfectly suited for the spectroscopic follow-up of the cluster members, including passive galaxies down to faint magnitudes ($K_{AB}$ ~ 23; see also Section 2.2).

- *Athena* will provide also large samples of AGN at 2 < z < 6, down to bolometric luminosities of 10$^{44}$ erg s$^{-1}$. A MOSAIC-like MOS will be ideal to:

1. identify large samples of these AGN at faint magnitudes (R> 26 mag) using the ~ 200 multi-plexing;

2. characterize the dynamical properties of the hosts of those AGN using the ~ 10 IFUs. In particular the IFU studies will allow one to extend the current studies of AGN outflows down to lower luminosity (~ $10^{44}$ erg s$^{-1}$) out to z ≈ 4 (see also Section 5).

## 13.3. 4MOST

4MOST (4-metre Multi-Object Spectroscopic Telescope) is a fibre-fed optical spectroscopic survey facility that will be mounted on the VISTA telescope with an expected start of science operations in 2021. It will observe a large fraction of the Southern Hemisphere in a few years. The facility will be able to simultaneously obtain spectra of ~ 2,400 objects distributed over an hexagonal FoV of 4 deg$^2$. The instrument will enable many science goals, but the design is especially intended to complement three key all-sky, space-based observatories of prime European interest: Gaia, EUCLID, and eROSITA, with all the surveys running in parallel. The science requires that, within a 5 year public survey, 20 (goal 30) million targets shall be observed at R ~ 5,000 and 2.0 (goal 3.0) million objects at R ~ 20,000. In this period an area of 15,000 (goal 20,000) deg$^2$ on the sky shall be covered at least twice (goal three times). In particular, VISTA/4MOST is expected to obtain good quality spectra down to r ~ 22 for 500,000 - 700,000 AGN/quasars and up to 1 million galaxies in about 50,000 clusters (z < 0.8 - 1) detected by eROSITA. For fainter sources characterized by the presence of emission lines, the spectra would be of sufficient quality to estimate the redshift.

## 13.4. ESO and ALMA archives

The ESO and ALMA (see Figure 16) science archives, as well as Astronomy as a whole, are seeing major changes as the amount of data and their complexity grow. In what follows, we will attempt a description of the data landscape of 2028 and will draw conclusions for *Athena* as well as for the ESO and ALMA archives.

By the end of the next decade the amount of data held by the ESO and ALMA archives will be enormous. We expect some 14PB/5PB of data and some 200/5 million observations for ESO's Science Archive Facility (SAF) and ALMA, respectively. At that time, and in contrast to today, all of those data will not only be available in raw form, but will be accompanied by science-grade data-products, i.e., fully calibrated and reduced products from which scientific measurements can be made directly.

Not less than 21 instruments will have or will be taking data on the VLT alone, many with AO, covering imaging, spectroscopy, IFU observations, polarimetry, interferometry and MOS. In addition, the other ESO telescopes will have taken surveys (VISTA, VST) and data in the mm/sub-mm regime (APEX, ALMA). And by 2028, the ELT will have been in operation for a while, providing the astronomical community with spectacular and transformational new data.

According to Tony Tyson (LSST Chief Scientist) astronomy is witnessing a transformation from a data-starved science to a science where data is overabundant. Certainly the amount of data will grow exponentially with the current and planned facilities and this will lead to a largely increased amount of "multiwavelength science". In this respect, it would be useful if *Athena* and ESO could look into more collaborations with other observatories to allow for multi-facility observations (see VLT–XMM-*Newton* for a good example).

We argue that the amount of data available to astronomers will be so substantial that the rare resource in astronomy will not be data any more but astronomers themselves. We roughly estimate that the amount of data per astronomer will grow from 70GB/year today (only considering VLT plus ALMA, MAGIC [https://magic.mppmu.mpg.de], HST) to about 1TB/year in 2028 (only considering VLT plus ELT plus ALMA, LSST, CTA, *JWST*). Including the SKA, approximately 180TB/year/astronomer of science data will be available waiting for analysis.

It is clear that by 2028 most of the pixels that get observed will never be looked at by humans. Three main strategies are possible. The first is to take less but higher quality data. Certainly *Athena* is in a very good position here. Secondly, observatories will need to work on producing even higher-level data products, i.e., source properties like light-curves, periods of variable phenomena, chemical composition, temperatures, densities, redshifts, etc. The first such efforts have already been started, e.g., with ALMA's development program (ADMIT) which allows automated post-processing of the science-grade data cubes and detects, identifies and classifies lines. Finally, more and more data analysis will be done using computers directly.

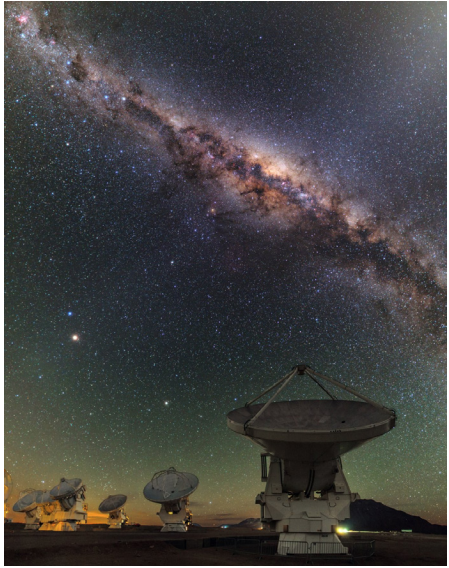

Figure 16: Several of the ALMA antennas under the central regions of the Milky Way. Credit: ESO/B. Tafreshi (twanight.org)

Machine-learning techniques are already being successfully applied to certain science cases, such as the classification of spectra or images. With the revolution in machine learning, in particular "deep learning", many more and novel techniques as well as smart tools will be available to astronomers at the end of the next decade. Such tools could, for example, automatically find scaling relation in the LSST data and uncover physical laws.

Following this trend where vast amounts of very high-level data are available, it can be expected that in 2028 many observational astronomers will not have written a single proposal. Archival astronomy is already a significant contribution to the scientific output of modern facilities (https://archive.stsci.edu/hst/bibliography/pubstat.html) but will be much more so in the future. This again puts larger responsibility on the observatories. Whereas today astronomers in most cases still can be considered being co-producers of the data, by 2028 they will mostly have transformed into consumers of data products provided by the observatories. And while the positive aspect of this is that they will then be able to concentrate on their science exclusively, it comes with the risk that astronomers will no longer understand the limitations and subtleties of the data well enough.

Observatories therefore will have to invest significant effort into providing data-reduction and data-analysis tools that are of such high quality that they can be fully trusted. In addition, far larger efforts than today on documentation as well as on user-support will be required. While the standards for both ESO, ALMA, and *Athena* are already relatively high in this respect, even more will be required in order to stay competitive in the world of 2028.

This in turn indicates that the cost ratio for data management versus the total cost of the facilities is very likely to rise. An extreme example certainly is LSST where 52% of the 1.25 billion dollar total survey cost are reserved for data management. Proper measures would need to be taken within ESO, ALMA, and *Athena* to prepare for this evolution, including raising the awareness with management and funding agencies but more importantly by providing the means to make the design and delivery of data reduction pipelines and data analysis as well as science archives an integral part of the design of each new instrument and facility right from the start.

Page 57

The likely evolution of the increased amounts of high-level data and increased demand by archival astronomers means for science archives at ESO–ALMA or for *Athena* that the challenge will be to change the mind-set from offering search only on properties of the observations (integration time, filter/band) to physical properties of the sources (fluxes, lines, object-types etc). It is expected that by 2028 data models, data discovery and data services will all be done through International Virtual Observatory Alliance standards and protocols. Although the Virtual Observatory layer will power all the services, astronomers might not even realise this fact as the tools they will be using will abstract the technical layer away from them.

As a conclusion, and while the evolution outlined above poses interesting challenges to facilities like ESO, ALMA, and *Athena*, the future for astronomers and astronomy as a whole is extremely bright with much facilitated and accelerated scientific discoveries.

This also puts *Athena* in a very different position as compared to previous X-ray missions like *Chandra* and XMM-*Newton*. Namely, while the latter had to secure the multiwavelength data necessary for the identification and astrophysical interpretation of the X-ray data after launch, *Athena* will have a treasure trove of observations already available in the ESO (and other) archives.

## 13.5. ALMA developments

ALMA, the most powerful mm/sub-mm interferometer in the world, is producing compelling results in all fields of astronomy and has already enhanced our knowledge in thematics ranging from protoplanetary disk science (e.g., ALMA partnership, 2015b, and other recently submitted works) to the high redshift Universe (e.g., ALMA partnership, 2015a), or chemistry of complex organic molecules (e.g., Öberg et al., 2015; Oya et al., 2016) just to name a few, while other breakthroughs are on the way, such as resolving the event horizon of nearby SMBHs. There is a number of possible developments currently being discussed (Bolato et al., 2015) with impact on the ALMA - *Athena* synergies, that could materialize in the late 2020s, namely:

- Larger bandwidths and better sensitivities. The capability to provide and process wider instantaneous bandwidths, together with better receiver sensitivity, can increase significantly the speed of ALMA observations. This will also improve the possibility of simultaneous line observations for variability and high accuracy line ratio studies. Enlarging the bandwidth will increase enormously the legacy value of the archive while, at the same time, enhancing the probability of serendipitous discoveries. Examples of scientific areas that will benefit from such a development are spectral scans of high-redshift galaxies, studies of the molecular complexity of disks, line surveys, as well as the study of the outer Solar System.

- Longer baselines. ALMA recently demonstrated the potential of >10 km baselines in mm interferometry by producing amazing images of, e.g., the SDP81 gravitational lens (ALMA partnership, 2015a) and the HL Tau protostellar disk (ALMA partnership, 2015b). Doubling the designed maximum baseline (16 km) to 32 km will provide an angular resolution of ~8 mas at 230 GHz (the size of the photosphere of $\alpha$ Centaurus A), equivalent to a resolution of 1 AU at a distance of ~140 pc or a resolution of a few tens of pc in the higher-redshift Universe. Longer baselines will contribute to advances in the imaging of disks on resolutions reaching closer to the scales of the habitable zone, in probing the centres of active galaxies using masers or in studying the processes that lead to mass loss in evolved stars.

- Increasing wide field mapping speed. One of the important limitations of ALMA is the small FoV (~1´ at Band 3, decreasing with $\nu^{-1}$), as determined by the diameter and the primary beam of the antennas. Enlarging the FoV, e.g., via focal-plane arrays, will enable faster wide field mapping, an improvement which large imaging and spectral surveys as well as studies of nearby galaxies, galactic molecular clouds and star-forming regions will greatly benefit from.

- Developments of the ALMA archive. The legacy of an observatory is its archive (see Section 13.4) and in order for an archive to be useful, it needs to be public, user-friendly, easy to mine, and it has to contain reduced, science-ready data, as well as automatically generated or post-publication user-submitted value-added products. ALMA data carry a wealth of information (high sensitivity, position, frequency information, large bandwidth), and have the potential for great science to be carried out, far more than what was suggested in the original proposals. Such an example are the lines beyond the targeted transitions that are detected and mapped. Future features, such as visualizations tools that will allow one to quickly browse the various datasets, will make the ALMA archive a powerful and easy-to-use tool that will enable new science and enhance the synergies with other archives. The projections for the ALMA and ESO archives around the time of *Athena* launch have been discussed in detail in Section 13.4.

Last but not least, a large single-dish telescope with cameras suitable for fast large-scale mapping, even though outside the scope of the envisioned ALMA development plans, would be an important scientific complement to ALMA and of great importance to the ESO - *Athena* synergies, as also outlined throughout this paper (e.g., in Section 2.3.3 for the SZ effect and in Section 3.5.2 on mapping the cool gas on large scales)[3].

## 13.6. Further ESO developments

Finally, it is worth mentioning that ESO is starting to consider the case for a large aperture (10 - 12m) optical spectroscopic survey telescope with a FoV comparable to that of the LSST (which is 9.6 deg$^2$) and a large multiplex (a few thousand)[4]. This facility would obviously be of great interest for all the *Athena* science cases in synergy with a MOS discussed in this White Paper. We refer the reader to de Zeeuw (2016) for a comprehensive view of ESO's long term perspective.

## 13.7. The ESO – *Athena* astronomical scene in the 2020s

To have a broad look at science areas where the synergetic use of the ESO facilities and *Athena* in the late 2020s might result in scientific added value, we took two complementary approaches by:

1. checking the science Top Level Requirements (TLRs[5]) for the three ELT instruments under construction (MICADO, HARMONI, and METIS) and the two, which are in Phase A (MOS and HIRES);

2. looking over the key science cases discussed in the Science Priorities at ESO document (ESO/STC-551, Section 4.3[6]), which discusses the likely astronomical landscape in the 2020s[7]. The overlap is quite significant, as the science cases in common with those discussed in this White Paper are:

■ ELT/MICADO: Galactic Centre, extragalactic transients, resolved structure and physical properties of high-redshift galaxies;

■ ELT/HARMONI: IMBHs, GRBs and their hosts, physics of high redshift galaxies;

■ ELT/METIS: extragalactic transients, AGN and the growth of SMBHs;

■ ELT/HIRES: IGM, extragalactic transients;

---

3 See the ESO Submm Single Dish Scientific Strategy Working Group Report available at https://www.eso.org/public/about-eso/committees/stc/stc-87th/public/STC-567_ESO_Submm_Single_Dish_Scientific_Strategy_WG_Report_87th_STC_Mtg.pdf

4 See the ESO Future of Multi-Object Spectroscopy Working Group Report available at https://www.eso.org/public/about-eso/committees/stc/stc-88th/public/STC_579_MOS_WG_Report_88th_STC_Meeting.pdf.

5 These are available at https://www.eso.org/sci/facilities/eelt/instrumentation/

6 Available at http://www.eso.org/public/about-eso/committees/stc/stc-85th/public/STC-551_Science_Priorities_at_ESO_85th_STC_Mtg_Public.pdf.

7 Both of these approaches have obviously to be taken with a grain of salt, given the notorious difficulty in predicting the astronomical landscape more than ten years in advance as testified, for example, by the discoveries made by HST, many of which were not only unpredicted but also unexpected.

- ELT/MOS: primordial galaxies and the reionization of the Universe, IGM tomography;

- ESO in the 2020s: large-scale structure of the Universe, structure and evolution of galaxies (including AGN), life cycle of interstellar matter, life cycle of stars, extreme states of matter, time-domain astronomy.

In short, ESO and *Athena* at this point in time appear poised to tackle many similar astronomical topics at the end of the next decade, which obviously will be beneficial to both.

# 14. Summary

Modern astronomy is a multiwavelength enterprise and will be even more so in the future (see, e.g., Padovani, 2016, for a forward look for radio astronomy, including the SKA, and the prominent role *Athena* will play). *Athena* will be no exception. This White Paper has highlighted the many synergies between *Athena* and the ESO facilities, current and future, which we summarize here[8]:

- Early groups and clusters

    » spectroscopic surveys with VISTA/4MOST, VLT/MOONS, and ELT/MOS: clusters of galaxies (Section 2.1)

    » spectroscopic surveys with VISTA/4MOST, VLT/MOONS, and ELT/MOS; imaging, galaxy SEDs, and stellar masses with ELT/MICADO and HARMONI; galaxy morphology and mergers with ELT/METIS; metal enrichment and gas in galaxies with ELT/HARMONI, ELT-MOS and VLT/MOONS; SFRs with ELT/METIS and HARMONI; molecular and ionized gas outflows with ELT/HARMONI; diffuse Ly$\alpha$ gas with either ELT/HARMONI or ELT/MICADO; shock tracers with ELT/METIS and HARMONI: proto-cluster formation (Section 2.2)

    » SZ mapping with ALMA: the SZ effect in galaxy clusters (Section 2.3.1)

- Physics of the ICM

    » SZ mapping with ALMA: combining X-ray and SZ data (Section 3.1)

    » spectra of heavy element isotopes to measure ICM velocities with ALMA: gas velocities (Section 3.2)

    » measurements of the cold gas velocities of the ICM in CO with ALMA and H$\alpha$ with VLT/X-Shooter: filaments (Section 3.3)

    » NIR observations of warm molecular gas with VLT–ELT; observations of cold molecular gas (CO) with ALMA; observations of ionized gas with VLT/MUSE; spectra of NIR ro-vibrational lines of molecular Hydrogen with ELT/HARMONI; BH masses of BCGs through AO-assisted IFU observations: molecular gas (Section 3.5.1)

- Missing baryons in cosmic filaments

    » redshift surveys at ±500 km s$^{-1}$ with respect to the filament's redshift with VLT/MOONS and VISTA/4MOST: physics of the IGM (Section 4.1)

    » SZ mapping with ALMA; spectra of rare isotopes and molecules associated with the WHIM cold interface with ALMA; deep UV/OIR imaging and IFU spectroscopy with ELT/HARMONI of galaxies; detection of atomic and molecular outflows with ALMA and ELT/HARMONI: observational studies of the WHIM (Section 4.2)

---

8 When reference is made to a current ESO instrument, possibly not available by the *Athena* launch, this should be read as, e.g., VLT/SINFONI-like instrument.

- SMBH history

    » characterization of galaxy properties with ELT/MICADO and ELT/METIS; observations of cold molecular gas (CO) with ALMA; characterization of the host galaxy kinematics with ELT/HARMONI and ELT/MOS: galaxy evolution and SMBH growth (Section 5.1.1)

    » spectroscopic surveys with VISTA/4MOST and ALMA; IFU spectra of CT and $z > 6$ AGN with VLT/MUSE and ELT/HARMONI: AGN identification (Section 5.2.1)

    » observations of the cold dust, obscured SFRs, and molecular gas mass and dynamics with ALMA; redshifts and BH masses with the ELT; stellar masses and SFRs through AO-assisted NIR and MIR imaging with ELT/MICADO and ELT/METIS: SMBHs at high redshift (Section 5.3)

- SMBH accretion disks

    » AGN BH mass estimates up to $z \sim 0.1$ with ELT NIR observations at the diffraction limit and through reverberation mapping with VLT/MOONS at larger distances; spectroscopic surveys with VLT/MOONS or ELT/MOS: BH masses (Section 6.2)

    » optical/UV (rest-frame) MOS follow-up of large X-ray AGN samples, extending into the NIR with, e.g., VLT/MOONS, to reach $z >1$; time-resolved optical/UV spectra with VLT/X-Shooter and ELT/HARMONI: AGN SEDs (Section 6.3)

- AGN feedback – molecular outflows

    » ionized gas mass estimates from the H$\alpha$ luminosity with VLT/SINFONI; gas excitation determination with VLT/MUSE, gas dynamics with VLT/SINFONI; shock tracing through the NIR H$_2$ line and IFU H$\alpha$ observations: ESO – *Athena* synergies (Section 7.1)

    » spectra to measure molecular abundances with ALMA: ALMA – *Athena* synergies (Section 7.2)

    » detection of atomic and molecular outflows with ALMA; observations of ionized outflows with VLT/SINFONI, ELT/HARMONI and MOS: high redshift synergies (Section 7.3)

- Ultra-fast outflows

    » connection between cold and hot AGN flows with VLT/ELT spectra (X-Shooter for nearby sources and HARMONI at higher redshifts); observations of cold molecular gas with ALMA; spatially resolved observations of the ionized atomic phase with VLT/SINFONI and MUSE; exploration of the presence of extended emission, bubbles and galaxy-scale winds with ALMA, ELT/MICADO, HARMONI, and MOS: ESO – *Athena* synergies (Section 8.2)

- Accretion Physics

    » simultaneous mm–IR–X-ray observations (APEX–ALMA–VLT–ELT) of XRBS and SMBHs; monitoring campaigns with *Athena* and ESO (8m or even 4m class) IR facilities; fast-timing observations on ELT/MICADO: jets (Section 9.1.1)

    » simultaneous X-ray, OIR, and (sub-)mm SED fitting of XRBs; ALMA observations of water maser sources: winds (Section 9.2.1)

    » high-resolution spectroscopic observations of XRBs with *Athena* and ELT/HARMONI; spec-

tral continuum measurements with ELT/HARMONI and ALMA: jet-disk-wind connection (Section 9.3.1)

» time-resolved OIR spectroscopy with ELT/HARMONI; observations of OIR counterparts of ULXs with ELT/MICADO: ULXs (Section 9.4.1)

- Transient Science

» OIR photometry and spectroscopy with ELT/MICADO and ELT/HARMONI; IR data with ELT/MICADO; polarization observations with VLT/FORS2 and, if feasible, on the ELT: GRB studies (Section 10.1.1)

» absorption system studies with ELT/HARMONI or HIRES; host galaxy studies with ELT/HARMONI; imaging with ELT/MICADO: high redshift GRBs (Section 9.2.1)

» deep imaging and spectroscopy follow-up with ELT/MICADO and VLT/FORS2 or with IFUs such as VLT/MUSE and ELT/HARMONI for sources with bad source localisation: TDEs (Section 9.3.1)

- Star Formation

» observations of $N_2H^+$ and $HCO^+$ and mapping ion emission on small scales with ALMA: star formation (Section 10.1.1)

» observations of $N_2H^+$, $HCO^+$, $DCO^+$, $H_2D^+$ with ALMA (simultaneous with *Athena*); high spatial resolution observations with ALMA: protoplanetary disks (Section 10.2.1)

» continuum and CO observations with the upgraded ALMA and high spectral resolution IR observations with ELT/METIS and HIRES of nearby star forming regions; high-resolution spectroscopy of planetary atmospheres with ELT/HIRES: YSOs (Section 10.3.1)

- Stars

» high-resolution optical spectra (e.g., VLT/UVES) or medium resolution broad-band spectra (e.g., VLT/X-Shooter); spectropolarimetry with ELT/HIRES: low mass stars (Section 11.1.1)

» high spatial resolution observations with ALMA of $HCO^+$, SiO, and $CH_3OH$; simultaneous X-ray– optical IFU observations with R = 10, 000 - 50, 000 (H$\alpha$ and He II $\lambda$4686 Å); high-resolution optical spectra with ELT/HIRES in the 3500 Å – 6000 Å range; spectropolarimetry with ELT: high mass stars (Section 11.2.1)

To all of the above we can also add the fact that the ESO archives will be filled with observations relevant to the interpretation of *Athena* data well before its launch (Section 13.4).

To summarize, the ESO facilities, which are most needed to exploit the synergies with *Athena* include, in approximate ranking order, IFUs (i.e., VLT/MUSE and ELT/HARMONI), ALMA, multi-object spectrographs (i.e., VISTA/4MOST, VLT/MOONS, and ELT/MOS), NIR imagers (mostly ELT/MICADO), and high-resolution spectrographs (i.e., VLT/UVES and ELT/HIRES).

Finally, two requirements on ESO facilities, which are still missing or under discussion have come out of this White Paper:

1. the need for a single dish (sub-)mm telescope (i.e., 40 - 50m) equipped with a wide FoV ( ~10´) photometric camera operating at least at ~2 mm, and, optimally, simultaneously at ~1 mm and ~850 $\mu$m (Section 2.3.3 and 3.5.2);

2. a polarimetric facility at the ELT (Section 10.1.1, 12.1.1, 12.2.1).

# 15. References

ALMA partnership; Vlahakis, C., Hunter, T. R., Hodge, J. A. et al. 2015a, ApJ, 808, L4

ALMA partnership; Brogan, C. L., Perez, L. M., Hunter, T. R. et al. 2015b, ApJ, 808, L3

Aalto, S., Muller, S., Sakamoto, K. et al. 2012, A&A 546, A68

Abbott, B. P., Abbott, R., Abbott, T. D., et al. 2016, Phys.Rev.D., 93, 122003

Adam, R., Comis, B., Macías-Pérez, J. F., et al. 2014, A&A, 569, A66

Adam, R., Comis, B., Macías-Pérez, J.-F., et al. 2015, A&A, 576, A12

Adam, R., Bartalucci, I., Pratt, G. W., et al. 2016, arXiv:1606.07721

Aird, J., Comastri, A., Brusa, M., et al. 2013, arXiv:1306.2325

Alexander, R., Pascucci, I., Andrews, S., Armitage, P., & Cieza, L. 2014, Protostars and Planets VI, 475

Ansdell, M., Williams, J. P., van der Marel, N. et al. 2016, ApJ, 828, 46

Antonucci, R., & Barvainis, R. 1988, ApJL, 332, L13

Armitage, P. J. 2011, ARA&A, 49, 195

Ballantyne, D. R. 2005, MNRAS, 362, 1183

Ballantyne, D. R. 2007, Modern Physics Letters A, 22, 2397

Bañados, E., Venemans, B. P., Decarli, R., et al. 2016, arXiv:1608.03279

Bandyopadhyay, R. M., Shahbaz, T., Charles, P. A., & Naylor, T. 1999, MNRAS, 306, 417

Basu, K., Sommer, M., Erler, J., et al. 2016, ApJL, 829, L23

Bauer, A. H., Baltay, C., Ellman, N., et al. 2012, ApJ, 749, 56

Bayliss, M. B., Ashby, M. L. N., Ruel, J., et al. 2014, ApJ, 794, 12

Behar, E., Baldi, R. D., Laor, A., et al. 2015, MNRAS, 451, 517

Bleem, L. E., Stalder, B., de Haan, T., et al. 2015, ApJS, 216, 27

Boirin, L., Méndez, M., Díaz Trigo, M., Parmar, A. N., & Kaastra, J. S. 2005, A&A, 436, 195

Bolato, A., Corder, S., Iono, D., Testi, L., Wootten, A. 2015, Pathways to developing ALMA

Bouvier, J., Grankin, K., Ellerbroek, L. E., Bouy, H., & Barrado, D. 2013, A&A, 557, A77

Branchini, E., Ursino, E., Corsi, A., et al. 2009, ApJ, 697, 328

Brandl, B. R., Agócs, T., Aitink-Kroes, G., et al. 2016, Proceedings of the SPIE, 9908, 990820

Brandt, W. N., Alexander, D. M., 2015, A&ARv, 23,1

Brenneman, L. W., & Reynolds, C. S. 2006, ApJ, 652, 1028

Brusa, M., Zamorani, G., Comastri, A., et al. 2007, ApJS, 172, 353

Brusa, M., Civano, F., Comastri, A., 2010, ApJ,716, 348

Budavári, T., Szalay,. A.S., 2008, ApJ, 679,301

Cano-Diaz, M., Maiolino, R., Marconi, A. et al. 2012, A&A 537, L8

Cappi, M., Done, C., Behar, E., et al. 2013, arXiv:1306.2330

Carniani, S., Marconi, A., Maiolino, R. et al. 2016, A&A 591, A28

Casares, J., & Jonker, P. G. 2014, SpaceScienceRev., 183, 223

Casella, P., Maccarone, T. J., O'Brien, K., et al. 2010, MNRAS, 404, 21

Ceccarelli, C., Dominik, C., López-Sepulcre, A., et al. 2014, ApJL, 790, L1

Cecil, G., Bland-Hawthorn, J., & Veilleux, S. 2002, ApJ, 576, 745

Chartas, G., Brandt, W. N., Gallagher, S. C., & Garmire, G. P. 2002, ApJ, 579, 169

Chartas, G., Saez, C., Brandt, W. N., Giustini, M., & Garmire, G. P. 2009, ApJ, 706, 644

Chatzikos, M., Ferland, G. J., Williams, R. J. R., & Fabian, A. C. 2014, ApJ 787, 96

Churazov, E., Sunyaev, R., Gilfanov, M., Forman, W., & Jones, C. 1998, MNRAS 297, 1274

Churazov, E., Forman, W., Jones, C., & Böhringer, H. 2000, A&A 356, 788

Church, S., Philhour, B., Runyan, M. C., et al. 1997, The Far Infrared and Submillimetre Universe., 401, 169

Ciaravella, A., Cecchi-Pestellini, C., Chen, Y.-J., et al. 2016, ApJ, 828, 29

Cicone, C., Maiolino, R., Sturm, E. et al. 2014, A&A 562, A21

Cirasuolo, M., Afonso, J., Carollo, M., et al. 2014, Proceedings of the SPIE, 9147, 91470N

Cleeves, L. I., Bergin, E. A., & Adams, F. C. 2014, ApJ, 794, 123

Colbert, E. J. M., & Mushotzky, R. F. 1999, ApJ, 519, 89

Combes, F., Garcia-Burillo, S., Casasola, V. et al. 2013, A&A 558, A124

Combes, F., García-Burillo, S., Casasola, V., et al. 2014, A&A, 565, A97

Croston, J. H., Sanders, J. S., Heinz, S., et al. 2013, arXiv:1306.2323

Dalgarno, A. 2006, Proceedings of the National Academy of Science, 103, 12269

Dasyra, K. M., Bostrom, A. C., Combes, F., Vlahakis, N. 2015, ApJ 815, 34

David, L. P., Lim, J., Forman, W., et al. 2014, ApJ 792, 94

Davies, R., Schubert, J., Hartl, M., et al. 2016, Proceedings of the SPIE, 9908, 99081Z

Désert, F.-X., Benoit, A., Gaertner, S., et al. 1998, New Astron., 3, 655

de Zeeuw, T. 2016, The Messenger, 166, 2 (arXiv:1701.01249)

Díaz Trigo, M., & Boirin, L. 2016, Astronomische Nachrichten, 337, 368

Díaz Trigo, M., Parmar, A. N., Boirin, L., Méndez, M., & Kaastra, J. S. 2006, A&A, 445, 179

Díaz Trigo, M., Migliari, S., Miller-Jones, J. C. A., & Guainazzi, M. 2014, A&A, 571, A76

Dicker, S. R., Abrahams, J. A., Ade, P. A. R., et al. 2006, Proceedings of the SPIE, 6275, 62751B

Diolaiti, E., Ciliegi, P., Abicca, R., et al. 2016, Proceedings of the SPIE, 9909, 99092D

Dovciak, M., Matt, G., Bianchi, S., et al. 2013, arXiv:1306.2331

Dutson, K. L., Edge, A. C., Hinton, J. A. et al. 2014, MNRAS 442, 2048

Dwelly, T., Salvato, M., Merloni, A., et al. 2017, MNRAS, in prep

Ebisuzaki, T., Makino, J., Tsuru, T. G., et al. 2001, ApJL, 562, L19

Ercolano, B., Rosotti, G. P., Picogna, G., & Testi, L. 2017, MNRAS, 464, L95

Ettori, S., Pratt, G. W., de Plaa, J., et al. 2013, arXiv:1306.2322

Evans, C.J., Puech, M., Rodrigues, M., et al. 2016, arXiv:1608.06542

Fabian, A. C. 2012, ARAA 50, 455

Falcke, H., Körding, E., & Markoff, S. 2004, A&A, 414, 895

Feldmeier, A., Kudritzki, R.-P., Palsa, R., Pauldrach, A. W. A., & Puls, J. 1997, A&A, 320, 899

Feigelson, E., Townsley, L., Güdel, M., & Stassun, K. 2007, Protostars and Planets V, 313

Feruglio, C., Maiolino, R., Piconcelli, E. et al. 2010, A&A 518, L155

Feruglio, C., Fiore, F., Maiolino, R., et al. 2013, A&A, 549, A51

Feruglio, C., Fiore, F., Carniani, S., et al. 2015, A&A, 583, A99
Fiore, F., Feruglio, C., Shankar, F., et al. arXiv:1702.04507
Flock, M., Ruge, J. P., Dzyurkevich, N., et al. 2015, A&A, 574, A68
Fossati, L., Castro, N., Schöller, M., et al. 2015, A&A, 582, A45
Fukumura, K., Kazanas, D., Contopoulos, I., & Behar, E. 2010, ApJ, 715, 636
Gaibler, V., Khochfar, S., Krause, M., Silk, J. 2012, MNRAS 425, 438
Galeev, A. A., 1979, ApJ, 229,318
Gandhi, P., Makishima, K., Durant, M., et al. 2008, MNRAS, 390, L29
Garcia-Burillo, S., Combes, F., Usero, A. et al. 2014, A&A, 567, 125
Georgakakis, A., Carrera, F., Lanzuisi, G., et al. 2013, arXiv:1306.2328
Giallongo, E., Grazian, A., Fiore, F., et al. 2014, A&A, 578, A83
Giardino, G., Favata, F., Micela, G., Sciortino, S., & Winston, E. 2007, A&A, 463, 275
Glassgold, A. E., Galli, D., & Padovani, M. 2012, ApJ, 756, 157
Glennie, A., Jonker, P. G., Fender, R. P., Nagayama, T., & Pretorius, M. L. 2015, MNRAS, 450, 3765
Gobat, R., Daddi, E., Onodera, M., et al. 2011, A&A, 526, A133
Gofford, J., Reeves, J. N., McLaughlin, D. E., et al. 2015, MNRAS, 451, 4169
Goosmann, R. W., Czerny, B., Mouchet, M., et al. 2006, A&A, 454, 741
Graham, A. W., Scott, N., 2013, ApJ, 764, 151
Grainge, K., Jones, M., Pooley, G., et al. 1996, MNRAS, 278, L17
Grandi, P., Urry, C. M., & Maraschi, L. 2002, New Astronomy Reviews, 46, 221
Greenhill, L., Kondratko, P. T., Lovell, J. E. J., et al. 2003, ApJ, 582, L11
Guelin, M., Langer, W. D., Snell, R. L., & Wootten, H. A. 1977, ApJL, 217, L165
Guillard, P., Boulanger, F., Pineau Des Forêts, G., & Appleton, P. N. 2009, A&A, 502, 515
Guo, Y., Ferguson, H. C., Giavalisco, M., 2013, ApJS, 207, 24
Hagiwara, Y., Henkel, C., Menten, K. M. & Nakai, N. 2001, ApJ, 560, L37
Hagiwara, Y., Miyoshi, M., Doi, A. & Horiuchi, S. 2013, ApJL, 768, L38
Haisch, K. E., Jr., Lada, E. A., & Lada, C. J. 2001, ApJL, 553, L153
Hamer, S., Salomé, P., Combes, F., & Salomé, Q. 2015, A&A, 575, L3
Hammer, F., Morris, S., Kaper, L., et al. 2016, Proceedings of the SPIE, 9908, 990824
Hasselfield, M., Hilton, M., Marriage, T. A., et al. 2013, J. Cosmology Astropart. Phys., 7, 008
Heida, M., Jonker, P. G., Torres, M. A. P., et al. 2014, MNRAS, 442, 1054
Heida, M., Torres, M. A. P., Jonker, P. G., et al. 2015, MNRAS, 453, 3510
Heida, M., Jonker, P. G., Torres, M. A. P., et al. 2016, MNRAS, 459, 771
Heinz, S., & Sunyaev, R. A. 2003, MNRAS, 343, L59
Hinshaw, G., Larson, D., Komatsu, E., et al. 2013, ApJS, 208, 19
Hogan, M. T., 2014, PhD thesis, University of Durham
Hogan, M. T., Edge, A. C., Hlavacek-Larrondo, J., et al. 2015a, MNRAS, 453, 1201
Hogan, M. T., Edge, A. C., Geach, J. E. et al. 2015b, MNRAS 453, 1223
Holzapfel, W. L., Arnaud, M., Ade, P. A. R., et al. 1997, ApJ, 480, 449
Hori, T., Ueda, Y., Shidatsu, M., et al. 2014, ApJ, 790, 20
Hubrig, S., Schöller, M., Ilyin, I., et al. 2013, A&A, 551, A33
Huenemoerder, D. P., Gayley, K. G., Hamann, W.-R., et al. 2015, ApJ, 815, 29
Ignace, R., Hole, K. T., Cassinelli, J. P., & Henson, G. D. 2011, A&A, 530, A82
Ilbert, O., Lauger, S., Tresse, L., et al. 2006, A&A, 453, 809
Inoue, Y., & Doi, A. 2014, PASJ, 66, L8
Jin, C., Ward, M., Done, C., Gelbord, J., 2012, MNRAS, 420, 1825
Jonker, P., O'Brien, P., Amati, L., et al. 2013a, arXiv:1306.2336
Jonker, P. G., Glennie, A., Heida, M., et al. 2013b, ApJ, 779, 14
Kaaret, P., Ward, M. J., & Zezas, A. 2004, MNRAS, 351, L83
Kaastra, J., Finoguenov, A., Nicastro, F., et al. 2013, arXiv:1306.2324
Kalamkar, M., Casella, P., Uttley, P., et al. 2016, MNRAS, 460, 3284
Kama, M., López-Sepulcre, A., Dominik, C., et al. 2013, A&A, 556, A57
Khatri, R., & Gaspari, M. 2016, MNRAS, 463, 655
Khedekar, S., Churazov, E., Kravtsov, A., et al. 2013, MNRAS 431, 954
Khedekar, S., Churazov, E., Sazonov, S., Sunyaev, R., & Emsellem, E. 2014, MNRAS 441, 1537
King, A. R., & Pounds, K. A. 2015, ARAA, 53, 115
Kitayama, T., Komatsu, E., Ota, N., et al. 2004, PASJ, 56, 17
Kitayama, T., Ueda, S., Takakuwa, S., et al. 2016, PASJ, 68, 88
Komatsu, E., Matsuo, H., Kitayama, T., et al. 2001, PASJ, 53, 57
Kormendy, J., Ho, L. C., 2013, ARA&A, 51, 511
Korngut, P. M., Dicker, S. R., Reese, E. D., et al. 2011, ApJ, 734, 10
Lamarre, J. M., Giard, M., Pointecouteau, E., et al. 1998, ApJL, 507, L5
LaMassa, S., Urry, C. M., Cappelluti, N., et al. 2016, ApJ, 817,172
Laor, A., & Behar, E. 2008, MNRAS, 390, 847
Lindner, R. R., Aguirre, P., Baker, A. J., et al. 2015, ApJ, 803, 79
Longinotti, A. L., Krongold, Y., Guainazzi, M., et al. 2015, ApJl, 813, L39
Lorenzani, A., & Palla, F. 2001, From Darkness to Light: Origin and Evolution of Young Stellar Clusters, 243, 745
Luo, B., Brandt, W. N., Xue, Y. Q., et al. 2010, ApJS, 187, 560
Luo, B., Brandt, N., & Bauer, F. 2014, The Astronomer's Telegram, 6541
Lusso, E., Risaliti, G., 2016, ApJ, 819, 154
Ma, Q., Maio, U., Ciardi, B., & Salvaterra, R. 2016, arXiv:1610.03594
MacLeod, M., Guillochon, J., Ramirez-Ruiz, E., Kasen, D., & Rosswog, S. 2016, ApJ, 819, 3
Madau, P., Dickinson, M., 2014, ARA&A, 52, 415
Maiolino, R., Haehnelt, M., Murphy, M. T., et al. 2013, arXiv:1310.3163
Malzac, J. 2013, MNRAS, 429, L20
Malzac, J. 2014, MNRAS, 443, 299
Mantz, A. B., Abdulla, Z., Carlstrom, J. E., et al. 2014, ApJ, 794, 157
Marchesi, S., Civano, F., Elvis, M., 2016, ApJ, 817, 34
Marconi, A., Di Marcantonio, P., D'Odorico, V., et al. 2016, Proceedings of the SPIE, 9908, 990823
Martens, P. C. H., & Kuin, N. P. M. 1989, Solar Physics, 122, 263
Massa, D., Oskinova, L., Fullerton, A. W., et al. 2014, MNRAS, 441, 2173
Mauerhan, J. C., Muno, M. P., Morris, M. R., Stolovy, S. R., & Cotera, A. 2010, ApJ, 710, 706
McGaugh, S. S., Schombert, J. M., de Blok, W. J. G., & Zagursky, M. J. 2010, ApJL, 708, L14

McGuire, J. T. W., Tanvir, N. R., Levan, A. J., et al. 2016, ApJ, 825, 135

McNamara, B. R., Russell, H. R., Nulsen, P. E. J., et al. 2014, ApJ 785, 44

Menanteau, F., Hughes, J. P., Sifón, C., et al. 2012, ApJ, 748, 7

Merloni, A., Heinz, S., & di Matteo, T. 2003, MNRAS, 345, 1057

Merloni, A., Predehl, P., Becker, W., et al. 2012, arXiv:1209.3114

Middleton, M. et al. 2016, New Astronomy Reviews, submitted

Monfardini, A., Swenson, L. J., Bideaud, A., et al. 2010, A&A, 521, A29

Morganti, R., Oosterloo, T., Oonk, J. B. R., Frieswijk, W., & Tadhunter, C. 2015, A&A, 580, A1

Morganti, R., Veilleux, S., Oosterloo, T., et al. 2016, A&A 593, A30

Motch, C., Pakull, M. W., Soria, R., Grisé, F., & Pietrzyński, G. 2014, Nature, 514, 198

Mullan, D. J. 1984, ApJ, 283, 303

Muñoz-Darias, T., Casares, J., Mata Sánchez, D., et al. 2016, Nature, 534, 75

Nandra, K., Barret, D., Barcons, X., et al. 2013, arXiv:1306.2307

Nandra K., Laird, E. S., Aird, J. A., et al. 2015, ApJ, 220, 10

Nardini, E., Wang, J, Fabbiano, G et al. 2013, ApJ 765, 141

Nardini, E., Reeves, J. N., Gofford, J., et al. 2015, Science, 347, 860

Naylor, T., Broos, P. S., Feigelson, E. D. 2013, ApJS, 209, 30

Nebot Gómez-Morán, A., Motch, C., Pineau, F.-X., et al. 2015, MNRAS, 452, 884

Neilsen, J., & Lee, J. C. 2009, Nature, 458, 481

Neufeld, D. A., Goicoechea, J. R., Sonnentrucker, P., et al. 2010, A&A, 521, L10

Nesvadba, N. P. H., Lehnert, M. D., De Breuck, C., Gilbert, A. M., & van Breugel, W. 2008, A&A, 491, 407

Newman, A. B., Ellis, R. S., Andreon, S., et al. 2014, ApJ, 788, 51

Nieva, M.-F., & Przybilla, N. 2012, A&A, 539, A143

Nord, M., Basu, K., Pacaud, F., et al. 2009, A&A, 506, 623

Öberg, Karin I., Guzman, Viviana V., Furuya, Kenji et al. 2015, Nature, 520, 198

Ogle, P. M., Davis, S. W., Antonucci, R. R. J., et al. 2005, ApJ, 618, 139

Orienti, M., D'Ammando, F., Giroletti, M., Giovannini, G., & Panessa, F. 2015, Advancing Astrophysics with the Square Kilo- metre Array (AASKA14), 87

Oskinova, L. M., Clarke, D., & Pollock, A. M. T. 2001, A&A, 378, L21

Oskinova, L. M., Hamann, W.-R., Feldmeier, A., Ignace, R., & Chu, Y.-H. 2009, ApJL, 693, L44

Oya, Y., Sakai, N., Lopez-Sepulcre, A. et al. 2016, ApJ, 824, 88

Padovani, P. 2016, A&AR, 24, #13

Panessa, F., & Giroletti, M. 2013, MNRAS, 432, 1138

Pfrommer, C., Enßlin, T. A., & Sarazin, C. L. 2005, A&A 430, 799

Pineau, F.-X., Derriere, S., Motch, C., et al. 2016, A&A

Pinto, C., Middleton, M. J., & Fabian, A. C. 2016, Nature, 533, 64

*Planck* Collaboration, Ade, P. A. R., Aghanim, N., et al. 2013a, A&A, 550, A131

*Planck* Collaboration, Ade, P. A. R., Aghanim, N., et al. 2013b, A&A, 550, A134

*Planck* Collaboration, Ade, P. A. R., Aghanim, N., et al. 2016, A&A, 594, A27

Plotkin, R. M., Markoff, S., Kelly, B. C., Körding, E., & Anderson, S. F. 2012, MNRAS, 419, 267

Pointecouteau, E., Giard, M., Benoit, A., et al. 2001, ApJ, 552, 42

Pointecouteau, E., Hattori, M., Neumann, D., et al. 2002, A&A, 387, 56

Pointecouteau, E., Reiprich, T. H., Adami, C., et al. 2013, arXiv:1306.2319

Polko, P., Meier, D. L., & Markoff, S. 2014, MNRAS, 438, 959

Ponti, G., Fender, R. P., Begelman, M. C., et al. 2012, MNRAS, 422, 11

Pounds, K. A., Reeves, J. N., King, A. R., et al. 2003, MNRAS, 345, 705

Proga, D., & Kallman, T. R. 2004, ApJ, 616, 688

Prokhorov, D. A., Antonuccio-Delogu, V., & Silk, J. 2010, A&A 520, A106

Raginski, I., & Laor, A. 2016, MNRAS, 459, 2082

Rahoui, F., Coriat, M., & Lee, J. C. 2014, MNRAS, 442, 1610

Rauw, G., & Nazé, Y. 2016, A&A, 594, A82

Reeves, J. N., Sambruna, R. M., Braito, V., & Eracleous, M. 2009, ApJl, 702, L187

Reeves, J. N., Gofford, J., Braito, V., & Sambruna, R. 2010, ApJ, 725, 803

Rodríguez-Baras, M., Rosales-Ortega, F. F., Díaz, A. I., Sánchez, S. F., & Pasquali, A. 2014, MNRAS, 442, 495

Ruppin, F., Adam, R., Comis, B., et al. 2016, arXiv:1607.07679

Russell, D. M., Fender, R. P., 2008, MNRAS, 387, 713

Russell, D. M., Fender, R. P., Hynes, R. I., et al. 2006, MNRAS, 371, 1334

Russell, T. D., Soria, R., Miller-Jones, J. C. A., et al. 2014, MNRAS, 439, 1390

Russell, H. R., McNamara, B. R., Fabian, A. C., et al. 2016, MNRAS 458, 3134

Salvaterra, R., Campana, S., Covino, S., et al. 2011, Swift and the Surprising Sky: The First Seven Years of Swift. Online at: http://www.brera.inaf.it/docM/OAB/Research/SWIFT/Swift7/?p=program

Salvaterra, R., Maio, U., Ciardi, B., & Campisi, M. A. 2013, MNRAS, 429, 2718

Salvato, M., Buchner, J., Budavari, T., et al. 2017, A&A in prep
Sambruna, R. M., Reeves, J. N., Braito, V., et al. 2009, ApJ, 700, 1473

Schwan, D., Kneissl, R., Ade, P., et al. 2012, The Messenger, 147, 7

Sciortino, S., Rauw, G., Audard, M., et al. 2013, arXiv:1306.2333

Shidatsu, M., Done, C., & Ueda, Y. 2016, ApJ, 823, 159

Shields, G. A., McKee, C. F., Lin, D. N. C., & Begelman, M. C. 1986, ApJ, 306, 90

Shull, J. M., Smith, B. D., & Danforth, C. W. 2012, ApJ, 759, 23

Sikora, M., Stawarz, Ł., & Lasota, J.-P. 2007, ApJ, 658, 815 Silk, J., & Rees, M. J. 1998, A&A, 331, L1

Sturm, E., Gonzalez-Alfonso, E., Veilleux, S. et al. 2011, ApJ 733, L16

Sunyaev, R. A., & Churazov, E. M. 1984, Soviet Astronomy Letters, 10, 201

Sunyaev, R. A., & Docenko, D. O. 2007, Astronomy Letters, 33, 67

Sutherland, W., Saunders, W., 1992, MNRAS, 259, 413

Szokoly, G.P., Bergeron, J., Hasinger, G., et al. 2004, ApJS, 155, 271

Tanvir, N. R., Levan, A. J., Fruchter, A. S., et al. 2012, ApJ, 754, 46

Tchernin, C., Eckert, D., Ettori, S., et al. 2016, arXiv:1606.05657

Tepper-García, T., Richter, P., Schaye, J., et al. 2012, MNRAS, 425, 1640

Thatte, N. A., Clarke, F., Bryson, I., et al. 2016, Proceedings of the SPIE, 9908, 99081X

Tinker, J., Kravtsov, A. V., Klypin, A., et al. 2008, ApJ, 688, 709-728

Tombesi, F., Cappi, M., Reeves, J. N., et al. 2010, A&A, 521, A57

Tombesi, F., Tazaki, F., Mushotzky, R. F., et al. 2014, MNRAS, 443, 2154

Tombesi, F., Melendez, M., Veilleux, S. et al. 2015, Nature 519, 436

Torresi, E., Grandi, P., Longinotti, A. L., et al. 2010, MNRAS, 401, L10

Torresi, E., Grandi, P., Costantini, E., & Palumbo, G. G. C. 2012, MNRAS, 419, 321

Treister, E., Bauer, F., & Schawinski, K. 2014, The Astronomer's Telegram, 6603

Tremblay, G. R., Oonk, J. B. R., Combes, F., et al. 2016, Nature 534, 218

Trenti, M., Perna, R., Levesque, E. M., Shull, J. M., & Stocke, J. T. 2012, ApJL, 749, L38

Tsuneta, S. 1997, ApJ, 483, 507

Tuthill, P., Monnier, J., Tanner, A., et al. 2006, Science, 313, 935

Ueda, Y., Murakami, H., Yamaoka, K., Dotani, T., & Ebisawa, K. 2004, ApJ, 609, 325

Ueda, Y., Yamaoka, K., & Remillard, R. 2009, ApJ, 695, 888

Valentino, F., Daddi, E., Strazzullo, V., et al. 2015, ApJ, 801, 132

Valentino, F., Daddi, E., Finoguenov, A., et al. 2016, ApJ, 829, 53

Veilleux, S., Meléndez, M., Sturm, E., et al. 2013, ApJ, 776, 27

Walton, D. J., Nardini, E., Fabian, A. C., Gallo, L. C., & Reis, R. C. 2013, MNRAS, 428, 2901

Wang, J.-M., Watarai, K.-Y., Mineshige, S., 2004 ApJ, 607, 107

Wang, T., Elbaz, D., Daddi, E., et al. 2016, ApJ, 828, 56

Wiersema, K., Covino, S., Toma, K., et al. 2014, Nat, 509, 201

Wu, K., Soria, R., Hunstead, R. W., & Johnston, H. M. 2001, MNRAS, 320, 177

Yamada, K., Kitayama, T., Takakuwa, S., et al. 2012, PASJ, 64, 102

Zhuravleva, I., Churazov, E. M., Schekochihin, A. A., et al. 2014, ApJL, 788, L13

Zubovas, K., & Nayakshin, S., 2014, MNRAS 440, 2625

# 16. Acronym list

| | |
|---|---|
| 4MOST | 4-metre Multi-Object Spectrograph Telescope (VISTA/ESO) |
| ACT | Atacama Cosmology Telescope |
| ADMIT | ALMA Data Mining Toolkit |
| AGN | Active Galactic Nuclei (RQ/RL AGN are Radio-Quiet/Radio-Loud AGN; LLAGN are Low Luminosity AGN) |
| ALMA | Atacama Large Millimetre/submillimetre Array |
| AO | Adaptive Optics |
| APEX | Atacama Pathfinder Experiment |
| ASKAP EMU | Australian SKA Pathfinder/Evolutionary Map of the Universe |
| ATCA | Australia Telescope Compact Array |
| *Athena* | Advanced Telescope for High ENergy Astrophysics |
| BCG | Brightest Cluster Galaxy |
| BH | Black Hole |
| BL Lacs | BL Lacertae objects, a class of AGN |
| CANDELS | Cosmic Assembly Near-infrared Deep Extragalactic Legacy Survey |
| CCD | Charge Coupled Device |
| CDFS | Chandra Deep Field South |
| CGM | Circum-galactic medium |
| CMB | Cosmic Microwave Background |
| CR | Cosmic Ray |
| CTA | Cherenkov Telescope Array |
| ELT | Extremely Large Telescope |
| EAST | ESO-*Athena* Synergy Team |
| eROSITA | extended ROentgen Survey with an Imaging Telescope Array |
| ESA | European Space Agency |
| ESO | European Southern Observatory |
| EW | Equivalent Width |
| FORS2 | FOcal Reducer/low dispersion Spectrograph 2 (VLT/ESO) |
| FoV | Field of View |
| FUV | Far Ultra-Violet |
| FWHM | Full-width at Half Maximum |
| GBT | Giant Binocular Telescope |
| GRB | Gamma-Ray Burst |
| GROND | Gamma-Ray Burst Optical/Near-Infrared Detector |
| GTC | Gran Telescopio Canarias |
| HARMONI | High Angular Resolution Monolithic Optical and Near-infrared Integral field spectrograph (ELT/ESO) |
| HAWK-I | High Acuity Wide field K-band Imager (VLT/ESO) |
| HIRES | HIgh REsolution Spectrograph (ELT/ESO) |
| HSC | Hyper Suprime-Cam (Subaru) |
| HST | Hubble Space Telescope |
| ICM | Intra-cluster medium |
| IFU | Integral Field Unit |
| IGM | Inter-Galactic Medium |
| IR | Infrared |
| IRAM | Institut de Radioastronomie Millimétrique |
| ISAAC | Infrared Spectrometer And Array Camera (VLT/ESO) |
| JVLA | Jansky Very Large Array |
| JWST | James Webb Space Telescope |
| KMOS | K-band Multi Object Spectrograph (VLT/ESO) |
| LABOCA | LArge BOlometer Camera (APEX/ESO) |
| LEdd | Eddington luminosity |
| LIGO | Laser Interferometer Gravitational-Wave Observatory |
| LINER | Low-Ionization Nuclear Emission-line Region |
| LSST | Large Synoptic Survey Telescope |
| MAORY | Multi-conjugate Adaptive Optics RelaY (ELT/ESO) |
| METIS | Mid-infrared ELT Imager and Spectrograph (ELT/ESO) |
| MHD | MagnetoHydroDynamics |
| MICADO | Multi-AO Imaging Camera for Deep Observations (ELT/ESO) |
| MIR | Mid-IR |
| MOONS | Multi Object Optical and Near-infrared Spectrograph (VLT/ESO) |
| MOS | Multi-object spectroscopy |
| MRS | Maximum Recoverable Scale |
| MUSE | Multi Unit Spectroscopic Explorer (VLT/ESO) |
| MUSTANG | The MUltiplexed SQUID/TES Array at Ninety GHz (GBT) |
| NIKA | New IRAM KID Arrays (IRAM) |
| NIR | Near-Infrared |
| NOEMA | NOrthern Extended Millimeter Array (Plateau de Bure, IRAM) |
| NS | Neutron Star |
| NTT | New Technology Telescope (ESO) |
| PCS | Planetary Camera and Spectrograph (ELT/ESO) |
| PESSTO | Public ESO Spectroscopic Survey for Transient Objects |
| PSF | Point Spread Function |
| RRM | Rapid Response Mode |
| RXTE | Rossi X-ray Timing Explorer |
| SED | Spectral Energy Distribution |
| SFR | Star Formation Rate |
| SINFONI | Spectrograph for INtegral Field Observations in the Near Infrared (VLT/ESO) |
| SKA | Square Kilometre Array |
| SMBH | Super Massive Black Hole |
| SN | Supernova |
| SNR | Signal-to-Noise Ratio |
| SOXS | Son Of X-Shooter |
| SPT | South Polar Telescope |
| SZ | Sunyaev-Zel'dovich |
| TDE | Tidal Disruption Event |
| THESEUS | Transient High Energy Sky and Early Universe Surveyor (a mission proposed to ESA) |
| UFO | Ultra-Fast Outflow |
| ULIRG | Ultra-Luminous IR Galaxy |
| VISTA | Visible and Infrared Survey Telescope for Astronomy (ESO) |
| VLBA | Very Long Baseline Array |
| WFI | Wide Field Imager |
| WFIRST | Wide Field Infrared Survey Telescope (a NASA mission) |
| WHIM | Warm-Hot Intergalactic Medium |
| WMAP | Wilkinson Microwave Anisotropy Probe |
| WODAN | Westerbork Observations of the Deep APERTIF Northern sky |
| WSRT | Westerbork Synthesis Radio Telescope |
| X-IFU | X-ray Integral Field Unit |
| XMM-Newton | X-ray Multi-Mirror Mission |

XRB ................. X-Ray Binary
XSPEC ............. X-Ray Spectral Fitting Package
YSO.................. Young Stellar Object
ZDI................... Zeeman Doppler imaging
ZTF .................. Zwicky Transient Facility